\documentclass[%
prx
,twocolumn%
,amsmath,amssymb,aps,nobibnotes
,notitlepage
,nofootinbib
,preprintnumbers
]{revtex4-1}

\usepackage[colorlinks,linkcolor=blue,citecolor=blue]{hyperref}
\usepackage[dvipdfmx]{graphicx}
\usepackage{bm}
\usepackage{braket}
\usepackage{color}
\usepackage{tabularx}
\usepackage{mathtools}

\newcommand{\omat}{p}
\newcommand{\intb}{\int_0^{\beta}}
\newcommand{\vd}{\hat{\mathcal{V}}}
\newcommand{\itime}{x}

\begin{document}
\title{Functional Renormalization Group Approach to Circuit Quantum Electrodynamics}

\author{Takeru Yokota}
\email{takeru.yokota@riken.jp}
\affiliation{Interdisciplinary Theoretical and Mathematical Sciences Program (iTHEMS), RIKEN, Wako, Saitama 351-0198, Japan}
\affiliation{Institute for Solid State Physics, The University of Tokyo, Kashiwa, Chiba 277-8581, Japan}

\author{Kanta Masuki}
\email{masuki@g.ecc.u-tokyo.ac.jp}
\affiliation{Department of Physics, University of Tokyo, 7-3-1 Hongo, Bunkyo-ku, Tokyo 113-0033, Japan}

\author{Yuto Ashida}
\email{ashida@phys.s.u-tokyo.ac.jp}
\affiliation{Department of Physics, University of Tokyo, 7-3-1 Hongo, Bunkyo-ku, Tokyo 113-0033, Japan}
\affiliation{Institute for Physics of Intelligence, University of Tokyo, 7-3-1 Hongo, Tokyo 113-0033, Japan}

\date{\today}

\preprint{RIKEN-iTHEMS-Report-22}

\begin{abstract}
A nonperturbative approach is developed to analyze superconducting circuits coupled to quantized electromagnetic continuum within the framework of the functional renormalization group. The formalism allows us to determine complete physical pictures of equilibrium properties in the circuit quantum electrodynamics (cQED) architectures with high-impedance waveguides, which have recently become accessible in experiments. We point out that nonperturbative effects can trigger breakdown of the supposedly effective descriptions, such as the spin-boson and boundary sine-Gordon models, and lead to qualitatively new phase diagrams. The origin of the failure of conventional understandings is traced to strong renormalizations of circuit parameters at low-energy scales. 
Our results indicate that a nonperturbative analysis is essential for a comprehensive understanding of cQED platforms consisting of superconducting circuits and long high-impedance transmission lines. 
\end{abstract}

\maketitle

\section{Introduction\label{sec:intro}}
A quantum many-body state arising from strong light-matter interaction holds the promise of exploring fundamental physics and advancing quantum information technologies. In these decades, there have been significant ongoing efforts for achieving and understanding unprecedentedly strong coupling regimes in a variety of fields, including quantum optics \cite{WA04,BLS09,AO10,HIC12,RD13,TJD13,vLA13,AMS14,YR14,MJA14,GA14,LP15,MM19,KB20,ZH10,GTA13,PB13,TS18,PR18,SBE19,NDD22},  condensed matter physics \cite{JC17,KS18,RS18,GA18,KJ20,CT20,MNS20,RM14,SJ15,HD17,SMA18,SF19,CJB19,RV19,MG19,YA20,PP20,SL21,DO21,YA21,CA21,Basovaag1992,BJ19,FJGV21,SF22,BJ22}, and quantum chemistry \cite{HJA12,GJ15,FJ15,HF16,FJ17,TA19,DM22}. Circuit quantum electrodynamics (cQED) provides a controllable platform to study interactions between an artificial atom consisting of Josephson junction (JJ) and quantized electromagnetic environment at microwave frequencies \cite{BA21,CAA20}. Recent experiments have realized the strong coupling between a nonlinear superconducting circuit and the continuum of photons propagating in a transmission line with a controllable impedance, pushing the light-matter coupling to new regimes \cite{FDP17,MJP19,LS19,KR19,PM20,KR21,LS22}. These remarkable developments have led to renewed recent interest in the field of quantum dissipative systems or, equivalently, bosonic quantum impurity systems \cite{Goldstein13,SBE14,SI15,LJ18,ML18,KLK18,GN18,HM20,YT21,BAm21,KK21,GGC21,YA22,KM22,Murami20,*Hakonen21,*Murani21}.

Understanding of bosonic quantum impurity systems has long been one of the central problems in quantum many-body physics \cite{Caldeira81,*Caldeira83a,*Caldeira83b,Leggett87,Kane92,Furusaki93,Affleck01,GT03,Werner05a,Oshikawa06,RG07,HUR20082208,Halperin11_2,Weiss12,KA19}. Two of the most common settings are the spin-boson model and the boundary sine-Gordon model, where a two-level system or a quantum particle subject to cosine potential is coupled to the bosonic environment represented as a collection of harmonic oscillators, respectively.  Early perturbative analyses have suggested the existence of quantum phase transitions, namely,  the localization-delocalization transition \cite{anderson_exact_1970,GuF85,Leggett87} or the superconductor-insulator transitions \cite{Schmid83,Bulgadaev84,Guinea85,Fisher85,Schon90}. These predictions were initially thought to be of direct relevance to recent cQED architectures realizing long high-impedance waveguides.

However, an accurate interpretation of those previous results in the context of cQED has so far remained a major challenge. The main reason for this is that it requires a careful investigation of the validity of several nontrivial simplifications which have often been made implicit in the literature. For instance, the spin-boson model could be derived in cQED settings only if one can safely replace the superconducting circuit (e.g., a fluxonium circuit) by the simplified, two-level system. Similarly, the boundary sine-Gordon model could provide an effective description of cQED systems only if a certain capacitance term can be neglected (see below). Crucially, the validity of these simplifying assumptions becomes ambiguous especially in nonperturbative regimes that have been accessible in recent experiments. Indeed, a qualitative modification from the Schmid-Bulgadaev diagram \cite{Schmid83,Bulgadaev84} due to nonperturbative effects has been recently reported \cite{KM22}.   
To comprehensively understand physics of cQED architectures emulating quantum impurity systems, it is thus highly desirable to develop a nonperturbative theoretical framework that avoids those ambiguities.

\begin{figure}[t]
  \begin{center}
    \includegraphics[width=1\columnwidth]{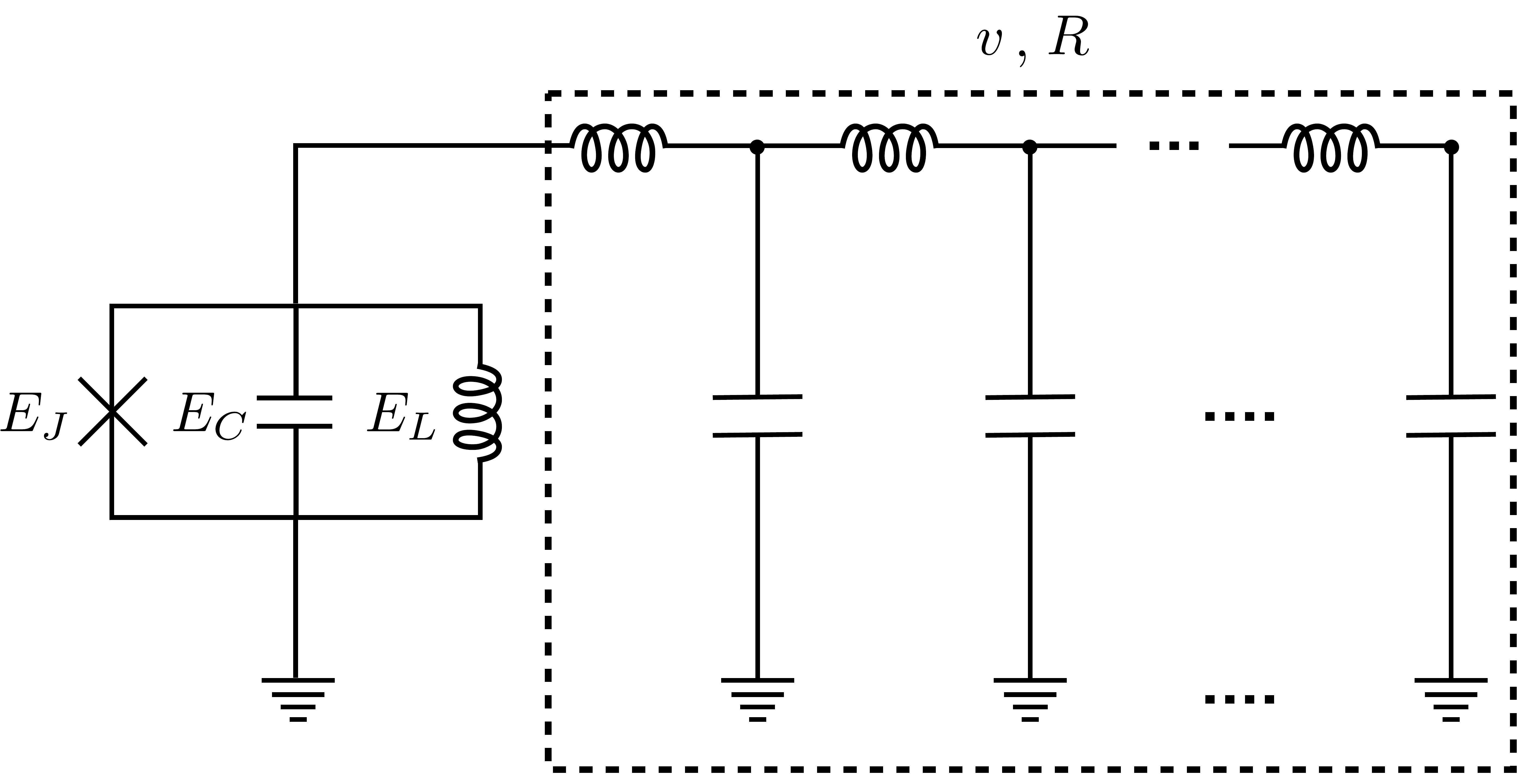}
    \caption{Schematic of a cQED architecture analyzed in this paper. A generic superconducting circuit with arbitrary Josephson energy $E_J$, charging energy $E_{\rm C}$, and inductive energy $E_L$ interacts with the continuum of microwave photons propagating through a long transmission line with the wave velocity $v$ and the impedance $R$ (dashed box).}
        \label{fig:circuit}
 \end{center}
\end{figure}

The aim of this paper is to develop a functional renormalization group (FRG) approach to cQED architectures and reveal their complete physical pictures including previously unexplored regimes.  Our nonperturbative framework provides a unified way to analyze a generic JJ circuit coupled to transmission line, in which an arbitrary phase potential can be included (Fig.~\ref{fig:circuit}). Importantly, we achieve this without resorting to simplifying approximations whose validity is not obvious in nonperturbative regimes. Indeed, by applying the present framework to several concrete examples,  we demonstrate that many of those simplifications cannot be justified due to nonperturbative effects. In particular, our analysis leads to the phase diagrams that are strikingly different from the ones predicted in the supposedly effective models, such as the spin-boson or boundary sine-Gordon models.  These results demonstrate that a nonperturbative analysis is crucial for comprehensively understanding the physics of recent cQED platforms.

The remainder of the paper is organized as follows. In Sec.~\ref{sec:frg}, we present an FRG framework for a JJ interacting with a one-dimensional continuum of photons propagating in a transmission line. In Sec.~\ref{sec:double}, we apply our general framework to the case of a JJ with double-well phase potential. We demonstrate the breakdown of the two-level approximation due to nonperturbative effects and determine the complete phase diagram which is modified from the one predicted in the spin-boson model. In Sec.~\ref{sec:cosine}, we next analyze the case of cosine potential corresponding to the problem of resistively shunted JJ. We point out the crucial role of the capacitance term, which has been overlooked in previous studies that rely on the boundary sine-Gordon model, and present the qualitatively modified phase diagram from the one originally predicted by Schmid and Bulgadaev. In Sec.~\ref{sec:conc}, we give a summary of results and suggest several interesting future directions.

\section{General formalism \label{sec:frg}}
This section presents an FRG formalism of cQED platforms recently realized in experiments, namely, a JJ coupled to a long microwave transmission line (see, e.g., Refs.~\cite{FDP17,MJP19,LS19,KR19,PM20,KR21,LS22}). The transmission waveguide itself typically consists of a superconducting chain of a large number of JJs, where low-energy degrees of freedom can be represented as a collection of harmonic oscillators. The JJ then acts as a quantum impurity particle that is subject to dissipation caused by the bosonic environment and moves through a nonlinear potential corresponding to Josephson energy. Specifically, the circuit Hamiltonian is given by
\begin{align}
	\label{eq:genham}
	\hat{H}
	=&
	E_{\rm C}
	\left(
	\hat{N}
	-
	\hat{n}_r
	\right)^2
	+
	V(\hat{\varphi})
	+
	\sum_{0<k\leq W/v}
	\omega_k 
	\hat{a}_k^\dagger
	\hat{a}_k,
	\\
	\hat{n}_r
	=&
	\frac{\sqrt{\gamma}}{2\pi}
	\sum_{0<k\leq W/v}
	\sqrt{\frac{2\pi}{kL}}
	\left(
	\hat{a}_k^\dagger
	+
	\hat{a}_k
	\right),
\end{align}
in units of $\hbar=1$. Here, $\hat{\varphi}$ ($\hat{N}$) is the JJ phase (charge) operator and $\hat{a}_k$ ($\hat{a}_k^\dagger$) is the bosonic annihilation (creation) operator of environmental mode $k$, which satisfy the following commutation relations:
\begin{align}
	\left[\hat{\varphi},\hat{N}\right]=i,\quad 
	\left[\hat{a}_k,\hat{a}_{k'}^\dagger\right]=\delta_{k,k'}.
\end{align}
The environmental frequency of mode $k=m\pi/L$ ($m=1,2,\ldots,M$) is given by $\omega_k=vk$ and $W=vM\pi/L$ is the frequency cutoff. The constants $v$ and $L$ represent the wave velocity and the length of the transmission line, respectively. The coupling strength is characterized by the dimensionless parameter $\gamma=R_Q/R$ with the quantum of resistance $R_Q=h/(4e^2)$ and the waveguide impedance $R$. 
The charging energy is denoted by $E_{\rm C}$, and the potential $V(\hat{\varphi})$ represents any phase-dependent terms such as Josephson energy $-E_J\cos(\hat{\varphi})$ and/or inductive energy $E_L\hat{\varphi}^2/2$. While our formalism developed in this section can be applied to a generic potential,  we will later focus on two important cases, namely,  double-well and cosine potentials, which are directly relevant to cQED experiments. 

\subsection{Path-integral formalism}
The path-integral representation of the partition function \cite{feynman_quantum_2010} is useful to describe the formalism of FRG. For the sake of completeness, we here briefly explain how the cQED system described by the Hamiltonian~\eqref{eq:genham} can be represented in the path-integral formalism. We then derive its effective action by integrating out the environmental degrees of freedom. 

To this end, we start from the partition function at temperature $T$,
\begin{align}
	Z=\sum_{i}\Braket{i|e^{-\beta \hat{H}}|i},
\end{align}
where $\lbrace\Ket{i}\rbrace$ are the eigenstates of $\hat{H}$ and $\beta=1/T$. To switch to the path-integral representation, we first decompose the operator $e^{-\beta \hat{H}}$ into $\mathcal{N}$ pieces of operator $\hat{K}_{\mathcal{N}}=\exp\left(-\delta\itime\hat{H}\right)$ with $\delta\itime =\beta/\mathcal{N}$:
\begin{align}
	Z=\sum_{i}\Braket{i|
	\underbrace{
	\hat{K}_{\mathcal{N}}
	\hat{K}_{\mathcal{N}}
	\cdots
	\hat{K}_{\mathcal{N}}
	}_{\mathcal{N}}
	|i}.
\end{align}
We next insert the following identity operator
after each $\hat{K}_{\mathcal{N}}$:
\begin{widetext}
\begin{align}
	\label{eq:id}
	\hat{1}
	=
	\int \frac{d\varphi^{(n)} dN^{(n)}}{2\pi}
	\prod_k
	\int
	\left(
	\frac{da^{(n)}_k da^{*(n)}_k}{\pi}
	\right)
	e^{i\varphi^{(n)} N^{(n)}-\sum_k a^{*(n)}_k a^{(n)}_k}
	\Ket{N^{(n)},a^{(n)}}\Bra{\varphi^{(n)},a^{(n)}}
	\quad
	(n=1,\ldots,\mathcal{N}).
\end{align}
Here, we have introduced
\begin{align*}
	\Ket{N^{(n)},a^{(n)}}
	=
	\Ket{N^{(n)}}\otimes\Ket{a^{(n)}},
	\quad
	\Ket{\varphi^{(n)},a^{(n)}}
	=
	\Ket{\varphi^{(n)}}\otimes\Ket{a^{(n)}},
\end{align*}
where $\Ket{\varphi^{(n)}}$ ($\Ket{N^{(n)}}$)
is an eigenstate of $\hat{\varphi}$ ($\hat{N}$),
and $\Ket{a^{(n)}}$ is an unnormalized coherent state of the bosons, i.e., they satisfy
\begin{align*}
	\hat{\varphi}\Ket{\varphi^{(n)}}
	=
	\varphi^{(n)}\Ket{\varphi^{(n)}},
	\quad
	\hat{N}
	\Ket{N^{(n)}}
	=
	N^{(n)}\Ket{N^{(n)}},
	\quad
	\hat{a}_k \Ket{a^{(n)}}
	=
	a_k^{(n)} \Ket{a^{(n)}}.
\end{align*}
The identity \eqref{eq:id} is obtained from the following completeness relations on the Hilbert spaces of the Josephson phase and the bosonic environment  \cite{altland_condensed_2010}:
\begin{align*}
	\hat{1}_{\varphi}
	=&
	\int d\varphi^{(n)}
	\Ket{\varphi^{(n)}}\Bra{\varphi^{(n)}}
	=
	\int d\varphi^{(n)}
	\int \frac{dN^{(n)}}{2\pi}
	\Ket{N^{(n)}}
	\Braket{N^{(n)}|\varphi^{(n)}}
	\Bra{\varphi^{(n)}}
	=
	\int \frac{d\varphi^{(n)} dN^{(n)}}{2\pi}e^{i\varphi^{(n)} N^{(n)}}\Ket{N}
	\Bra{\varphi},
	\\
	\hat{1}_{\rm env}
	=&
	\prod_k
	\int
	\left(
	\frac{da_k^{(n)} da_k^{*(n)}}{\pi}
	\right)
	e^{-\sum_k a_k^{*(n)} a_k^{(n)}}
	\Ket{a}\Bra{a}.
\end{align*}
Then the partition function $Z$ is rewritten as
\begin{align}
	\label{eq:zN}
	Z
	=&
	\int \mathcal{D}\varphi_{\mathcal{N}}
	\int \mathcal{D}N_{\mathcal{N}}
	\int \mathcal{D}(a,a^{*})_{\mathcal{N}}
	e^{\sum_{n=1}^{\mathcal{N}}i\varphi^{(n)} N^{(n)}
	-
	\sum_{n=1}^{\mathcal{N}}
	\sum_k a^{*(n)}_k a^{(n)}_k}
	\notag
	\\
	&\times
	\sum_{i}\Braket{i|
	\hat{K}_{\mathcal{N}}|N^{(\mathcal{N})},a^{(\mathcal{N})}}
	\left(
	\prod_{i=1}^{\mathcal{N}-1}
	K_{\mathcal{N}}(\varphi^{(i+1)},N^{(i)},a^{*(i+1)},a^{(i)})
	\right)
	\Braket{
	\varphi^{(1)},a^{(1)}
	|i}
	\notag
	\\
	=&
	\int \mathcal{D}\varphi_{\mathcal{N}}
	\int \mathcal{D}N_{\mathcal{N}}
	\int \mathcal{D}(a,a^{*})_{\mathcal{N}}
	\prod_{n=1}^{\mathcal{N}}
	\left(
	e^{-i\varphi^{(n)} N^{(n)}-\sum_k a^{*(n)}_k a^{(n)}_k}
	K_{\mathcal{N}}(\varphi^{(n+1)},N^{(n)},a^{*(n+1)},a^{(n)})
	\right),
\end{align}
where we have introduced the boundary conditions
\begin{align}
	\label{eq:bound}
	\varphi^{(\mathcal{N}+1)}=\varphi^{(1)},\quad
	N^{(\mathcal{N}+1)}=N^{(1)},\quad
	a^{(\mathcal{N}+1)}=a^{(1)},\quad
	a^{*(\mathcal{N}+1)}=a^{*(1)},
\end{align}
and adopted the following notation:
\begin{align}
	K_{\mathcal{N}}(\varphi,N,a^{*},a)
	=&
	\Braket{\varphi,a|\hat{K}_{\mathcal{N}}|N,a},
	\\
	\int \mathcal{D}\varphi_{\mathcal{N}}
	\int \mathcal{D}N_{\mathcal{N}}
	\int \mathcal{D}(a,a^{*})_{\mathcal{N}}
	=&
	\left(
	\prod_{n=1}^{\mathcal{N}}
	\int d\varphi^{(n)}
	\right)
	\left(
	\prod_{n=1}^{\mathcal{N}}
	\int \frac{dN^{(n)}}{2\pi}
	\right)
	\left(
	\prod_{n=1}^{\mathcal{N}}
	\prod_k
	\int
	\frac{da^{(n)}_k da^{*(n)}_k}{\pi}
	\right).
\end{align}
To derive Eq.~\eqref{eq:zN} we have used $\sum_i \Ket{i}\Bra{i}=1$. For small $\delta \itime$, $K_{\mathcal{N}}(\varphi,N,a^{*},a)$ can be evaluated as
\begin{align}
	K_{\mathcal{N}}(\varphi,N,a^{*},a)
	=
	\Braket{\varphi,a|
	\left(
	1-\delta \itime \hat{H}	
	\right)
	|N,a}
	+
	O(\delta \itime^2).
	\label{eq_K}
\end{align}
The right-hand side of Eq.~\eqref{eq_K} can straightforwardly be evaluated after rewriting the Hamiltonian \eqref{eq:genham} such that $\hat{a}_k$ and $\hat{a}_k^\dagger$ satisfy the normal ordering. While the quadratic term $E_{\rm C}\hat{n}_r^2$, which appears when the first term on the right-hand side of Eq.~\eqref{eq:genham} is expanded, does not satisfy the normal ordering, the difference to the normal-ordered one is only a constant part and thus physically irrelevant; we ignore it hereafter. Then we have
\begin{align}
	\label{eq:Knexpand}
	K_{\mathcal{N}}(\varphi,N,a^{*},a)
	=
	\left(
	1-\delta \itime H\left(\varphi, N, a, a^*\right)
	\right)
	\Braket{\varphi,a|N,a}
	+
	O(\delta \itime^2),
\end{align}
where
\begin{align}
	H\left(\varphi, N, a, a^*\right)
	=&
	E_{\rm C}
	\left(
	N
	-
	n_r
	\right)^2
	+
	V\left(\varphi\right)
	+
	\sum_{0<k\leq W/v}
	\omega_k 
	a_k^*
	a_k,
	\\
	n_r
	=&
	\frac{\sqrt{\gamma}}{2\pi}
	\sum_{0<k\leq W/v}
	\sqrt{\frac{2\pi}{kL}}
	\left(
	a^*_k
	+
	a_k
	\right),
\end{align}
with $a$ ($a^{*}$) being the shorthand of $\lbrace a_k\rbrace_k$
($\lbrace a_k^*\rbrace_k$).
By using
$
\Braket{\varphi,a|N,a}
=
\Braket{\varphi|N}	
\Braket{a|a}
=
e^{-i\varphi N + \sum_k a_k^{*}a_k}
$,
Eq.~\eqref{eq:Knexpand} is rewritten as
\begin{align}
	K_{\mathcal{N}}(\varphi,N,a^{*},a)
	=
	e^{-i\varphi N + \sum_k a_k^* a_k
	-
	\delta \itime H\left(\varphi, N, a, a^*\right)
	}
	+
	O(\delta \itime^2).
\end{align}
Plugging this into Eq.~\eqref{eq:zN}, we obtain
\begin{align}
	Z
	=&
	\int \mathcal{D}\varphi_{\mathcal{N}}
	\int \mathcal{D}N_{\mathcal{N}}
	\int \mathcal{D}(a,a^{*})_{\mathcal{N}}
	\notag
	\\
	&\times
	\exp\left(
	-\delta \itime
	\sum_{n=1}^{\mathcal{N}}
	\left[
	iN^{(n)}\frac{\varphi^{(n+1)}-\varphi^{(n)}}{\delta \itime} 
	+
	\sum_k a_k^{*(n+1)}\frac{a_k^{(n+1)}-a_k^{(n)}}{\delta \itime}
	+
	H\left(\varphi^{(n+1)},N^{(n)},a^{*(n+1)},a^{(n)}\right)
	\right]
	\right)+O(\delta \itime).
\end{align}
Replacing the index $n$ with the imaginary time 
$\itime=n\delta \itime = (n/\mathcal{N})\beta$
and taking the limit $\mathcal{N}\to \infty$,
we finally arrive at the path integral representation:
\begin{align}
\label{eq_Z}
	Z=
	\int \mathcal{D}\varphi
	\int \mathcal{D}N
	\int \mathcal{D}(a,a^{*})
	e^{-S_{\rm all}\left[\varphi, N, a, a^*\right]},
\end{align}
where 
$	\int \mathcal{D}\varphi
	\int \mathcal{D}N
	\int \mathcal{D}(a,a^{*})
	=
	\lim_{\mathcal{N}\to\infty}
	\int \mathcal{D}\varphi_{\mathcal{N}}
	\int \mathcal{D}N_{\mathcal{N}}
	\int \mathcal{D}(a,a^{*})_{\mathcal{N}}$,
and the action $S_{\rm all}$ is given by
\begin{align}
	S_{\rm all}\left[\varphi, N, a, a^*\right]
	=&
	\intb d\itime
	\left[
	iN(\itime)\partial_\itime \varphi(\itime)
	+
	\sum_k a^*_k(\itime)\partial_\itime a_k(\itime)
	+
	H\left(
	\varphi(\itime), N(\itime), a(\itime), a^*(\itime)
	\right)
	\right],
\end{align}
which is a functional of $\varphi(\itime)$, $N(\itime)$, 
$a(\itime)$, and $a^*(\itime)$.
Note that the integral variables satisfy the following boundary conditions:
\begin{align}
	\varphi(\beta)=\varphi(0),\quad
	N(\beta)=N(0),\quad
	a(\beta)=a(0),\quad
	a^{*}(\beta)=a^{*}(0),
\end{align}
which follow from Eq.~\eqref{eq:bound}.
\end{widetext}

We can significantly simplify the expression~\eqref{eq_Z} by performing the integrals with respect to $N$, $a$, and $a^{*}$, which are the Gaussian integrals. The result is 
\begin{align}
	Z
	=&
	\int \mathcal{D}\varphi \int \mathcal{D}N
	\int \mathcal{D}(a,a^*)
	e^{-S_{\rm all}\left[\varphi, N, a,a^*\right]}\label{eq:Sphi}\\
	\equiv&\int \mathcal{D}\varphi\, e^{-S[\varphi]},
\end{align}
where the action $S[\varphi]$ only contains $\varphi$ and is given by
\begin{align}
	S[\varphi]
	=&
	\frac{1}{\beta}
	\sum_{n}
	\left(
	\frac{\omat_n^2}{4E_{\rm C}}
	+
	\frac{\gamma v}{2\pi L}
	\sum_k
	\frac{1}{1+\left(\frac{\omega_k}{\omat_n}\right)^2}
	\right)
	\tilde{\varphi}_n
	\tilde{\varphi}_{-n}
	\notag
	\\
	&+
	\intb d\itime V\left(\varphi(\itime)\right),
	\label{eq:sgaussint}
\end{align}
as shown in Appendix~\ref{sec:deri}. Here, we have introduced the Fourier components
\begin{align}
	\tilde{\varphi}_n=\intb d\itime e^{-i\omat_n \itime}\varphi(\itime)
\end{align}
with the Matsubara frequency $p_n=\frac{2\pi}{\beta} n$ ($n\in \mathbb{Z}$). For the sake of simplicity, hereafter we take the thermodynamic and wideband limits, $L\to\infty$ and $W\to\infty$, in Eq.~\eqref{eq:sgaussint}, which allows us to arrive at the familiar expression:
\begin{align}
	\label{eq:phi_action}
	S[\varphi]
	\!=\!
	\frac{1}{2\beta}
	\sum_{n}
	\left(
	\frac{\omat_n^2}{2E_{\rm C}}
	+
	\frac{\gamma|\omat_n|}{2\pi}
	\right)
	\tilde{\varphi}_n
	\tilde{\varphi}_{-n}
	+
	\intb d\itime V\left(\varphi(\itime)\right).
\end{align}
Nevertheless, we note that the formalism developed below is equally applicable to the exact cQED action~\eqref{eq:sgaussint} with a finite cutoff $W$ and size $L$ of the environment. Indeed, recent studies have revealed that, when a cutoff/size of the transmission line is finite and small, the ground-state phase diagram can be qualitatively modified (see, e.g., Refs.~\cite{YA22,KM22}).

\subsection{Effective action}
Previous studies often resort to a simplified analysis that is perturbative with respect to the coupling $\gamma$ or/and neglects the capacitance term (the term proportional to $1/E_{\rm C}$ in Eq.~\eqref{eq:phi_action}) as it is expected to be irrelevant from the scaling dimensional analysis. Another common simplification is to replace the JJ by a two-level system as done in the spin-boson model.  However, the validity of these simplifications needs to be carefully examined in nonperturbative regimes that have recently become accessible in experiments.  
To accurately determine the equilibrium properties of the cQED system~\eqref{eq:phi_action} in such strong coupling regimes, we thus choose to employ a nonperturbative approach known as FRG without relying on those approximations. For readers from the cQED community, we give a brief review on FRG approach below. 

We first need to introduce the effective action $\Gamma[\varphi]$ that plays a central role in FRG. To this end, we start from the partition function in the presence of an external field $J$ coupled to $\varphi$:
\begin{align}
	Z[J]=\int \mathcal{D}\varphi\,e^{-S[\varphi]+\intb d\itime J(\itime)\varphi(\itime)}.
\end{align}
We note that $Z[J]$ is a generating functional of all the imaginary-time correlation functions:
\begin{align}
	\Braket{
	\varphi(\itime_1)\varphi(\itime_2)\cdots \varphi(\itime_n)
	}
	=
	\frac{1}{Z[0]}
	\frac{\delta^n Z}{\delta J(\itime_1)\cdots \delta J(\itime_n)}
	[0]
\end{align}
with
\begin{align}
	\Braket{O}
	=
	\frac{1}{Z[0]}
	\int \mathcal{D}\varphi\, O\, e^{-S[\varphi]}.
\end{align}
The following Legendre transform then defines the effective action:
\begin{align}
	\label{eq:ea}
	\Gamma[\varphi]=\sup_{J}\left(\intb d\itime J(\itime)\varphi(\itime)-W[J]\right),
\end{align}
where $W[J]=\ln Z[J]$ is the generating functional of connected correlation functions, i.e., the $n$-point connected correlation function (i.e., cumulant) $G^{(n)}(\itime_1,\ldots,\itime_n)$ is given by
\begin{align}
	G^{(n)}(\itime_1,\ldots,\itime_n)
	=
	\frac{\delta^n W}{\delta J(\itime_1)\cdots \delta J(\itime_n)}
	[0].
\end{align}
As a nature of the Legendre transformation, $\Gamma[\varphi]$ satisfies convexity.

One of the essential properties of $\Gamma[\varphi]$ is that it satisfies the variational principle. From Eq.~\eqref{eq:ea}, the derivative of $\Gamma[\varphi]$ with respect to $\varphi(\itime)$ is given by
\begin{align}
	\frac{\delta \Gamma[\varphi]}{\delta \varphi(\itime)}
	=&
	J_{\rm sup}[\varphi](\itime),
	\label{eq:gam1}
\end{align}
where $J_{\rm sup}[\varphi](\itime)$ satisfies
\begin{align}
	\label{eq:qave}
	\varphi(\itime)=\frac{\delta W}{\delta J(\itime)}[J_{\rm sup}[\varphi]]
	=
	\Braket{\varphi(\itime)}_{J_{\rm sup}[\varphi]}
\end{align}
with
\begin{align}
	\Braket{O}_{J}
	=
	\frac{1}{Z[J]}
	\int \mathcal{D}\varphi'\,
	O\, e^{-S[\varphi']+\intb d\itime' J(\itime')\varphi'(\itime')}.
\end{align}
In particular, we have $\varphi(\itime)=\Braket{\varphi(\itime)}$ in the case of $J_{\rm sup}[\varphi](\itime)=0$. This fact together with  Eq.~\eqref{eq:qave}  means that the quantum average $\varphi(\itime)=\Braket{\varphi(\itime)}$ is obtained by the variational equation
\begin{align}
	\frac{\delta \Gamma[\varphi]}{\delta \varphi(\itime)}=0.
\end{align}

The effective action $\Gamma[\varphi]$ plays the role of generating functional of one-particle irreducible (1PI) correlation functions \cite{peskin_introduction_2019}. In other words, the $n$th derivative
\begin{align*}
	\Gamma^{(n)}[\varphi](\itime_1,\ldots,\itime_n)
	=
	\frac{\delta^n \Gamma[\varphi]}
	{\delta \varphi(\itime_1)\cdots \delta \varphi(\itime_n)}
\end{align*}
is the $n$-point 1PI correlation function, which is given by a sum of diagrams that cannot be separated by eliminating one internal propagator. In particular, the second derivative of $\Gamma[\varphi]$ gives the inverse of $G^{(2)}(\itime_1,\itime_2)$; to see this, we first use Eq.~\eqref{eq:gam1} to obtain
\begin{align}
	\label{eq:gam2}
	\Gamma^{(2)}[\varphi](\itime_1,\itime_2)
	=
	\frac{\delta J_{\rm sup}[\varphi](\itime_1)}{\delta \varphi(\itime_2)},
\end{align}
while we note that the derivative of Eq.~\eqref{eq:qave} with respect to $\varphi(\itime')$ is
\begin{align}
	\label{eq:w2}	
	\delta(\itime-\itime')
	=
	\intb d\itime'' \frac{\delta J_{\rm sup}[\varphi](\itime'')}{\delta \varphi(\itime')}
	G^{(2)}[J_{\rm sup}[\varphi]](\itime'',\itime),
\end{align}
where $G^{(2)}[J](\itime'',\itime)=\frac{\delta^2 W}{\delta J(\itime'')\delta J(\itime)}[J]$ is the two-point correlation function in the presence of the external field $J$:
\begin{align*}
	G^{(2)}[J](\itime,\itime')
	=&
	\Braket{\varphi(\itime)\varphi(\itime')}_J-\Braket{\varphi(\itime)}_J\Braket{\varphi(\itime')}_J.
\end{align*}
Equations \eqref{eq:gam2} and \eqref{eq:w2} then give the following relation:
\begin{align}
	\label{eq:g2gam2}
	G^{(2)}[J_{\rm sup}[\varphi]](\itime,\itime')
	=
	\Gamma^{(2)-1}[\varphi](\itime,\itime'),
\end{align}
where $\Gamma^{(2)-1}[\varphi](\itime,\itime')$ is the functional inverse of $ \Gamma^{(2)}[\varphi](\itime, \itime')$, i.e., it satisfies
\begin{align}
	\intb d\itime''
	\Gamma^{(2)-1}[\varphi](\itime,\itime'')
	\Gamma^{(2)}[\varphi](\itime'',\itime')
	=
	\delta(\itime-\itime').
\end{align}
Importantly, when $\varphi(\itime)=\Braket{\varphi(\itime)}$,  Eq.~\eqref{eq:g2gam2} shows that the inverse of the second derivative, $\Gamma^{(2)-1}[\varphi](x,x')$, is equal to the two-point connected correlation function $G^{(2)}(\itime,\itime')=G^{(2)}[0](\itime,\itime')$.

\subsection{Flow equation}

\begin{figure}[b]
  \begin{center}
  	\includegraphics[width=0.67\columnwidth]{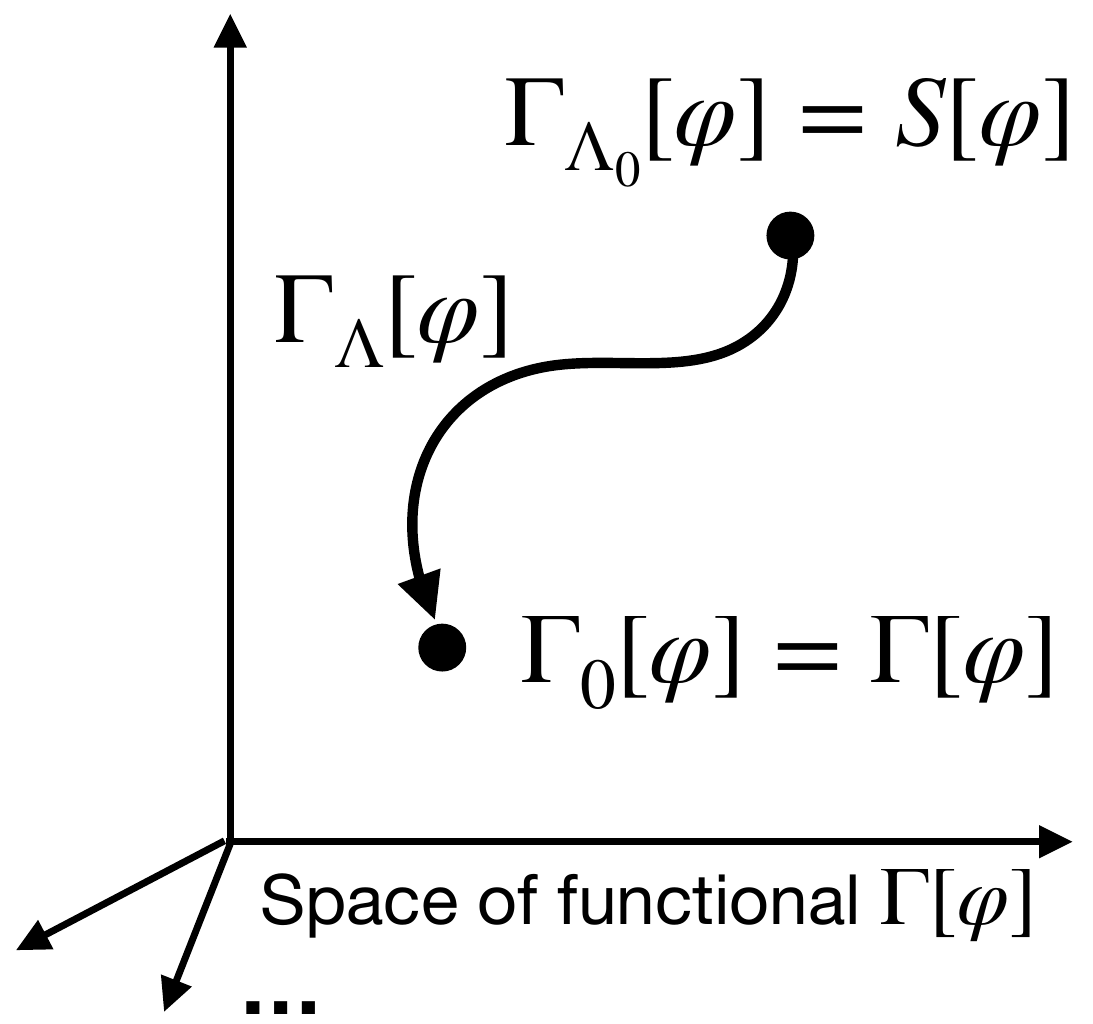}
    \caption{Schematic of RG Flow of $\Gamma_\Lambda[\varphi]$ in the functional space.}
    \label{fig:gam_flow}
  \end{center}
\end{figure}
Following the FRG formalism \`a la Wetterich \cite{wetterich_exact_1993}, we derive an exact renormalization group (RG) equation of the effective action. To describe the RG flow, we introduce the effective average action $\Gamma_{\Lambda}[\varphi]$, which depends on the energy scale parameter $\Lambda$ and interpolates the bare action $S[\varphi]$ and the effective action $\Gamma[\varphi]$:
\begin{align}
	\label{eq:gam_bound}
	\Gamma_{\Lambda_0}[\varphi]=S[\varphi]+\mathrm{const.},
	\quad
	\Gamma_{0}[\varphi]=\Gamma[\varphi]
\end{align}
with a large UV scale $\Lambda_0$ as illustrated in Fig.~\ref{fig:gam_flow}. The definition of $\Gamma_{\Lambda}[\varphi]$ is given by
\begin{align}
	\Gamma_\Lambda[\varphi]
	\!=\!&
	\sup_J
	\left(
	\intb d\itime J(\itime)\varphi(\itime)
	-
	W_\Lambda[J]
	\right)
	-
	\Delta S_\Lambda[\varphi]
	\notag
	\\
	\!=\!&
	\intb d\itime J_{\rm sup,\Lambda}[\varphi](\itime)\varphi(\itime)
	\!-\!
	W_\Lambda[J_{\rm sup,\Lambda}[\varphi]]
	\!-\!
	\Delta S_\Lambda[\varphi],
	\label{eq:eaa}
\end{align}
where we have introduced the scale-dependent generating functionals
\begin{align}
	\label{eq:zlam}
	Z_\Lambda[J]
	=
	e^{W_\Lambda[J]}
	=
	\int \mathcal{D}\varphi\, e^{-S[\varphi]-\Delta S_\Lambda[\varphi]+\intb d\itime J(\itime)\varphi(\itime)},
\end{align}
and $J_{\rm sup,\Lambda}[\varphi](\itime)$ satisfying
\begin{align}
	\label{eq:jsup}
	\left.
	\frac{\delta}{\delta J(\itime)}
	\left(
	\intb d\itime J(\itime)\varphi(\itime)
	-
	W_\Lambda[J]
	\right)
	\right|_{J=J_{\rm sup,\Lambda}[\varphi]}
	=
	0.
\end{align}
The regulator term $\Delta S_\Lambda[\varphi]$, which is a key quantity to realize the RG flow, has the following form:
\begin{align}
	\label{eq:dS}
	\Delta S_\Lambda[\varphi]
	=
	\frac{1}{2}
	\intb d\itime 
	\intb d\itime'
	R_{\Lambda}(\itime-\itime')
	\varphi(\itime)\varphi(\itime').
\end{align}
Here, the coefficient $R_{\Lambda}(\itime-\itime')$ is called a regulator function. To attain the RG procedure in Eq.~\eqref{eq:gam_bound}, the regulator $R_{\Lambda}(\itime-\itime')$ should satisfy the following conditions:
\begin{itemize}
	\item[(a)] $\lim_{\Lambda^2/p_n^2 \to 0}\tilde{R}_\Lambda(p_n)=0$,
	\item[(b)] $\lim_{p_n^2/\Lambda^2 \to 0}\tilde{R}_\Lambda(p_n)=\infty$,
\end{itemize}
where $\tilde{R}_\Lambda(p_n)$ is the Fourier coefficient of $R_{\Lambda}(\itime)$ and $p_n$ is the Matsubara frequency. It is obvious that the condition (a) realizes $\Gamma_0[\varphi]=\Gamma[\varphi]$ since we have $\Delta S_{\Lambda\to 0}[\varphi]\to 0$ and then Eq.~\eqref{eq:eaa} at $\Lambda\to 0$ becomes equivalent to Eq.~\eqref{eq:ea}. The condition (b) justifies the use of the saddle point approximation to evaluate the path integral in Eq.~\eqref{eq:zlam} at $\Lambda=\Lambda_0\to\infty$:
\begin{align}
	W_{\Lambda_0}[J]
	\approx&
	-S[\varphi_0[J]]
	-\Delta S_{\Lambda_0}[\varphi_0[J]]
	\notag
	\\
	&
	+
	\intb d\itime J(\itime)\varphi_0[J]
	+
	\mathrm{const.},
	\label{eq:wsta}
\end{align}
where $\varphi_0[J]$ satisfies the saddle point equation:
\begin{align}
	\label{eq:qsta}
	J(\itime)
	=&
	\left.
	\frac{\delta}{\delta \varphi(\itime)}
	\left(
	S[\varphi]+\Delta S_{\Lambda_0}[\varphi]
	\right)
	\right|_{\varphi=\varphi_0[J]}.
\end{align}
Plugging Eq.~\eqref{eq:wsta} into Eq.~\eqref{eq:eaa}, we obtain
\begin{align}
\Gamma_{\Lambda_0}[\varphi]=S[\varphi]+\mathrm{const.}
\end{align}
Here, we use Eqs.~\eqref{eq:jsup}, \eqref{eq:qsta}, and the relation $\varphi(\itime)=\varphi_0[J_{\rm sup,\Lambda}[\varphi]](\itime)$.

\begin{figure}[b]
  \begin{center}
  	\includegraphics[width=0.7\columnwidth]{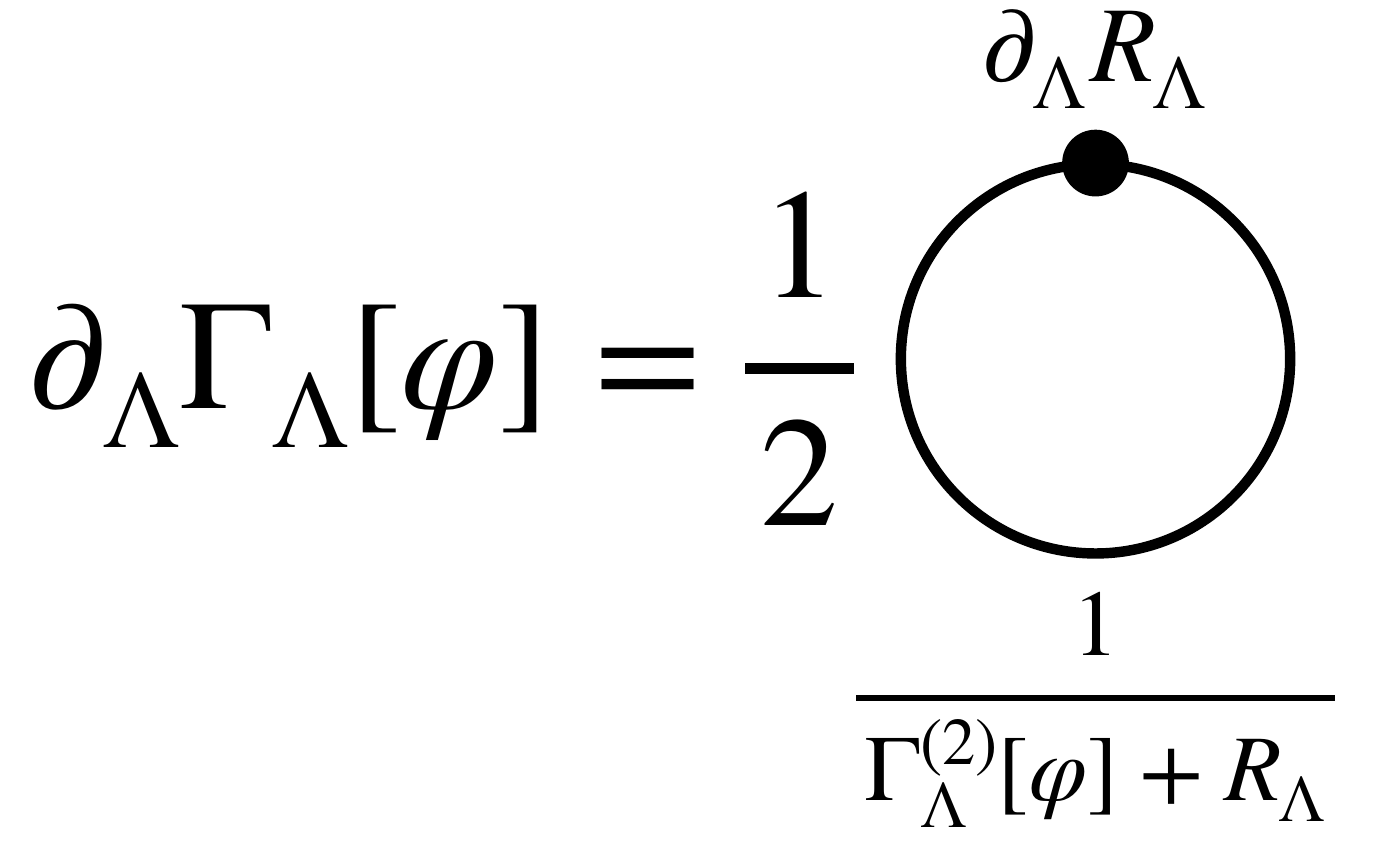}
    \caption{Diagrammatic representation of the Wetterich equation.}
    \label{fig:wetterich}
  \end{center}
\end{figure}

Now we derive the RG flow equation of $\Gamma_\Lambda[\varphi]$. By use of Eq.~\eqref{eq:jsup}, the derivative of Eq.~\eqref{eq:gam_bound} with respect to $\Lambda$ is written as
\begin{align}
	\label{eq:dea}
	\partial_\Lambda
	\Gamma_\Lambda[\varphi]
	=
	-
	\left(\partial_\Lambda W_\Lambda\right)[J_{\rm sup,\Lambda}[\varphi]]
	-
	\partial_\Lambda
	\Delta S_\Lambda[\varphi].
\end{align}
Using Eq.~\eqref{eq:zlam}, the first term on the right-hand side is evaluated as
\begin{widetext}
\begin{align}
	\label{eq:dW}
	(\partial_\Lambda W_\Lambda)[J_{\rm sup,\Lambda}[\varphi]]
	=&
	-
	\frac{1}{Z_\Lambda[J_{\rm sup,\Lambda}[\varphi]]}
	\int \mathcal{D}\varphi\,
	\partial_\Lambda \Delta S_\Lambda[\varphi]
	e^{-S[\varphi]-\Delta S_\Lambda[\varphi]
	+
	\intb d\itime J_{\rm sup,\Lambda}[\varphi](\itime)\varphi(\itime)}
	\notag
	\\
	=&
	-
	\frac{1}{2}
	\intb d\itime
	\intb d\itime'
	\partial_\Lambda R_\Lambda(\itime-\itime')
	\Braket{\varphi(\itime)\varphi(\itime')}_{\Lambda}
	\notag
	\\
	=&
	-
	\frac{1}{2}
	\intb d\itime
	\intb d\itime'
	\partial_\Lambda R_\Lambda(\itime-\itime')
	\left[
	G_\Lambda^{(2)}(\itime,\itime')
	+
	\Braket{\varphi(\itime)}_{\Lambda}\Braket{\varphi(\itime')}_{\Lambda}
	\right],
\end{align}
\end{widetext}
where $G^{(2)}_\Lambda(\itime,\itime')$ is the two-point correlation function
\begin{align}
	\label{eq:g2w}
	G^{(2)}_\Lambda(\itime,\itime')
	=
	\frac{\delta^2 W_\Lambda}{\delta J(\itime)\delta J(\itime')}
	[J_{\rm sup,\Lambda}[\varphi]]
\end{align}
and the average $\Braket{O}_\Lambda$ is defined by
\begin{align*}
	\Braket{O}_\Lambda
	=&
	\frac{1}{Z_\Lambda[J_{\rm sup,\Lambda}[\varphi]]}
	\notag
	\\
	&\times
	\int \mathcal{D}\varphi\,
	O\,
	e^{-S[\varphi]-\Delta S_\Lambda[\varphi]
	+
	\intb d\itime J_{\rm sup,\Lambda}[\varphi](\itime)\varphi(\itime)}.
\end{align*}
By use of Eqs.~\eqref{eq:dW} and \eqref{eq:dS} together with $\varphi(\itime)=\Braket{\varphi(\itime)}_{\Lambda}$, which is obtained in the same manner as Eq.~\eqref{eq:qave}, Eq.~\eqref{eq:dea} can be rewritten as
\begin{align}
	\label{eq:dgrg}
	\partial_\Lambda \Gamma_\Lambda[\varphi]
	=&
	\frac{1}{2}
	\intb d\itime
	\intb d\itime'
	\partial_\Lambda R_\Lambda(\itime-\itime')
	G^{(2)}_\Lambda(\itime, \itime').
\end{align}
In the similar manner as in Eq.~\eqref{eq:g2gam2}, the two-point correlation function $G^{(2)}_\Lambda(\itime, \itime')$ can be related to the second-order derivative of $\Gamma_\Lambda[\varphi]$; evaluating the second-order derivative of Eq.~\eqref{eq:eaa} with respect to $\varphi$, we obtain
\begin{align}
	\label{eq:djdq}
	\frac{\delta J_{\rm sup,\Lambda}[\varphi](\itime)}{\delta \varphi(\itime')}
	=
	\frac{\delta^2 \Gamma_\Lambda[\varphi]}{\delta \varphi(\itime)\delta \varphi(\itime')}
	+
	R_{\Lambda}(\itime-\itime').
\end{align}
By use of Eq.~\eqref{eq:g2w} and
\begin{align}
	\varphi(\itime)=\frac{\delta W_\Lambda}{\delta J(\itime)}[J_{\rm sup,\Lambda}[\varphi]],
\end{align}
which is derived from Eq.~\eqref{eq:jsup}, we also have
\begin{align}
	\label{eq:dqdj}
	G^{(2)}_\Lambda(\itime,\itime')
	=
	\frac{\delta \varphi(\itime)}{\delta J_{\rm sup,\Lambda}[\varphi](\itime')}.
\end{align}
Equations \eqref{eq:djdq} and \eqref{eq:dqdj} lead to the following relation:
\begin{align}
	\label{eq:g2g2inv}
	G_{\Lambda}^{(2)}(\itime,\itime')
	=
	\left[
	\frac{1}{\Gamma^{(2)}_\Lambda[\varphi]+R_\Lambda}
	\right]
	(\itime,\itime'),
\end{align}
where the right-hand side is the functional inverse of
$\Gamma^{(2)}_\Lambda[\varphi](\itime,\itime')
+
R_{\Lambda}(\itime-\itime')$.
Plugging Eq.~\eqref{eq:g2g2inv} into Eq.~\eqref{eq:dgrg}, we finally obtain the RG equation in the form of a functional derivative equation of $\Gamma_\Lambda[\varphi]$:
\begin{align}
	\partial_\Lambda \Gamma_\Lambda[\varphi]
	=&
	\frac{1}{2}
	\intb d\itime 
	\intb d\itime'
	\notag
	\\
	&
	\times
	\left[
	\partial_\Lambda R_\Lambda(\itime-\itime')
	\frac{1}{\Gamma^{(2)}_\Lambda[\varphi]+R_\Lambda}
	(\itime,\itime')
	\right],
	\label{eq:wet}
\end{align}
which is known as the Wetterich equation \cite{wetterich_exact_1993}. This equation has a one-loop structure and can be diagrammatically represented as in Fig.~\ref{fig:wetterich}, where $[\Gamma^{(2)}_\Lambda[\varphi]+R_\Lambda]^{-1}$ is regarded as a dressed propagator including nonperturbative effects.

While Eq.~\eqref{eq:wet} is an exact equation, it is a functional equation and hard to solve without any approximations. A conventional way is to rewrite Eq.~\eqref{eq:wet} as infinite series of hierarchical equations by expanding  $\Gamma_\Lambda[\varphi]$ and truncate the series at certain order. Perturbation theory can be reproduced in the current formalism by expanding $\Gamma_\Lambda[\varphi]$ with respect to the Plank constant $\hbar$, i.e., applying the loop expansion. The powerfulness of Eq.~\eqref{eq:wet} is that nonperturbative analysis can be performed by introducing other expansion methods. Such expansions include a functional Taylor expansion around $\varphi$ of interest, which is called the vertex expansion, or the derivative expansion, which is an expansion with respect to $\partial_x$. In the remaining parts of this paper, we shall show several case studies using the latter expansion.

\section{Double-well potential\label{sec:double}}

\begin{figure}[t]
  \begin{center}
  	\includegraphics[width=\columnwidth]{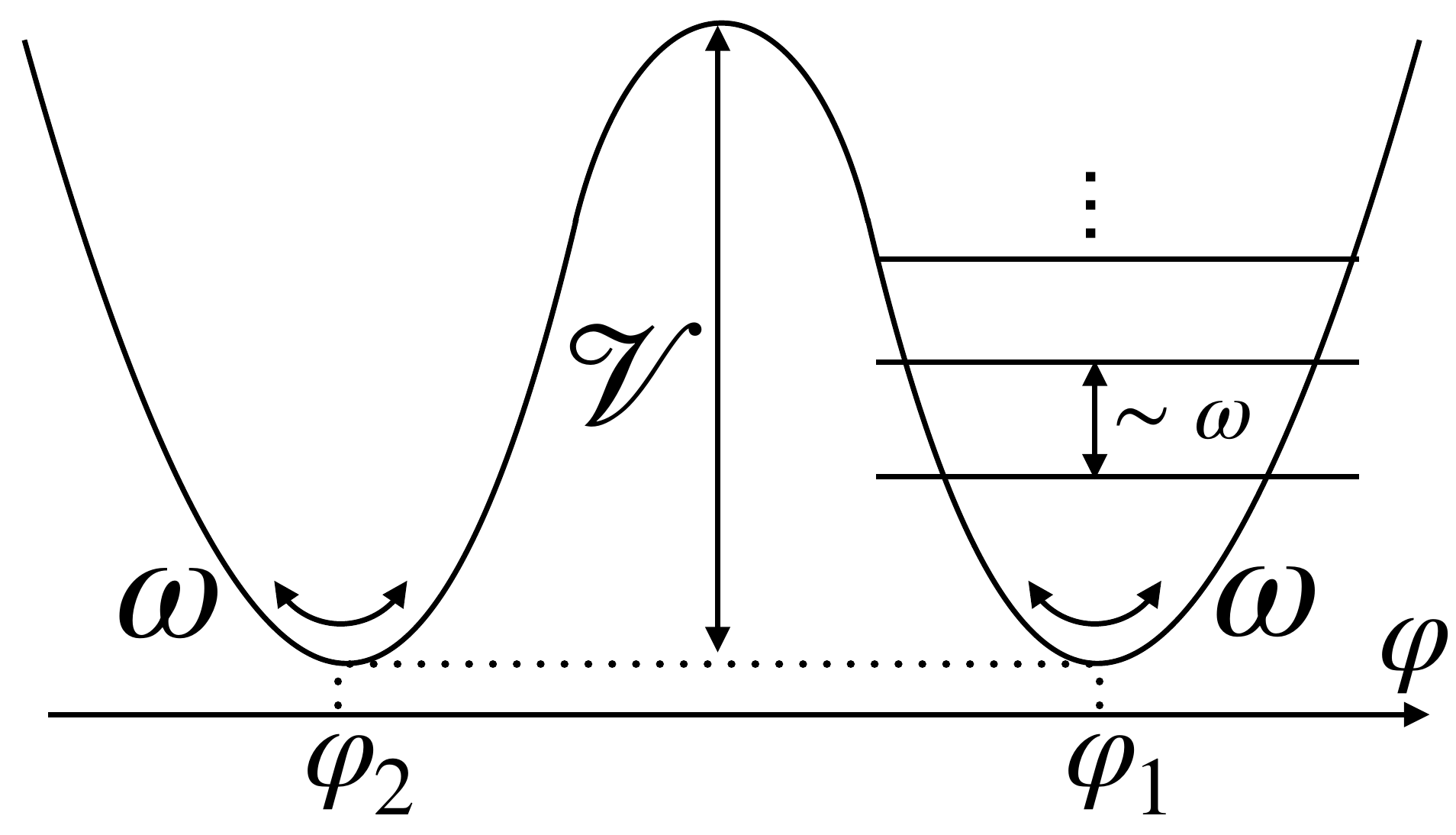}
    \caption{Double-well potential. The potential depth is denoted by $\mathcal{V}$. The oscillation frequency $\omega$, which is assumed to be the same for both wells, characterizes the level intervals in each well. The two-level approximation, which truncates energy states higher than the two lowest levels, can be justified when $\mathcal{V}\gg\omega$.}
    \label{fig:double}
  \end{center}
\end{figure}

\begin{figure*}[t]
  \begin{center}
  	\includegraphics[width=2\columnwidth]{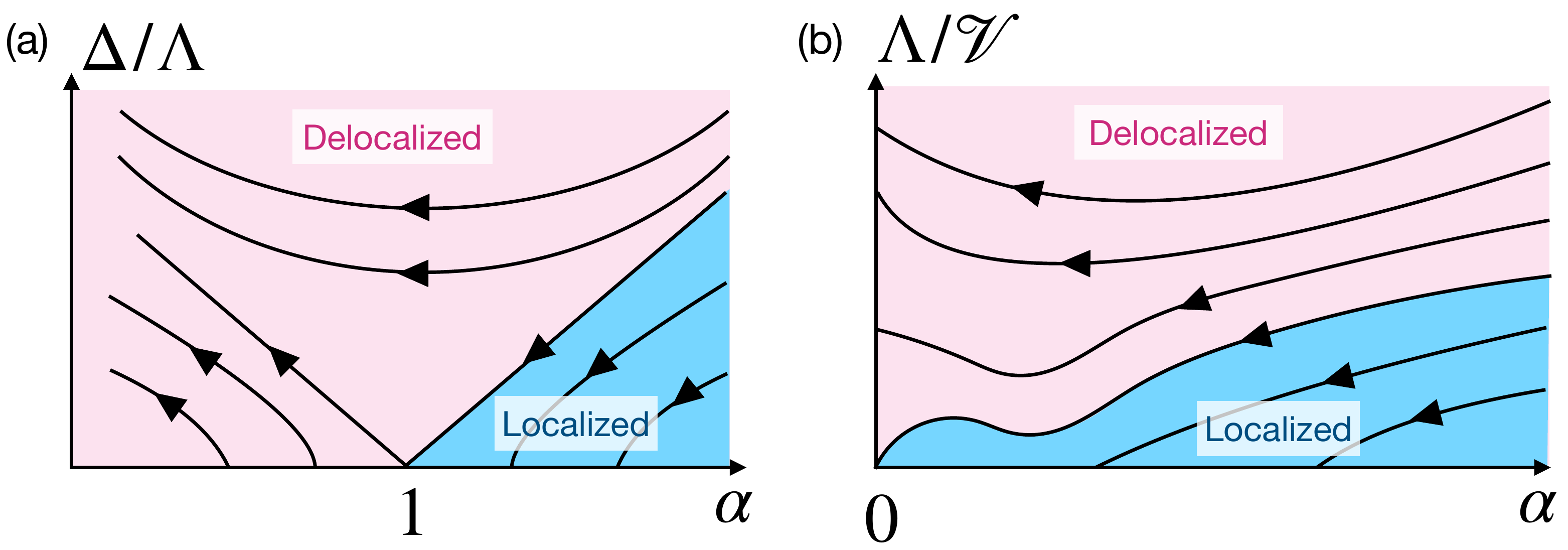}
    \caption{Schematic figures illustrating the flow diagrams obtained from (a) the perturbative analysis of the spin-boson model and (b) the FRG analysis of the exact cQED action~\eqref{eq:phi_action} without the two-level approximation. The origin of the discrepancy between (a) and (b) is traced to strong renormalizations of the potential barrier $\mathcal{V}$ and excitation frequency $\omega$ in nonperturbative regimes, which invalidates the necessary condition~\eqref{eq:sb_cond} for the two-level approximation.}
    \label{fig:double_concept}
  \end{center}
\end{figure*}
To be concrete, we first consider the case of a double-well potential in the  cQED setup described by Eq.~\eqref{eq:phi_action}. Such potential profile is routinely realized in, e.g., flux qubits, where one combines a flux-tunable Josephson energy $-E_J\cos(\varphi-\varphi_{\rm ext})$ with an inductive energy $E_L\varphi^2/2$. A common simplification is to take into account only the two lowest energy levels of JJ (corresponding to the minima of both wells) and then reduce the original Hamiltonian to the spin-boson model. On the one hand, it is well-established that the localization-delocalization phase transition is predicted to occur in the spin-boson model. On the other hand, however, we emphasize that this two-level treatment can be justified only if the following condition is satisfied \cite{Leggett87}:
	\begin{align}
		\label{eq:sb_cond}
		\mathcal{V} \gg \omega,
	\end{align}
where $\mathcal{V}$ and $\omega$ are the potential depth and the excitation frequency in each well, respectively, as illustrated in Fig.~\ref{fig:double}. The condition~\eqref{eq:sb_cond} ensures that the level interval between the two lowest states determined by the tunneling rate is exponentially small compared to $\omega$, which should allow one to truncate high-lying levels. Nevertheless, as demonstrated below, the  condition~\eqref{eq:sb_cond} is in general violated especially in nonperturbative regimes due to strong renormalizations caused by the environment, indicating the need of a more comprehensive analysis beyond the spin-boson description.

This section aims to carefully reexamine this fundamental problem without relying on the two-level approximation. Specifically, we analyze the cQED action~\eqref{eq:phi_action} with the double-well potential on the basis of our FRG formalism. We find that the localization-delocalization transition is still present, but surprisingly the obtained phase diagram is qualitatively modified from the one expected in the spin-boson model \cite{anderson_exact_1970,Leggett87}. 
As detailed below, this is because the renormalized potential barrier ${\cal v}_\Lambda$ and excitation frequency $\omega_\Lambda$ are strongly modified during FRG flows, and the condition~\eqref{eq:sb_cond} eventually breaks down in most of the regions in the flow diagram.
Figure~\ref{fig:double_concept}(a) illustrates the flow diagram on the $(\alpha, \Delta/\Lambda)$ plane obtained in the spin-boson model, where $\Delta$ is the tunneling rate between the wells and the dimensionless coupling strength $\alpha$ is defined by
\begin{align}
	\label{eq:def_alpha}
	\alpha=\gamma \left(\frac{\varphi_{1}-\varphi_{2}}{2\pi} \right)^2
\end{align}
with $\varphi_{1,2}$ being the positions of the two wells (cf. Fig.~\ref{fig:double}). As shown in Fig.~\ref{fig:double_concept}(a), the system always flows to the delocalized phase in $\alpha<1$, while it flows to the localized phase at smaller values of $\Delta/\Lambda$ in $\alpha>1$. In contrast, the present FRG analysis suggests that the localized phase is extended to the regions in $\alpha<1$ as shown in Fig.~\ref{fig:double_concept}(b), where the flow diagram on $\alpha$ and $\Lambda/\mathcal{V}$ is illustrated. 

We now describe in detail how our FRG approach can be applied to the present case of the double-well potential. The equilibrium phase diagram is determined by the minimum point of the effective potential
\begin{align}
	V_\Lambda(\varphi_{\rm s})=\frac{1}{\int d\itime}\Gamma_\Lambda[\varphi_{\rm s}],
\end{align}
where $\varphi_{\rm s}$ stands for $\varphi$ independent of $\itime$, as its minimum determines the equilibrium value of $\varphi$. We calculate  $V_\Lambda(\varphi_{\rm s})$ by employing the local potential approximation (LPA):
\begin{align}
	\Gamma_\Lambda[\varphi]
	\approx
	\Gamma_\Lambda^{\rm LPA}[\varphi]
	=
	&
	\frac{1}{2}
	\int_{-\infty}^{\infty}
	\frac{d\omat}{2\pi}
	\left(
	\frac{\omat^2}{2E_{\rm C}}
	+\frac{\gamma |\omat|}{2\pi}\right)
	\tilde{\varphi}(p)\tilde{\varphi}(-p)
	\notag
	\\
	&
	+\int_0^{\infty} d\itime V_\Lambda\left(\varphi(\itime)
	\right).
\end{align}
Here, we note that the evolution of higher-order terms in the derivative expansion is ignored:
\begin{align*}
	\Gamma_{\Lambda}[\varphi]=\Gamma_{\Lambda}^{\rm LPA}[\varphi]+O(\partial_x).
\end{align*}
Plugging this ansatz into Eq.~\eqref{eq:wet}, we obtain the flow equation of $V_\Lambda\left(\varphi(\itime)\right)$:
\begin{align}
	\label{eq:vflow}
	\frac{\partial}{\partial \Lambda}
	V_\Lambda(\varphi)
	=
	\frac{1}{2}
	\int_{-\infty}^{\infty}
	\frac{d\omat}{2\pi}
	\partial_{\Lambda} \tilde{R}_{\Lambda}(\omat)
	G_\Lambda(\varphi; \omat)
\end{align}
with the regulated propagator
\begin{align}
	G_\Lambda(\varphi; \omat)
	=&
	\frac{1}{
	\frac{1}{2E_{C}}
	\omat^2
	+
	\frac{\gamma |\omat|}{2\pi}
	+
	\partial_\varphi^2 V_\Lambda(\varphi)
	+
	\tilde{R}_\Lambda(\omat)}.
\end{align}

We can further simplify the flow equation by using the Taylor expansion. Specifically, we assume that the potential is symmetric under $\varphi\leftrightarrow -\varphi$ and denote the positions of the potential minima as $\varphi_1=-\varphi_2=\varphi_\Lambda$. In this case, the potential becomes a function of $\rho=\frac{1}{2}\varphi^2$ as  $V_\Lambda(\varphi)=V_\Lambda(\rho)$. Then the Taylor expansion around $\rho_\Lambda=\frac{1}{2}\varphi_\Lambda^2$ with respect to $\rho$ is described by
\begin{align}
	V_\Lambda(\rho)
	=
	\begin{cases}
		\frac{1}{2}a_\Lambda
		\left(\rho-\rho_{\Lambda}\right)^2
		+
		O\left(
		\left(\rho-\rho_{\Lambda}\right)^3
		\right)
		&
		(\rho_{\Lambda}>0)
		,		
		\\		
		c_\Lambda \rho
		+
		\frac{1}{2}a_\Lambda \rho^2
		+
		O\left(\rho^3\right)
		&
		(\rho_{\Lambda}=0),	
	\end{cases}	
\end{align}
up to the second order without the constant terms. Here, $a_\Lambda$ and $c_\Lambda$ are the nonnegative Taylor coefficients. Note that the first-order term in the case of the localized phase $\rho_\Lambda>0$ vanishes due to $\frac{\partial}{\partial \varphi} V(\varphi_\Lambda)=0$. For the sake of convenience, we introduce the dimensionless parameters $\overline{\rho}_\Lambda$ and $\vd_\Lambda$ defined by
\begin{align}
	\label{eq:rhobar}
	\overline{\rho}_\Lambda
	=
	\begin{cases}
		\rho_\Lambda & (\rho_\Lambda>0)
		\\
		-c_\Lambda/a_\Lambda & (\rho_\Lambda=0)
	\end{cases},
	\quad
	\vd_\Lambda
	=
	\frac{\overline{\rho}_\Lambda^2 a_\Lambda}{2\Lambda}.
\end{align}
Then our expansion is concisely reparametrized as
\begin{align}
	\label{eq:vexp}
	&
	V_\Lambda(\rho)
	=
	\Lambda \vd_\Lambda
	\left(1-\frac{\rho}{\overline{\rho}_{\Lambda}}\right)^2
	+
	O\left(
	\left(\rho-\rho_{\Lambda}\right)^3
	\right)
\end{align}
up to a constant.
\begin{figure*}[t]
  \begin{center}
  	\includegraphics[width=2\columnwidth]{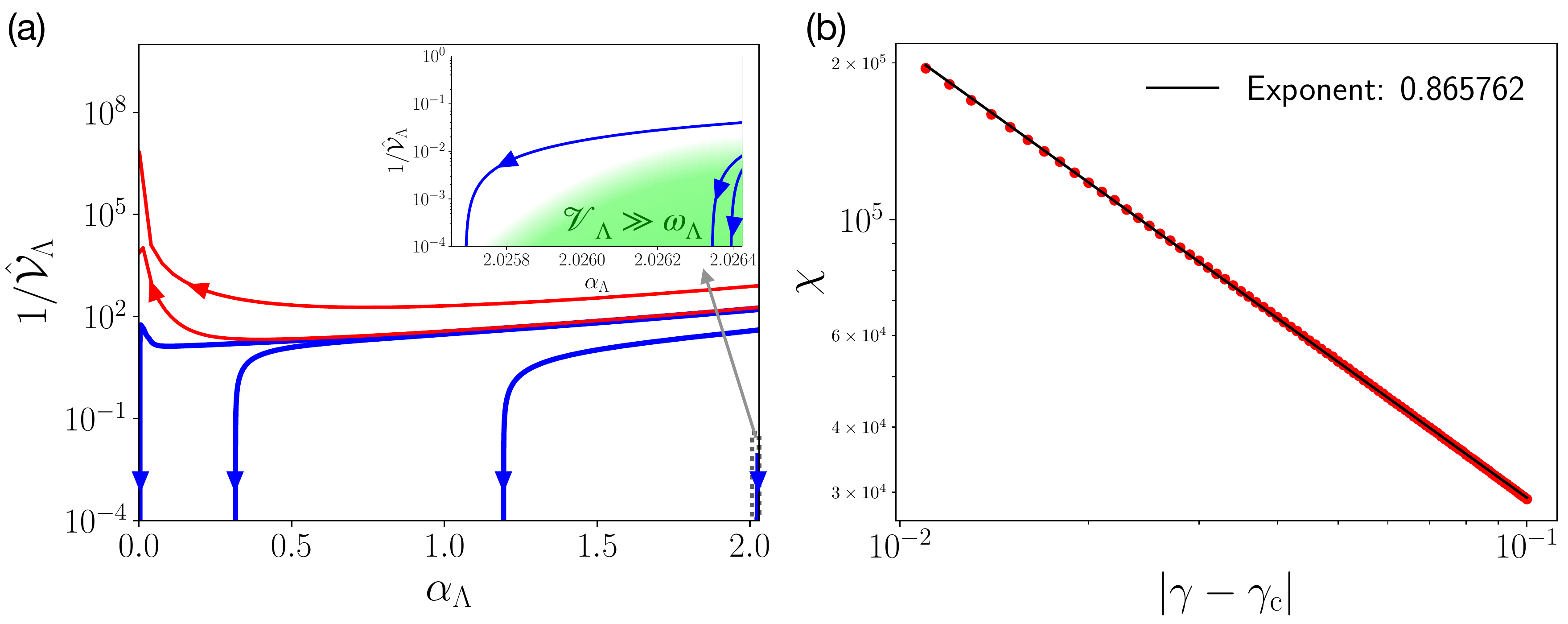}
    \caption{(a) Flow diagram of $1/\vd_\Lambda$ and $\alpha_\Lambda$. Each curve is the result with the initial condition $1/\vd_{\Lambda_0}=0.008,\ 40,\ 160,\ 176.62,\ 184,\ 800$ from below.  The blue and red curves denote the cases of the localized state ($\overline{\rho}_{0}>0$) and the delocalized state ($\overline{\rho}_{0}<0$), respectively. For other quantities, we have set $\gamma=20$, $\overline{\rho}_{\Lambda_0}=0.5$ and $\epsilon_{C,\Lambda_0}=1$, which corresponds to $\alpha_{\Lambda_0}=2.026$ (cf.~Eq.~\eqref{eq:alphadef}). In the inset, the curves with $1/\vd_{\Lambda_0}=0.004,\ 0.008,\ 0.04$ are shown. The region where $\mathcal{V}_\Lambda \gg \omega_\Lambda$ is satisfied, i.e., the two-level approximation can be justified, is restricted to a right-bottom part of the diagram and indicated as the green-shaded region. (b) $\gamma$ dependence of $\chi$ near the critical point $\gamma_c=20$ in the delocalized phase. We fix $1/\vd_{\Lambda_0}=176.62$. The red points are the results of FRG. Linear regression gives the black line.}
    \label{fig:double_result}
  \end{center}
\end{figure*}
Note that a positive (negative) value of $\overline{\rho}_\Lambda$ corresponds to the localized (delocalized) phase, as one can see from Eq.~\eqref{eq:rhobar}. In addition, $\vd_\Lambda$ can be regarded as the dimensionless potential depth in the case of $\rho_\Lambda > 0$ since we have
\begin{align*}
	\vd_\Lambda = \frac{V_\Lambda(0)-V_\Lambda(\rho_\Lambda)}{\Lambda}
\end{align*}
within the accuracy of the expansion. Then the flow equation reduces to those of $\vd_\Lambda$ and $\overline{\rho}_\Lambda$, which are obtained from the first and second derivatives of Eq.~\eqref{eq:vflow} with respect to $\varphi$:
\begin{align}
	\label{eq:flow_rho}
	\frac{\partial}{\partial l}
	\overline{\rho}_{\Lambda}
	=&
	-
	\frac{3}{2}
	I_{2,\Lambda}
	-
	18
	\frac{\vd_\Lambda}{|\overline{\rho}_\Lambda|}
	\theta(-\overline{\rho}_\Lambda)I_{3,\Lambda},
	\\
	\label{eq:flow_v}
	\frac{\partial}{\partial l}
	\vd_\Lambda
	=&
	\vd_\Lambda
	\left(
	1
	-
	\frac{3}{\overline{\rho}_\Lambda}
	I_{2,\Lambda}
	-
	18
	\frac{\vd_\Lambda}{\overline{\rho}_\Lambda^2}
	\theta(\overline{\rho}_\Lambda)
	I_{3,\Lambda}
	\right),
\end{align}
where $l=\ln(\Lambda_0/\Lambda)$ is a logarithmic RG scale with an ultraviolet scale $\Lambda_0$ and we introduce the dimensionless quantities
\begin{align}
	\label{eq:In}
	I_{n,\Lambda}=
	\int_{-\infty}^{\infty}
	\frac{d(\omat/\Lambda)}{2\pi}
	\partial_\Lambda \tilde{R}_\Lambda(p)
	\left[\Lambda G_\Lambda(\varphi_\Lambda;\omat)
	\right]^n
	\quad (n=2,3).
\end{align}
We note that, when $\overline{\rho}_\Lambda\geq 0$, Eq.~\eqref{eq:flow_rho} describes the RG flow of the dimensionless coupling strength $\alpha$:
\begin{align}
	\label{eq:flow_alpha}
	\frac{\partial}{\partial l}
	\alpha_{\Lambda}
	=&
	-
	\frac{3\gamma}{\pi^2}
	I_{2,\Lambda}
\end{align}
with 
\begin{align}
	\label{eq:alphadef}
	\alpha_\Lambda=\frac{2\gamma }{\pi^2}\rho_\Lambda.
\end{align}
We choose to use a simple regulator $\tilde{R}_\Lambda=\Lambda$, which allows us to evaluate Eq.~\eqref{eq:In} analytically:
\begin{align*}
	I_{n,\Lambda}
	=
	\frac{(-1)^{n-1}}{(n-1)!}\frac{d^{n-1}}{dr^{n-1}}I_{\Lambda}(1),
\end{align*}
where we define
\begin{align}
	I_{\Lambda}(r)
	=&
	\begin{cases}
		\frac{2}{\gamma}
		\frac{\pi-2\arctan\left(\frac{1}{\sqrt{y-1}}\right)}{\sqrt{y-1}}
		& (y\geq 1)
	\\
	\frac{2}{\gamma}
	\frac{1}{\sqrt{1-y}}
	\ln\left(
	\frac{1+\sqrt{1-y}}{1-\sqrt{1-y}}
	\right)
	& (1 > y >0)
	\end{cases},
	\notag
	\\
	y
	=&
	\frac{16\pi^2}{\gamma^2 \epsilon_{C,\Lambda}}
	\left[
	2(3\theta(\overline{\rho}_\Lambda)-1)
	\frac{\vd_\Lambda}{\overline{\rho}_\Lambda}
	+
	r
	\right],
	\notag
\end{align}
and $\epsilon_{C,\Lambda}=E_{C}/\Lambda$.

Before showing our numerical results, we describe how the condition \eqref{eq:sb_cond} is represented in our formalism. The excitation frequency within both wells $\omega_\Lambda$ can be related to the curvature of the potential as follows:
\begin{align}
	\omega_\Lambda = \sqrt{2E_{C}\frac{d^2}{d \varphi^2}V(\varphi_\Lambda)}
	=
	2\sqrt{\frac{E_{C}\Lambda \vd_\Lambda}{\overline{\rho}_\Lambda}},
\end{align}
where we have used Eq.~\eqref{eq:vexp} to obtain the last equality for $\overline{\rho}_\Lambda>0$. Thus, from Eq.~\eqref{eq:sb_cond}, the criterion to justify the two-level approximation can be understood as
\begin{align}
	\label{eq:sb_cond_pot}
	\frac{\mathcal{V}_\Lambda}{\omega_\Lambda}
	=
	\sqrt{\frac{\overline{\rho}_\Lambda \vd_\Lambda}{4\epsilon_{C,\Lambda}}}\gg 1,
\end{align}
where we recall that $\mathcal{V}_\Lambda=\Lambda \vd_\Lambda$ is the dimensionful potential depth.

Figure~\ref{fig:double_result}(a) shows the numerical result of the RG flow of $1/\vd_\Lambda$ and $\alpha_\Lambda$ with the initial conditions $\overline{\rho}_{\Lambda_0}=0.5$, $\epsilon_{C,\Lambda_0}=1$, $\gamma=20$ (corresponding to $\alpha_{\Lambda_0}=2.026$, see Eq.~\eqref{eq:alphadef}), and varying $1/\vd_{\Lambda_0}$. The fixed point is identified from the renormalized value of $\alpha_{\Lambda}$ at IR scale. As shown in Fig.~\ref{fig:double_result}(a), the critical point of the localization-delocalization  transition is $1/\vd_{\Lambda_0}=176.62$ at $\alpha_{\Lambda_0}\simeq 2$; for larger (smaller) values of $1/\vd_{\Lambda_0}$, the system flows to the delocalized (localized) phase. The corresponding schematic flow diagram is shown in  Fig.~\ref{fig:double_concept}(b). Surprisingly, the results demonstrate that the localized phase is extended to the regions in $\alpha<1$, thus indicating that the ground-state phase diagram must be qualitatively modified from the one expected in the spin-boson model. 
The origin of this striking discrepancy is traced to strong renormalizations of $\mathcal{V}_{\Lambda}$ and $\omega_{\Lambda}$ during the RG flows in nonperturbative regimes, which eventually invalidates the two-level approximation.  Indeed, we numerically observe that, except for the narrow area indicated by the green-shaded region in the inset of Fig.~\ref{fig:double_result}(a), the condition~\eqref{eq:sb_cond_pot} is not satisfied. Thus, the spin-boson description, which is supposed to be an effective model of cQED systems, can in general be invalid due to nonperturbative effects when the deep IR scale is reached. Since finite size/temperature effectively sets an IR cutoff in RG flows, this fact indicates that a nonperturbative analysis is essential when the waveguide is sufficiently long and the temperature is low enough as being relevant to recent experiments.

Finally, we also analyze the critical exponent associated with this new type of the localization-delocalization transition. The critical point is characterized by the divergence of the localization susceptibility defined by
\begin{align}
	\chi
	=
	\left(
	\frac{d^2}{d\varphi^2}
	V_{0}(\varphi_0)
	\right)^{-1}.
\end{align}
We introduce the critical exponent $\kappa$ associated with the divergence when $\gamma$ approaches the critical value $\gamma_c$ with a fixed $\vd_{\Lambda_0}$:
\begin{align}
	\chi\sim \frac{1}{|\gamma-\gamma_c|^{\kappa}}.
\end{align}
Figure~\ref{fig:double_result}(b) shows the divergent behavior of $\chi$ when $\gamma$ approaches $\gamma_{c}=20$ from the delocalized phase with $1/\vd_{\Lambda_0}=176.62$. The result of the linear fitting of the numerical data is also shown as the solid black line, which suggests $\kappa\approx 0.865862$. We note that this value is much smaller than those predicted in Refs.~\cite{aoki_nonperturbative_2002,kovacs_quantum-classical_2017}, where different type of regulator is used.

\section{Cosine potential\label{sec:cosine}}

\begin{figure*}[!t]
  \begin{center}
  	\includegraphics[width=2\columnwidth]{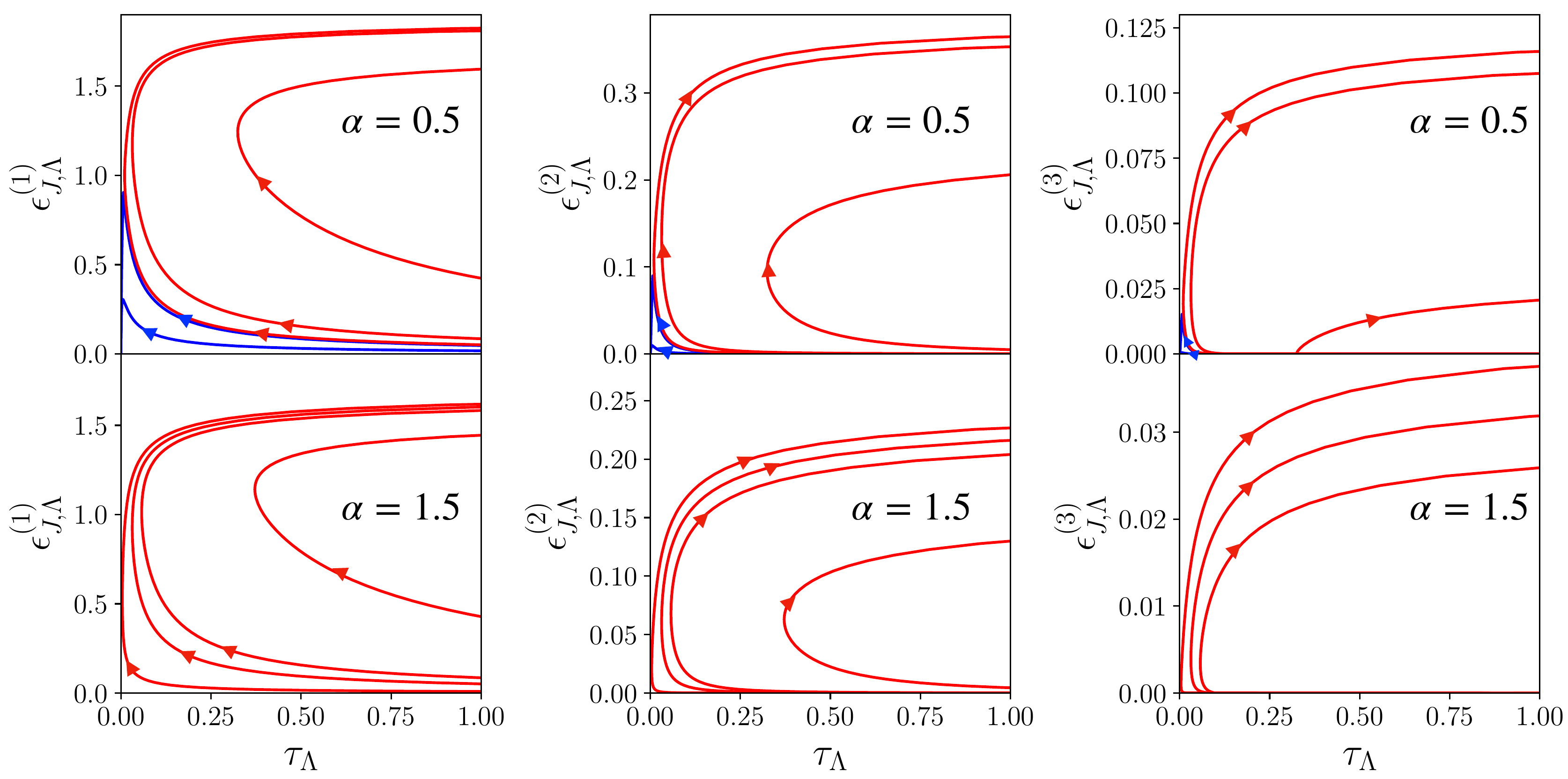}
    \caption{RG flow of $\tau_\Lambda$ and $\epsilon^{(1,2,3)}_{J,\Lambda}$ at $\alpha=0.5$ and $\alpha=1.5$ with $E_{C}/\Lambda_0=1/20$. In the case of $\alpha=0.5$, the flow approaches the fixed point of the insulating phase when $E_J/\Lambda_0=0.002$ and $0.0055$ (blue curves) or the superconducting phase when $E_J/\Lambda_0=0.006$, $0.01$, and $0.05$ (red curves). The results with $E_J/\Lambda_0=0.002$, $0.006$, $0.01$, and $0.05$ are shown in the case of $\alpha=1.5$, where the system always flows to the superconducting fixed point.}
    \label{fig:flow_cos}
  \end{center}
\end{figure*}
We next apply our general formalism to yet another fundamental problem, namely, the resistively shunted JJ, which corresponds to the case of a cosine potential
\begin{align}
	V(\varphi)=-E_{J}\cos(\varphi).
\end{align}
Early studies  \cite{Guinea85,aslangul_quantum_1985,Fisher85,zaikin_dynamics_1987} have neglected the perturbatively irrelevant $1/E_{C}$ term in Eq.~\eqref{eq:phi_action} and simplified the theory to the boundary sine-Gordon model. In this case, the ground state is expected to exhibit the superconducting phase in $\alpha>1$ and the insulator phase in $\alpha<1$ regardless of the value of $E_{J}/E_{C}$. Here, we note that $\alpha$ is defined in the same manner as in Eq.~\eqref{eq:def_alpha} and satisfies $\alpha=\gamma=R_{Q}/R$ since the distance between the nearest-neighbor potential minima is $2\pi$. However, a qualitatively new theoretical understanding has recently been revealed by nonperturbative analyses \cite{KM22}. Specifically, the previously overlooked $1/E_{C}$ term is found to turn into relevant at low energy due to nonperturbative effects, which leads to the suppression of the insulator phase into the deep charge regime $E_{J}/E_{C}\ll 1$ \cite{YN99,MY01}. 

The aim of this section is to provide further examinations of this nonperturbative picture on the basis of an advanced FRG approach developed here. To this end, we introduce the following LPA' ansatz, where the evolution of not only the lowest-order term, but also terms up to the second order in the derivative expansion are taken into account:
\begin{align}
	\Gamma_\Lambda[\varphi]
	\approx
	&
	\frac{1}{2}
	\int_{-\infty}^{\infty}
	\frac{d\omat}{2\pi}
	\left(
	\frac{\tau_{\Lambda}}{\Lambda}\omat^2
	+
	\frac{\gamma |\omat|}{2\pi}\right)
	\tilde{\varphi}(p)\tilde{\varphi}(-p)
	\notag
	\\
	&
	+\int_0^{\infty} d\itime 
	V_\Lambda\left(\varphi(\itime)\right),
	\label{eq:gam_tranc_cos}
\end{align}
where $V_\Lambda\left(\varphi\right)$ is an effective potential and $\tau_\Lambda$ is a dimensionless coefficient. These quantities satisfy
\begin{align}
	V_{\Lambda_0}\left(\varphi\right)=&V(\varphi),
	\\
	\tau_{\Lambda_{0}}=&\frac{\Lambda_{0}}{2E_{C}},
\end{align}
at the initial UV scale $\Lambda=\Lambda_0$.
 Because of the periodicity, $V_\Lambda\left(\varphi\right)$ can be expanded as
\begin{align}
	V_\Lambda\left(\varphi\right)
	=
	\Lambda
	\sum_{n=1}^{\infty}
	\epsilon_{J,\Lambda}^{(n)}\cos\left(n\varphi\right),
\end{align}
where $\epsilon_{J,\Lambda}^{(n)}$ is the dimensionless Fourier coefficient of order $n$. We here go beyond the analysis given in Ref.~\cite{KM22} by including the higher components with $n\geq 2$. As shown below, the most important conclusion drawn from our analysis is that the perturbatively irrelevant term proportional to $\tau_{\Lambda}$ can exhibit nonmonotonic RG flows due to nonperturbative corrections and qualitatively modify the phase diagram from the celebrated Schmid-Bulgadaev diagram. 

The flow equation of the effective potential is obtained in the same manner as in Eq.~\eqref{eq:vflow}:
\begin{align}
	\label{eq:flow_lamv_cos}
	&
	\partial_\Lambda V_\Lambda(\varphi)
	=
	\frac{1}{2}
	\int_{-\infty}^{\infty}\frac{d\omat}{2\pi}
	\partial_\Lambda \tilde{R}_{\Lambda}(\omat)
	G_\Lambda(\omat).
\end{align}
where
\begin{align}
	G_\Lambda(\omat)
	=
	\frac{1}{
	\frac{\gamma}{2\pi}
	|\omat|
	+
	\tau_{\Lambda}
	\frac{\omat^2}{\Lambda}	
	+
	V_{\Lambda}''(\varphi)
	+
	\tilde{R}_{\Lambda}(\omat)}.
\end{align}
In addition, plugging Eq.~\eqref{eq:gam_tranc_cos} into the second derivative of Eq.~\eqref{eq:wet} and extracting the zeroth-order Fourier component, we obtain
\begin{widetext}
\begin{align}
	\label{eq:flow_2nd_cos}
	\partial_\Lambda
	\left[
	\frac{\gamma}{2\pi}
	|\omat|
	+
	\tau_{\Lambda}
	\frac{\omat^2}{\Lambda}	
	\right]
	=
	\int_0^{2\pi}\frac{d\varphi}{2\pi}
	V_\Lambda^{(3)}(\varphi)^2
	\int_{-\infty}^{\infty}\frac{d\omat'}{2\pi}
	\partial_\Lambda \tilde{R}_{\Lambda}(\omat')	
	G_\Lambda(\omat')^2
	G_\Lambda(\omat+\omat').
\end{align}
To derive the flow equation of $\tau_\Lambda$, we expand the right-hand side of Eq.~\eqref{eq:flow_2nd_cos} with respect to $p$. The first-order term vanishes because $\lim_{p\to 0}\partial_p G_\Lambda(p+p')$ is an odd function of $p'$ and thus the momentum integral is evaluated as zero. This means that $\gamma=\alpha$ is not renormalized, which is consistent with previous analyses. From the second order of the expansion with respect to $p$, we obtain
\begin{align}
	\partial_\Lambda
	\frac{\tau_{\Lambda}}{\Lambda}	
	=&
	\frac{1}{2}
	\int_0^{2\pi}\frac{d\varphi}{2\pi}
	V_\Lambda^{(3)}(\varphi)^2
	\int_{-\infty}^{\infty}\frac{d\omat'}{2\pi}
	\partial_\Lambda \tilde{R}_{\Lambda}(\omat')	
	G_\Lambda(\omat')^2
	\lim_{\omat \to 0}
	\partial_{\omat}^2
	G_\Lambda(\omat+\omat').
	\label{eq:flow_taulam_cos}
\end{align}
By introducing $l=\ln(\Lambda_0/\Lambda)$ and rewriting Eqs.~\eqref{eq:flow_lamv_cos} and \eqref{eq:flow_taulam_cos}, our flow equations are summarized as
\begin{align}
	\label{eq:flow_v_cos}
	\partial_l
	\overline{V}_\Lambda(\varphi)
	=&
	\overline{V}_\Lambda(\varphi)
	-
	\frac{1}{2}
	\int_{-\infty}^{\infty}\frac{d(\omat/\Lambda)}{2\pi}
	\partial_\Lambda \tilde{R}_{\Lambda}(\omat)
	\overline{G}_\Lambda(\omat),
	\\
	\partial_l
	\tau_{\Lambda}
	=&
	-
	\tau_{\Lambda}
	-
	\frac{1}{2}
	\int_0^{2\pi}\frac{d\varphi}{2\pi}
	\overline{V}_\Lambda^{(3)}(\varphi)^2
	\int_{-\infty}^{\infty}\frac{d(\omat'/\Lambda)}{2\pi}
	\partial_\Lambda \tilde{R}_{\Lambda}(\omat')	
	\overline{G}_\Lambda(\omat')^2
	\lim_{\omat \to 0}
	\partial_{\omat}^2
	\overline{G}_\Lambda(\omat+\omat'),
	\label{eq:flow_tau_cos}
\end{align}
where $\overline{G}_\Lambda(\omat)=\Lambda G_\Lambda(\omat)$ and $\overline{V}_\Lambda(\varphi)=V_\Lambda(\varphi)/\Lambda$ are dimensionless. Hereafter, we set the regulator as $\tilde{R}_\Lambda(\omat)=\Lambda$.
\end{widetext}

In previous studies, the contribution of $\tau_\Lambda$ is often neglected as this term is expected to be irrelevant in perturbative regimes, as inferred from the minus sign in the first term of the right-hand side of Eq.~\eqref{eq:flow_tau_cos}. However, we find that nonperturbative corrections can make $\tau_\Lambda$ grow at low-energy scale, which drastically changes the phase diagram. To see this explicitly, when $|\overline{V}_{\Lambda}''(\varphi)|\ll 1$, we can use Eq.~\eqref{eq:flow_v_cos} to simplify the flow equation of $\epsilon_{J,\Lambda}^{(n)}$ in $\tau_\Lambda\gg 1$ and $\tau_\Lambda\to 0$ as follows:
\begin{align}
	\label{eq:lnej}
	\partial_l 
	\ln 
	\epsilon_{J,\Lambda}^{(n)}
	\approx
	\begin{cases}
		1
		-
		\frac{\pi}{4\sqrt{\tau_\Lambda}}
		&
		(\tau_\Lambda\gg 1)
		\\
		1
		-
		\frac{n^2}{\alpha}
		&
		(\tau_\Lambda\to 0)
	\end{cases}.
\end{align}
On the one hand, the latter indicates the phase transition at $\alpha=1$, which is consistent with previous perturbative studies, suggesting that all the Fourier components are irrelevant in $\alpha<1$ while $\epsilon_{J,\Lambda}^{(1)}$ is relevant in $\alpha>1$. On the other hand, the former indicates that all the Fourier components can in fact be relevant, which suggests the favoring of the superconducting phase even in $\alpha<1$.

We determine the fixed point which the RG flow approaches by solving Eqs.~\eqref{eq:flow_v_cos} and \eqref{eq:flow_tau_cos}. For the purpose of including the higher-order Fourier components of $\overline{V}_\Lambda(\varphi)$, we use the grid method, where $\overline{V}_\Lambda(\varphi)$ is evaluated on grid points in $\varphi \in [0,\pi]$. Our calculation is carried out on 128 grid points with equal intervals. Figure~\ref{fig:flow_cos} shows the RG flow of $\tau_\Lambda$ and $\epsilon^{(n\leq 3)}_{J,\Lambda}$ at $\alpha=0.5$ and $\alpha=1.5$ with different $E_J/\Lambda_0$. One observes that the superconducting phase is present not only in $\alpha>1$, but also in $\alpha<1$ with relatively large $E_J/\Lambda_0$, while the insulator phase appears in  $\alpha<1$ with small $E_J/\Lambda_0$. In particular, at small $\epsilon_{J,\Lambda}^{(n)}$, the region of the insulator phase spreads as $\tau_\Lambda$ decreases, which is consistent with the behavior expected from Eq.~\eqref{eq:lnej}.

\begin{figure}[t]
  \begin{center}
  	\includegraphics[width=\columnwidth]{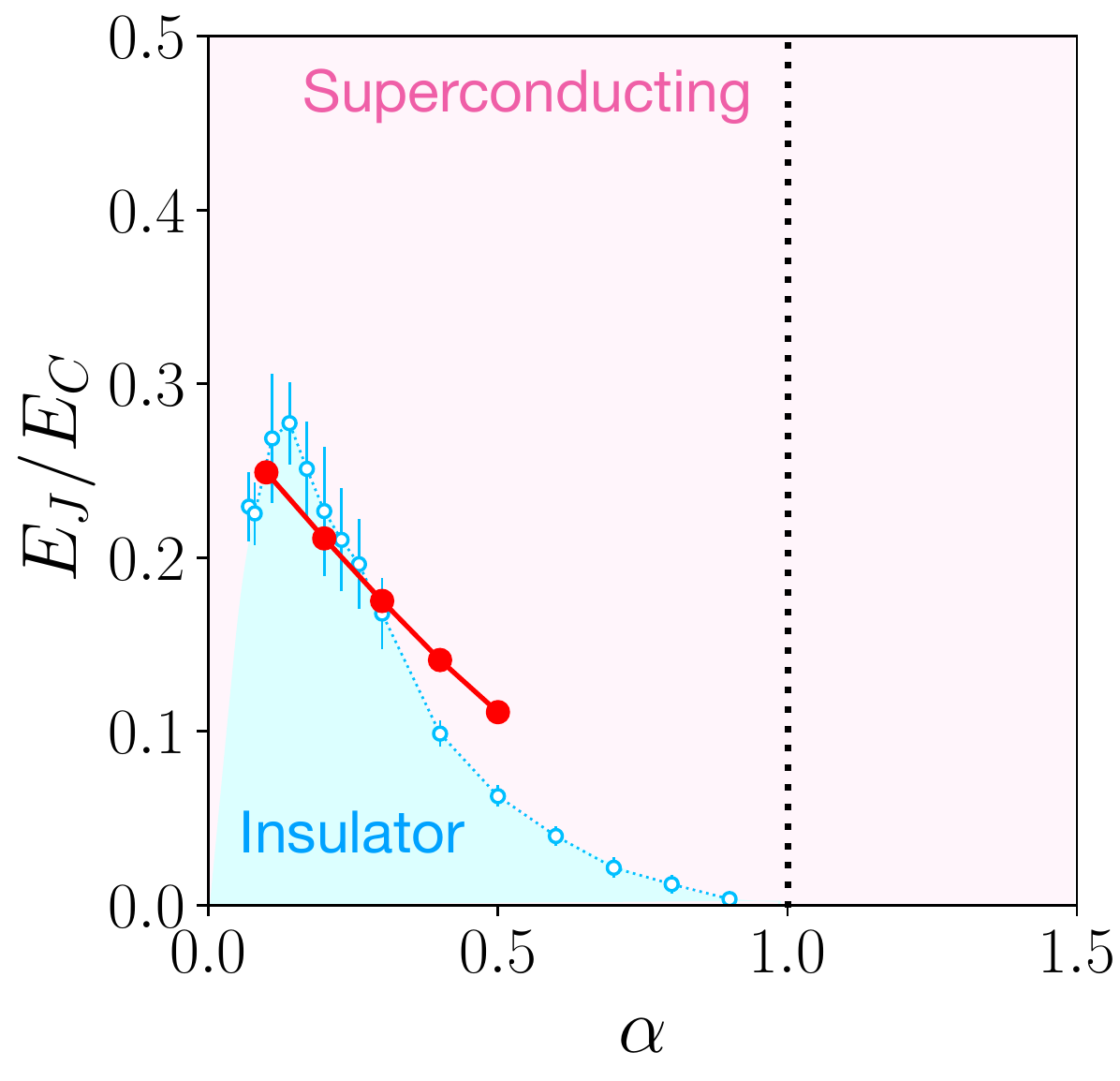}
    \caption{Phase diagram on the $(\alpha, E_J/E_C)$ plane. The red points (blue circles) are the results of FRG (NRG). In the FRG analysis, we fix $E_{C}/\Lambda_0=1/20$ and vary $E_{J}/\Lambda_0$ to find the transition points.}
    \label{fig:phase_cos}
  \end{center}
\end{figure}

Locating the critical value of $E_J/\Lambda_0$, we determine the phase boundary on the $(\alpha, E_J/E_C)$ plane. In Fig.~\ref{fig:phase_cos}, the FRG results at several values of $\alpha$, for which the calculation was performed without numerical instability, are shown as the red points. We also show the results obtained by the numerical RG (NRG) analysis as the blue circles \cite{KM22}. One can see the qualitative (and even nearly quantitative) agreement between the NRG and FRG results; both of them indicate that there exists the phase transition at finite values of $E_J/E_{\rm C}$ in $\alpha<1$ and the insulator phase shrinks as $\alpha$ approaches $\alpha=1$, while the superconducting phase appears at arbitrary $E_J/E_{\rm C}$ in $\alpha>1$. This makes a sharp contrast to what has been expected from the previous perturbative analyses, where the system was predicted to exhibit the insulator phase in $\alpha<1$ regardless of the value of $E_J/E_{\rm C}$.

\section{Conclusion\label{sec:conc}}

We developed a general framework to analyze superconducting circuits coupled to quantized electromagnetic continuum in nonperturbative regimes that have recently become accessible in cQED experiments. Our approach is based on the nonperturbative RG method, namely, FRG, and puts no limitations on the coupling strengths, enabling us to systematically explore the whole range of circuit parameters. In particular, the present FRG formalism provided the complete pictures of two representative cases that were previously supposed to be described by such effective models as the spin-boson model and the boundary sine-Gordon model. 

Specifically, we first applied the framework to the case of double-well potential in which the superconducting circuit has long been supposed to be well-approximated by a two-level system. Surprisingly, in contrast to the prediction from the spin-boson model, we found that the localized phase can be extended to the regions with a coupling strength $\alpha<1$ (Fig.~\ref{fig:double_concept}). The origin of this discrepancy was traced to strong renormalizations of both potential barrier and excitation frequency due to nonperturbative effects, which eventually invalidates the two-level description during RG flows. 

We next analyze the case of cosine potential that is also known as the resistively shunted JJ. In contrast to conventional perturbative analyses, our FRG approach now fully retained the previously overlooked capacitance term and led to the phase diagram that was strikingly different from the Schmid-Bulgadaev diagram (Fig.~\ref{fig:phase_cos}). We again identified nonperturbative effects as the origin of the failure of conventional understandings, where the nonmonotonic renormalization of the charging energy was found.  All in all, our results clearly indicate that a nonperturbative analysis is essential to develop an accurate understanding of recent cQED architectures realizing long high-impedance transmission lines.

Several further questions remain for future studies. First, while we focused on zero-frequency quantities relevant to dc measurements, it is intriguing to study what are finite-frequency signatures expected from the modified phase diagrams predicted in the present study. This consideration should have direct experimental relevance in light of recent developments in measuring finite-frequency response of cQED systems \cite{FDP17,MJP19,LS19,KR19,KR21,LS22}. Second, it merits further study to figure out how finite temperature and finite size of the transmission line can affect the present results. While these effects are expected to introduce effective IR cutoffs, one can address this question in a concrete manner on the basis of our FRG framework. In particular, one can start from the action~\eqref{eq:sgaussint} at finite size/temperature, and then follow the FRG procedure presented in this paper. Third,  it is also interesting to explore how decoherence or loss of microwave photons in a transmission line can lead to physical phenomena beyond the present findings. We envision that one can address this problem by combining our FRG framework with a method of Markovian open quantum systems \cite{YAreview}. Finally, while we focused on the Ohmic dissipation as appropriate for usual transmission lines, an extension of our analysis to the case of sub- or super-Ohmic dissipations merits further study \cite{AFB07,RB08}.

\acknowledgments
We are grateful to Eugene Demler, Atac Imamoglu, and Masaki Oshikawa for useful discussions. T.Y.~was supported by the RIKEN Special Postdoctoral Researchers Program. Y.A. acknowledges support from the Japan Society for the Promotion of Science through Grant No.~JP19K23424.

\begin{widetext}
\appendix
\section{Derivation of Eq.~\eqref{eq:sgaussint} \label{sec:deri}}

We present the derivation of Eq.~\eqref{eq:sgaussint} by evaluating the path integral in Eq.~\eqref{eq:Sphi}. First, we shift the integral variable as $N(\itime)\to N'(\itime)+n_{r}(\itime)-i\partial_{\itime}\varphi(\itime)/(2E_C)$ with
\begin{align}
	n_r(\itime)
	=&
	\frac{\sqrt{\gamma}}{2\pi}
	\sum_{0<k\leq W/v}
	\sqrt{\frac{2\pi}{kL}}
	\left(
	a^*_k(\itime)
	+
	a_k(\itime)
	\right)
\end{align}
and rewrite $S_{\rm all}$ as
\begin{align}
	\label{eq:sall_Np}
	S_{\rm all}\left[\varphi, N, a, a^*\right]
	=&
	\intb d\itime
	\left[
	in_r(\itime)
	\partial_\itime \varphi(\itime)
	+
	\sum_{0<k\leq W/v}
	a^*_k(\itime)\left(\partial_\itime+ \omega_k\right)a_k(\itime)
	+
	V\left(\varphi(\itime)\right)
	+
	\frac{\left(\partial_{\itime}\varphi(\itime)\right)^2}{4E_C}
	+
	E_{C}N'(\itime)^2
	\right].
\end{align}
In order to perform the integral with respect to $a_k(\itime)$ and $a_k^{*}(\itime)$, we introduce the Fourier transform:
\begin{align*}
	\tilde{\varphi}(\omat_n)
	=
	\intb d\itime e^{i\omat_n\itime}\varphi(\itime),
	\quad
	\tilde{a}_{k}(\omat_n)
	=
	\intb d\itime e^{i\omat_n\itime} a_k(\itime),
	\quad
	\tilde{a}_{k}^*(\omat_n)
	=
	\intb d\itime e^{i\omat_n\itime} a_k^*(\itime).
\end{align*}
Plugging these expressions into Eq.~\eqref{eq:sall_Np} and shifting the variables as
\begin{align*}
	\tilde{a}_k(\omat_n)
	\to
	\tilde{a}_k'(\omat_n)+\sqrt{\frac{\gamma}{2\pi kL}}\frac{\omat_n \tilde{\varphi}(\omat_n)}{i\omat_n+ \omega_k},
	\quad
	\tilde{a}_k^{*}(\omat_n)
	\to
	\tilde{a}_k^{\prime *}(\omat_n)+\sqrt{\frac{\gamma}{2\pi kL}}\frac{\omat_n \tilde{\varphi}(\omat_n)}{-i\omat_n+ \omega_k},
\end{align*}
$S_{\rm all}$ is rewritten as
\begin{align}
	S_{\rm all}\left[\varphi, N, a, a^*\right]
	=&
	\frac{1}{\beta}\sum_n
	\tilde{\varphi}(-\omat_n)
	\left(
	\frac{\omega_n^2}{4E_C}
	+
	\sum_{0<k\leq W/v}
	\frac{\gamma}{2\pi kL}
	\frac{\omat_n^2(\omega_k-i\omat_n)}{\omat_n^2+ \omega_k^2}
	\right)
	\tilde{\varphi}(\omat_n)
	+
	\intb d\itime 	V\left(\varphi(\itime)\right)
	\notag
	\\
	&+
	\sum_{0<k\leq W/v}
	\frac{1}{\beta}\sum_{n}
	\left(i\omat_n+ \omega_k\right)
	\tilde{a}^{\prime *}_k(-\omat_n)
	\tilde{a}'_k(\omat_n)
	+
	E_{C}
	\intb d\itime N'(\itime)^2.
\end{align}
In this expression, the imaginary part of the first term on the right-hand side
vanishes since it is an odd function of $\omat_n$. Plugging this into Eq.~\eqref{eq:Sphi}, integrating $\tilde{a}'$, $\tilde{a}^{\prime *}$, and $N'(\itime)$ out, and using $\omega_k=vk$, we obtain Eq.~\eqref{eq:sgaussint} up to a constant term.
\end{widetext}

\bibliography{Manuscript}

\begin{thebibliography}{115}%
\makeatletter
\providecommand \@ifxundefined [1]{%
 \@ifx{#1\undefined}
}%
\providecommand \@ifnum [1]{%
 \ifnum #1\expandafter \@firstoftwo
 \else \expandafter \@secondoftwo
 \fi
}%
\providecommand \@ifx [1]{%
 \ifx #1\expandafter \@firstoftwo
 \else \expandafter \@secondoftwo
 \fi
}%
\providecommand \natexlab [1]{#1}%
\providecommand \enquote  [1]{``#1''}%
\providecommand \bibnamefont  [1]{#1}%
\providecommand \bibfnamefont [1]{#1}%
\providecommand \citenamefont [1]{#1}%
\providecommand \href@noop [0]{\@secondoftwo}%
\providecommand \href [0]{\begingroup \@sanitize@url \@href}%
\providecommand \@href[1]{\@@startlink{#1}\@@href}%
\providecommand \@@href[1]{\endgroup#1\@@endlink}%
\providecommand \@sanitize@url [0]{\catcode `\\12\catcode `\$12\catcode
  `\&12\catcode `\#12\catcode `\^12\catcode `\_12\catcode `\%12\relax}%
\providecommand \@@startlink[1]{}%
\providecommand \@@endlink[0]{}%
\providecommand \url  [0]{\begingroup\@sanitize@url \@url }%
\providecommand \@url [1]{\endgroup\@href {#1}{\urlprefix }}%
\providecommand \urlprefix  [0]{URL }%
\providecommand \Eprint [0]{\href }%
\providecommand \doibase [0]{http://dx.doi.org/}%
\providecommand \selectlanguage [0]{\@gobble}%
\providecommand \bibinfo  [0]{\@secondoftwo}%
\providecommand \bibfield  [0]{\@secondoftwo}%
\providecommand \translation [1]{[#1]}%
\providecommand \BibitemOpen [0]{}%
\providecommand \bibitemStop [0]{}%
\providecommand \bibitemNoStop [0]{.\EOS\space}%
\providecommand \EOS [0]{\spacefactor3000\relax}%
\providecommand \BibitemShut  [1]{\csname bibitem#1\endcsname}%
\let\auto@bib@innerbib\@empty
\bibitem [{\citenamefont {Wallraff}\ \emph {et~al.}(2004)\citenamefont
  {Wallraff}, \citenamefont {Schuster}, \citenamefont {Blais}, \citenamefont
  {Frunzio}, \citenamefont {Huang}, \citenamefont {Majer}, \citenamefont
  {Kumar}, \citenamefont {Girvin},\ and\ \citenamefont {Schoelkopf}}]{WA04}%
  \BibitemOpen
  \bibfield  {author} {\bibinfo {author} {\bibfnamefont {A.}~\bibnamefont
  {Wallraff}}, \bibinfo {author} {\bibfnamefont {D.~I.}\ \bibnamefont
  {Schuster}}, \bibinfo {author} {\bibfnamefont {A.}~\bibnamefont {Blais}},
  \bibinfo {author} {\bibfnamefont {L.}~\bibnamefont {Frunzio}}, \bibinfo
  {author} {\bibfnamefont {R.-S.}\ \bibnamefont {Huang}}, \bibinfo {author}
  {\bibfnamefont {J.}~\bibnamefont {Majer}}, \bibinfo {author} {\bibfnamefont
  {S.}~\bibnamefont {Kumar}}, \bibinfo {author} {\bibfnamefont {S.~M.}\
  \bibnamefont {Girvin}}, \ and\ \bibinfo {author} {\bibfnamefont {R.~J.}\
  \bibnamefont {Schoelkopf}},\ }\href
  {https://www.nature.com/articles/nature02851} {\bibfield  {journal} {\bibinfo
   {journal} {Nature}\ }\textbf {\bibinfo {volume} {431}},\ \bibinfo {pages}
  {162} (\bibinfo {year} {2004})}\BibitemShut {NoStop}%
\bibitem [{\citenamefont {Bishop}\ \emph {et~al.}(2009)\citenamefont {Bishop},
  \citenamefont {Chow}, \citenamefont {Koch}, \citenamefont {Houck},
  \citenamefont {Devoret}, \citenamefont {Thuneberg}, \citenamefont {Girvin},\
  and\ \citenamefont {Schoelkopf}}]{BLS09}%
  \BibitemOpen
  \bibfield  {author} {\bibinfo {author} {\bibfnamefont {L.~S.}\ \bibnamefont
  {Bishop}}, \bibinfo {author} {\bibfnamefont {J.}~\bibnamefont {Chow}},
  \bibinfo {author} {\bibfnamefont {J.}~\bibnamefont {Koch}}, \bibinfo {author}
  {\bibfnamefont {A.}~\bibnamefont {Houck}}, \bibinfo {author} {\bibfnamefont
  {M.}~\bibnamefont {Devoret}}, \bibinfo {author} {\bibfnamefont
  {E.}~\bibnamefont {Thuneberg}}, \bibinfo {author} {\bibfnamefont
  {S.}~\bibnamefont {Girvin}}, \ and\ \bibinfo {author} {\bibfnamefont
  {R.}~\bibnamefont {Schoelkopf}},\ }\href
  {https://www.nature.com/articles/nphys1154} {\bibfield  {journal} {\bibinfo
  {journal} {Nat. Phys.}\ }\textbf {\bibinfo {volume} {5}},\ \bibinfo {pages}
  {105} (\bibinfo {year} {2009})}\BibitemShut {NoStop}%
\bibitem [{\citenamefont {Astafiev}\ \emph {et~al.}(2010)\citenamefont
  {Astafiev}, \citenamefont {Zagoskin}, \citenamefont {Abdumalikov},
  \citenamefont {Pashkin}, \citenamefont {Yamamoto}, \citenamefont {Inomata},
  \citenamefont {Nakamura},\ and\ \citenamefont {Tsai}}]{AO10}%
  \BibitemOpen
  \bibfield  {author} {\bibinfo {author} {\bibfnamefont {O.}~\bibnamefont
  {Astafiev}}, \bibinfo {author} {\bibfnamefont {A.~M.}\ \bibnamefont
  {Zagoskin}}, \bibinfo {author} {\bibfnamefont {A.~A.}\ \bibnamefont
  {Abdumalikov}}, \bibinfo {author} {\bibfnamefont {Y.~A.}\ \bibnamefont
  {Pashkin}}, \bibinfo {author} {\bibfnamefont {T.}~\bibnamefont {Yamamoto}},
  \bibinfo {author} {\bibfnamefont {K.}~\bibnamefont {Inomata}}, \bibinfo
  {author} {\bibfnamefont {Y.}~\bibnamefont {Nakamura}}, \ and\ \bibinfo
  {author} {\bibfnamefont {J.~S.}\ \bibnamefont {Tsai}},\ }\href {\doibase
  10.1126/science.1181918} {\bibfield  {journal} {\bibinfo  {journal}
  {Science}\ }\textbf {\bibinfo {volume} {327}},\ \bibinfo {pages} {840}
  (\bibinfo {year} {2010})}\BibitemShut {NoStop}%
\bibitem [{\citenamefont {Hoi}\ \emph {et~al.}(2012)\citenamefont {Hoi},
  \citenamefont {Palomaki}, \citenamefont {Lindkvist}, \citenamefont
  {Johansson}, \citenamefont {Delsing},\ and\ \citenamefont {Wilson}}]{HIC12}%
  \BibitemOpen
  \bibfield  {author} {\bibinfo {author} {\bibfnamefont {I.-C.}\ \bibnamefont
  {Hoi}}, \bibinfo {author} {\bibfnamefont {T.}~\bibnamefont {Palomaki}},
  \bibinfo {author} {\bibfnamefont {J.}~\bibnamefont {Lindkvist}}, \bibinfo
  {author} {\bibfnamefont {G.}~\bibnamefont {Johansson}}, \bibinfo {author}
  {\bibfnamefont {P.}~\bibnamefont {Delsing}}, \ and\ \bibinfo {author}
  {\bibfnamefont {C.~M.}\ \bibnamefont {Wilson}},\ }\href {\doibase
  10.1103/PhysRevLett.108.263601} {\bibfield  {journal} {\bibinfo  {journal}
  {Phys. Rev. Lett.}\ }\textbf {\bibinfo {volume} {108}},\ \bibinfo {pages}
  {263601} (\bibinfo {year} {2012})}\BibitemShut {NoStop}%
\bibitem [{\citenamefont {Reitz}\ \emph {et~al.}(2013)\citenamefont {Reitz},
  \citenamefont {Sayrin}, \citenamefont {Mitsch}, \citenamefont {Schneeweiss},\
  and\ \citenamefont {Rauschenbeutel}}]{RD13}%
  \BibitemOpen
  \bibfield  {author} {\bibinfo {author} {\bibfnamefont {D.}~\bibnamefont
  {Reitz}}, \bibinfo {author} {\bibfnamefont {C.}~\bibnamefont {Sayrin}},
  \bibinfo {author} {\bibfnamefont {R.}~\bibnamefont {Mitsch}}, \bibinfo
  {author} {\bibfnamefont {P.}~\bibnamefont {Schneeweiss}}, \ and\ \bibinfo
  {author} {\bibfnamefont {A.}~\bibnamefont {Rauschenbeutel}},\ }\href
  {\doibase 10.1103/PhysRevLett.110.243603} {\bibfield  {journal} {\bibinfo
  {journal} {Phys. Rev. Lett.}\ }\textbf {\bibinfo {volume} {110}},\ \bibinfo
  {pages} {243603} (\bibinfo {year} {2013})}\BibitemShut {NoStop}%
\bibitem [{\citenamefont {Thompson}\ \emph {et~al.}(2013)\citenamefont
  {Thompson}, \citenamefont {Tiecke}, \citenamefont {de~Leon}, \citenamefont
  {Feist}, \citenamefont {Akimov}, \citenamefont {Gullans}, \citenamefont
  {Zibrov}, \citenamefont {Vuleti{\'c}},\ and\ \citenamefont {Lukin}}]{TJD13}%
  \BibitemOpen
  \bibfield  {author} {\bibinfo {author} {\bibfnamefont {J.~D.}\ \bibnamefont
  {Thompson}}, \bibinfo {author} {\bibfnamefont {T.~G.}\ \bibnamefont
  {Tiecke}}, \bibinfo {author} {\bibfnamefont {N.~P.}\ \bibnamefont {de~Leon}},
  \bibinfo {author} {\bibfnamefont {J.}~\bibnamefont {Feist}}, \bibinfo
  {author} {\bibfnamefont {A.~V.}\ \bibnamefont {Akimov}}, \bibinfo {author}
  {\bibfnamefont {M.}~\bibnamefont {Gullans}}, \bibinfo {author} {\bibfnamefont
  {A.~S.}\ \bibnamefont {Zibrov}}, \bibinfo {author} {\bibfnamefont
  {V.}~\bibnamefont {Vuleti{\'c}}}, \ and\ \bibinfo {author} {\bibfnamefont
  {M.~D.}\ \bibnamefont {Lukin}},\ }\href {\doibase 10.1126/science.1237125}
  {\bibfield  {journal} {\bibinfo  {journal} {Science}\ }\textbf {\bibinfo
  {volume} {340}},\ \bibinfo {pages} {1202} (\bibinfo {year}
  {2013})}\BibitemShut {NoStop}%
\bibitem [{\citenamefont {van Loo}\ \emph {et~al.}(2013)\citenamefont {van
  Loo}, \citenamefont {Fedorov}, \citenamefont {Lalumi{\`e}re}, \citenamefont
  {Sanders}, \citenamefont {Blais},\ and\ \citenamefont {Wallraff}}]{vLA13}%
  \BibitemOpen
  \bibfield  {author} {\bibinfo {author} {\bibfnamefont {A.~F.}\ \bibnamefont
  {van Loo}}, \bibinfo {author} {\bibfnamefont {A.}~\bibnamefont {Fedorov}},
  \bibinfo {author} {\bibfnamefont {K.}~\bibnamefont {Lalumi{\`e}re}}, \bibinfo
  {author} {\bibfnamefont {B.~C.}\ \bibnamefont {Sanders}}, \bibinfo {author}
  {\bibfnamefont {A.}~\bibnamefont {Blais}}, \ and\ \bibinfo {author}
  {\bibfnamefont {A.}~\bibnamefont {Wallraff}},\ }\href {\doibase
  10.1126/science.1244324} {\bibfield  {journal} {\bibinfo  {journal}
  {Science}\ }\textbf {\bibinfo {volume} {342}},\ \bibinfo {pages} {1494}
  (\bibinfo {year} {2013})}\BibitemShut {NoStop}%
\bibitem [{\citenamefont {Arcari}\ \emph {et~al.}(2014)\citenamefont {Arcari},
  \citenamefont {S\"ollner}, \citenamefont {Javadi}, \citenamefont
  {Lindskov~Hansen}, \citenamefont {Mahmoodian}, \citenamefont {Liu},
  \citenamefont {Thyrrestrup}, \citenamefont {Lee}, \citenamefont {Song},
  \citenamefont {Stobbe},\ and\ \citenamefont {Lodahl}}]{AMS14}%
  \BibitemOpen
  \bibfield  {author} {\bibinfo {author} {\bibfnamefont {M.}~\bibnamefont
  {Arcari}}, \bibinfo {author} {\bibfnamefont {I.}~\bibnamefont {S\"ollner}},
  \bibinfo {author} {\bibfnamefont {A.}~\bibnamefont {Javadi}}, \bibinfo
  {author} {\bibfnamefont {S.}~\bibnamefont {Lindskov~Hansen}}, \bibinfo
  {author} {\bibfnamefont {S.}~\bibnamefont {Mahmoodian}}, \bibinfo {author}
  {\bibfnamefont {J.}~\bibnamefont {Liu}}, \bibinfo {author} {\bibfnamefont
  {H.}~\bibnamefont {Thyrrestrup}}, \bibinfo {author} {\bibfnamefont {E.~H.}\
  \bibnamefont {Lee}}, \bibinfo {author} {\bibfnamefont {J.~D.}\ \bibnamefont
  {Song}}, \bibinfo {author} {\bibfnamefont {S.}~\bibnamefont {Stobbe}}, \ and\
  \bibinfo {author} {\bibfnamefont {P.}~\bibnamefont {Lodahl}},\ }\href
  {\doibase 10.1103/PhysRevLett.113.093603} {\bibfield  {journal} {\bibinfo
  {journal} {Phys. Rev. Lett.}\ }\textbf {\bibinfo {volume} {113}},\ \bibinfo
  {pages} {093603} (\bibinfo {year} {2014})}\BibitemShut {NoStop}%
\bibitem [{\citenamefont {Yalla}\ \emph {et~al.}(2014)\citenamefont {Yalla},
  \citenamefont {Sadgrove}, \citenamefont {Nayak},\ and\ \citenamefont
  {Hakuta}}]{YR14}%
  \BibitemOpen
  \bibfield  {author} {\bibinfo {author} {\bibfnamefont {R.}~\bibnamefont
  {Yalla}}, \bibinfo {author} {\bibfnamefont {M.}~\bibnamefont {Sadgrove}},
  \bibinfo {author} {\bibfnamefont {K.~P.}\ \bibnamefont {Nayak}}, \ and\
  \bibinfo {author} {\bibfnamefont {K.}~\bibnamefont {Hakuta}},\ }\href
  {\doibase 10.1103/PhysRevLett.113.143601} {\bibfield  {journal} {\bibinfo
  {journal} {Phys. Rev. Lett.}\ }\textbf {\bibinfo {volume} {113}},\ \bibinfo
  {pages} {143601} (\bibinfo {year} {2014})}\BibitemShut {NoStop}%
\bibitem [{\citenamefont {Mlynek}\ \emph {et~al.}(2014)\citenamefont {Mlynek},
  \citenamefont {Abdumalikov}, \citenamefont {Eichler},\ and\ \citenamefont
  {Wallraff}}]{MJA14}%
  \BibitemOpen
  \bibfield  {author} {\bibinfo {author} {\bibfnamefont {J.~A.}\ \bibnamefont
  {Mlynek}}, \bibinfo {author} {\bibfnamefont {A.~A.}\ \bibnamefont
  {Abdumalikov}}, \bibinfo {author} {\bibfnamefont {C.}~\bibnamefont
  {Eichler}}, \ and\ \bibinfo {author} {\bibfnamefont {A.}~\bibnamefont
  {Wallraff}},\ }\href {https://www.nature.com/articles/ncomms6186} {\bibfield
  {journal} {\bibinfo  {journal} {Nat. Commun.}\ }\textbf {\bibinfo {volume}
  {5}},\ \bibinfo {pages} {1} (\bibinfo {year} {2014})}\BibitemShut {NoStop}%
\bibitem [{\citenamefont {Goban}\ \emph {et~al.}(2014)\citenamefont {Goban},
  \citenamefont {Hung}, \citenamefont {Yu}, \citenamefont {Hood}, \citenamefont
  {Muniz}, \citenamefont {Lee}, \citenamefont {Martin}, \citenamefont
  {McClung}, \citenamefont {Choi}, \citenamefont {Chang}, \citenamefont
  {Painter},\ and\ \citenamefont {Kimble}}]{GA14}%
  \BibitemOpen
  \bibfield  {author} {\bibinfo {author} {\bibfnamefont {A.}~\bibnamefont
  {Goban}}, \bibinfo {author} {\bibfnamefont {C.-L.}\ \bibnamefont {Hung}},
  \bibinfo {author} {\bibfnamefont {S.-P.}\ \bibnamefont {Yu}}, \bibinfo
  {author} {\bibfnamefont {J.}~\bibnamefont {Hood}}, \bibinfo {author}
  {\bibfnamefont {J.}~\bibnamefont {Muniz}}, \bibinfo {author} {\bibfnamefont
  {J.}~\bibnamefont {Lee}}, \bibinfo {author} {\bibfnamefont {M.}~\bibnamefont
  {Martin}}, \bibinfo {author} {\bibfnamefont {A.}~\bibnamefont {McClung}},
  \bibinfo {author} {\bibfnamefont {K.}~\bibnamefont {Choi}}, \bibinfo {author}
  {\bibfnamefont {D.~E.}\ \bibnamefont {Chang}}, \bibinfo {author}
  {\bibfnamefont {O.}~\bibnamefont {Painter}}, \ and\ \bibinfo {author}
  {\bibfnamefont {J.}~\bibnamefont {Kimble}},\ }\href
  {https://www.nature.com/articles/ncomms4808} {\bibfield  {journal} {\bibinfo
  {journal} {Nat. Commun.}\ }\textbf {\bibinfo {volume} {5}},\ \bibinfo {pages}
  {1} (\bibinfo {year} {2014})}\BibitemShut {NoStop}%
\bibitem [{\citenamefont {Lodahl}\ \emph {et~al.}(2015)\citenamefont {Lodahl},
  \citenamefont {Mahmoodian},\ and\ \citenamefont {Stobbe}}]{LP15}%
  \BibitemOpen
  \bibfield  {author} {\bibinfo {author} {\bibfnamefont {P.}~\bibnamefont
  {Lodahl}}, \bibinfo {author} {\bibfnamefont {S.}~\bibnamefont {Mahmoodian}},
  \ and\ \bibinfo {author} {\bibfnamefont {S.}~\bibnamefont {Stobbe}},\ }\href
  {\doibase 10.1103/RevModPhys.87.347} {\bibfield  {journal} {\bibinfo
  {journal} {Rev. Mod. Phys.}\ }\textbf {\bibinfo {volume} {87}},\ \bibinfo
  {pages} {347} (\bibinfo {year} {2015})}\BibitemShut {NoStop}%
\bibitem [{\citenamefont {Mirhosseini}\ \emph {et~al.}(2019)\citenamefont
  {Mirhosseini}, \citenamefont {Kim}, \citenamefont {Zhang}, \citenamefont
  {Sipahigil}, \citenamefont {Dieterle}, \citenamefont {Keller}, \citenamefont
  {Asenjo-Garcia}, \citenamefont {Chang},\ and\ \citenamefont
  {Painter}}]{MM19}%
  \BibitemOpen
  \bibfield  {author} {\bibinfo {author} {\bibfnamefont {M.}~\bibnamefont
  {Mirhosseini}}, \bibinfo {author} {\bibfnamefont {E.}~\bibnamefont {Kim}},
  \bibinfo {author} {\bibfnamefont {X.}~\bibnamefont {Zhang}}, \bibinfo
  {author} {\bibfnamefont {A.}~\bibnamefont {Sipahigil}}, \bibinfo {author}
  {\bibfnamefont {P.~B.}\ \bibnamefont {Dieterle}}, \bibinfo {author}
  {\bibfnamefont {A.~J.}\ \bibnamefont {Keller}}, \bibinfo {author}
  {\bibfnamefont {A.}~\bibnamefont {Asenjo-Garcia}}, \bibinfo {author}
  {\bibfnamefont {D.~E.}\ \bibnamefont {Chang}}, \ and\ \bibinfo {author}
  {\bibfnamefont {O.}~\bibnamefont {Painter}},\ }\href
  {https://www.nature.com/articles/s41586-019-1196-1} {\bibfield  {journal}
  {\bibinfo  {journal} {Nature}\ }\textbf {\bibinfo {volume} {569}},\ \bibinfo
  {pages} {692} (\bibinfo {year} {2019})}\BibitemShut {NoStop}%
\bibitem [{\citenamefont {Kannan}\ \emph {et~al.}(2020)\citenamefont {Kannan},
  \citenamefont {Ruckriegel}, \citenamefont {Campbell}, \citenamefont {Kockum},
  \citenamefont {Braum{\"u}ller}, \citenamefont {Kim}, \citenamefont
  {Kjaergaard}, \citenamefont {Krantz}, \citenamefont {Melville}, \citenamefont
  {Niedzielski}, \citenamefont {Vepsalainen}, \citenamefont {Winik},
  \citenamefont {Yoder}, \citenamefont {Nori}, \citenamefont {Orlando},
  \citenamefont {Gustavsson},\ and\ \citenamefont {Oliver}}]{KB20}%
  \BibitemOpen
  \bibfield  {author} {\bibinfo {author} {\bibfnamefont {B.}~\bibnamefont
  {Kannan}}, \bibinfo {author} {\bibfnamefont {M.~J.}\ \bibnamefont
  {Ruckriegel}}, \bibinfo {author} {\bibfnamefont {D.~L.}\ \bibnamefont
  {Campbell}}, \bibinfo {author} {\bibfnamefont {A.~F.}\ \bibnamefont
  {Kockum}}, \bibinfo {author} {\bibfnamefont {J.}~\bibnamefont
  {Braum{\"u}ller}}, \bibinfo {author} {\bibfnamefont {D.~K.}\ \bibnamefont
  {Kim}}, \bibinfo {author} {\bibfnamefont {M.}~\bibnamefont {Kjaergaard}},
  \bibinfo {author} {\bibfnamefont {P.}~\bibnamefont {Krantz}}, \bibinfo
  {author} {\bibfnamefont {A.}~\bibnamefont {Melville}}, \bibinfo {author}
  {\bibfnamefont {B.~M.}\ \bibnamefont {Niedzielski}}, \bibinfo {author}
  {\bibfnamefont {A.}~\bibnamefont {Vepsalainen}}, \bibinfo {author}
  {\bibfnamefont {R.}~\bibnamefont {Winik}}, \bibinfo {author} {\bibfnamefont
  {J.~L.}\ \bibnamefont {Yoder}}, \bibinfo {author} {\bibfnamefont
  {F.}~\bibnamefont {Nori}}, \bibinfo {author} {\bibfnamefont {T.~P.}\
  \bibnamefont {Orlando}}, \bibinfo {author} {\bibfnamefont {S.}~\bibnamefont
  {Gustavsson}}, \ and\ \bibinfo {author} {\bibfnamefont {W.~D.}\ \bibnamefont
  {Oliver}},\ }\href {https://www.nature.com/articles/s41586-020-2529-9}
  {\bibfield  {journal} {\bibinfo  {journal} {Nature}\ }\textbf {\bibinfo
  {volume} {583}},\ \bibinfo {pages} {775} (\bibinfo {year}
  {2020})}\BibitemShut {NoStop}%
\bibitem [{\citenamefont {Zheng}\ \emph {et~al.}(2010)\citenamefont {Zheng},
  \citenamefont {Gauthier},\ and\ \citenamefont {Baranger}}]{ZH10}%
  \BibitemOpen
  \bibfield  {author} {\bibinfo {author} {\bibfnamefont {H.}~\bibnamefont
  {Zheng}}, \bibinfo {author} {\bibfnamefont {D.~J.}\ \bibnamefont {Gauthier}},
  \ and\ \bibinfo {author} {\bibfnamefont {H.~U.}\ \bibnamefont {Baranger}},\
  }\href {\doibase 10.1103/PhysRevA.82.063816} {\bibfield  {journal} {\bibinfo
  {journal} {Phys. Rev. A}\ }\textbf {\bibinfo {volume} {82}},\ \bibinfo
  {pages} {063816} (\bibinfo {year} {2010})}\BibitemShut {NoStop}%
\bibitem [{\citenamefont {Gonz\'alez-Tudela}\ and\ \citenamefont
  {Porras}(2013)}]{GTA13}%
  \BibitemOpen
  \bibfield  {author} {\bibinfo {author} {\bibfnamefont {A.}~\bibnamefont
  {Gonz\'alez-Tudela}}\ and\ \bibinfo {author} {\bibfnamefont {D.}~\bibnamefont
  {Porras}},\ }\href {\doibase 10.1103/PhysRevLett.110.080502} {\bibfield
  {journal} {\bibinfo  {journal} {Phys. Rev. Lett.}\ }\textbf {\bibinfo
  {volume} {110}},\ \bibinfo {pages} {080502} (\bibinfo {year}
  {2013})}\BibitemShut {NoStop}%
\bibitem [{\citenamefont {Peropadre}\ \emph {et~al.}(2013)\citenamefont
  {Peropadre}, \citenamefont {Zueco}, \citenamefont {Porras},\ and\
  \citenamefont {Garc\'{\i}a-Ripoll}}]{PB13}%
  \BibitemOpen
  \bibfield  {author} {\bibinfo {author} {\bibfnamefont {B.}~\bibnamefont
  {Peropadre}}, \bibinfo {author} {\bibfnamefont {D.}~\bibnamefont {Zueco}},
  \bibinfo {author} {\bibfnamefont {D.}~\bibnamefont {Porras}}, \ and\ \bibinfo
  {author} {\bibfnamefont {J.~J.}\ \bibnamefont {Garc\'{\i}a-Ripoll}},\ }\href
  {\doibase 10.1103/PhysRevLett.111.243602} {\bibfield  {journal} {\bibinfo
  {journal} {Phys. Rev. Lett.}\ }\textbf {\bibinfo {volume} {111}},\ \bibinfo
  {pages} {243602} (\bibinfo {year} {2013})}\BibitemShut {NoStop}%
\bibitem [{\citenamefont {Shi}\ \emph {et~al.}(2018)\citenamefont {Shi},
  \citenamefont {Chang},\ and\ \citenamefont {Garc\'{\i}a-Ripoll}}]{TS18}%
  \BibitemOpen
  \bibfield  {author} {\bibinfo {author} {\bibfnamefont {T.}~\bibnamefont
  {Shi}}, \bibinfo {author} {\bibfnamefont {Y.}~\bibnamefont {Chang}}, \ and\
  \bibinfo {author} {\bibfnamefont {J.~J.}\ \bibnamefont
  {Garc\'{\i}a-Ripoll}},\ }\href {\doibase 10.1103/PhysRevLett.120.153602}
  {\bibfield  {journal} {\bibinfo  {journal} {Phys. Rev. Lett.}\ }\textbf
  {\bibinfo {volume} {120}},\ \bibinfo {pages} {153602} (\bibinfo {year}
  {2018})}\BibitemShut {NoStop}%
\bibitem [{\citenamefont {Parra-Rodriguez}\ \emph {et~al.}(2018)\citenamefont
  {Parra-Rodriguez}, \citenamefont {Rico}, \citenamefont {Solano},\ and\
  \citenamefont {Egusquiza}}]{PR18}%
  \BibitemOpen
  \bibfield  {author} {\bibinfo {author} {\bibfnamefont {A.}~\bibnamefont
  {Parra-Rodriguez}}, \bibinfo {author} {\bibfnamefont {E.}~\bibnamefont
  {Rico}}, \bibinfo {author} {\bibfnamefont {E.}~\bibnamefont {Solano}}, \ and\
  \bibinfo {author} {\bibfnamefont {I.~L.}\ \bibnamefont {Egusquiza}},\ }\href
  {\doibase 10.1088/2058-9565/aab1ba} {\bibfield  {journal} {\bibinfo
  {journal} {Quant. Sci. Tech.}\ }\textbf {\bibinfo {volume} {3}},\ \bibinfo
  {pages} {024012} (\bibinfo {year} {2018})}\BibitemShut {NoStop}%
\bibitem [{\citenamefont {S\'anchez-Burillo}\ \emph {et~al.}(2019)\citenamefont
  {S\'anchez-Burillo}, \citenamefont {Mart\'{\i}n-Moreno}, \citenamefont
  {Garc\'{\i}a-Ripoll},\ and\ \citenamefont {Zueco}}]{SBE19}%
  \BibitemOpen
  \bibfield  {author} {\bibinfo {author} {\bibfnamefont {E.}~\bibnamefont
  {S\'anchez-Burillo}}, \bibinfo {author} {\bibfnamefont {L.}~\bibnamefont
  {Mart\'{\i}n-Moreno}}, \bibinfo {author} {\bibfnamefont {J.~J.}\ \bibnamefont
  {Garc\'{\i}a-Ripoll}}, \ and\ \bibinfo {author} {\bibfnamefont
  {D.}~\bibnamefont {Zueco}},\ }\href {\doibase 10.1103/PhysRevLett.123.013601}
  {\bibfield  {journal} {\bibinfo  {journal} {Phys. Rev. Lett.}\ }\textbf
  {\bibinfo {volume} {123}},\ \bibinfo {pages} {013601} (\bibinfo {year}
  {2019})}\BibitemShut {NoStop}%
\bibitem [{\citenamefont {Noachtar}\ \emph {et~al.}(2022)\citenamefont
  {Noachtar}, \citenamefont {Kn\"orzer},\ and\ \citenamefont
  {Jonsson}}]{NDD22}%
  \BibitemOpen
  \bibfield  {author} {\bibinfo {author} {\bibfnamefont {D.~D.}\ \bibnamefont
  {Noachtar}}, \bibinfo {author} {\bibfnamefont {J.}~\bibnamefont {Kn\"orzer}},
  \ and\ \bibinfo {author} {\bibfnamefont {R.~H.}\ \bibnamefont {Jonsson}},\
  }\href {\doibase 10.1103/PhysRevA.106.013702} {\bibfield  {journal} {\bibinfo
   {journal} {Phys. Rev. A}\ }\textbf {\bibinfo {volume} {106}},\ \bibinfo
  {pages} {013702} (\bibinfo {year} {2022})}\BibitemShut {NoStop}%
\bibitem [{\citenamefont {Jin}\ \emph {et~al.}(2017)\citenamefont {Jin},
  \citenamefont {Kim}, \citenamefont {Suh}, \citenamefont {Shi}, \citenamefont
  {Chen}, \citenamefont {Fan}, \citenamefont {Kam}, \citenamefont {Watanabe},
  \citenamefont {Taniguchi}, \citenamefont {Tongay}, \citenamefont {Zettl},
  \citenamefont {Wu},\ and\ \citenamefont {Wang}}]{JC17}%
  \BibitemOpen
  \bibfield  {author} {\bibinfo {author} {\bibfnamefont {C.}~\bibnamefont
  {Jin}}, \bibinfo {author} {\bibfnamefont {J.}~\bibnamefont {Kim}}, \bibinfo
  {author} {\bibfnamefont {J.}~\bibnamefont {Suh}}, \bibinfo {author}
  {\bibfnamefont {Z.}~\bibnamefont {Shi}}, \bibinfo {author} {\bibfnamefont
  {B.}~\bibnamefont {Chen}}, \bibinfo {author} {\bibfnamefont {X.}~\bibnamefont
  {Fan}}, \bibinfo {author} {\bibfnamefont {M.}~\bibnamefont {Kam}}, \bibinfo
  {author} {\bibfnamefont {K.}~\bibnamefont {Watanabe}}, \bibinfo {author}
  {\bibfnamefont {T.}~\bibnamefont {Taniguchi}}, \bibinfo {author}
  {\bibfnamefont {S.}~\bibnamefont {Tongay}}, \bibinfo {author} {\bibfnamefont
  {A.}~\bibnamefont {Zettl}}, \bibinfo {author} {\bibfnamefont
  {J.}~\bibnamefont {Wu}}, \ and\ \bibinfo {author} {\bibfnamefont
  {F.}~\bibnamefont {Wang}},\ }\href
  {https://www.nature.com/articles/nphys3928} {\bibfield  {journal} {\bibinfo
  {journal} {Nat. Phys.}\ }\textbf {\bibinfo {volume} {13}},\ \bibinfo {pages}
  {127} (\bibinfo {year} {2017})}\BibitemShut {NoStop}%
\bibitem [{\citenamefont {Klembt}\ \emph {et~al.}(2018)\citenamefont {Klembt},
  \citenamefont {Harder}, \citenamefont {Egorov}, \citenamefont {Winkler},
  \citenamefont {Ge}, \citenamefont {Bandres}, \citenamefont {Emmerling},
  \citenamefont {Worschech}, \citenamefont {Liew}, \citenamefont {Segev},
  \citenamefont {Schneider},\ and\ \citenamefont {Hoefling}}]{KS18}%
  \BibitemOpen
  \bibfield  {author} {\bibinfo {author} {\bibfnamefont {S.}~\bibnamefont
  {Klembt}}, \bibinfo {author} {\bibfnamefont {T.}~\bibnamefont {Harder}},
  \bibinfo {author} {\bibfnamefont {O.}~\bibnamefont {Egorov}}, \bibinfo
  {author} {\bibfnamefont {K.}~\bibnamefont {Winkler}}, \bibinfo {author}
  {\bibfnamefont {R.}~\bibnamefont {Ge}}, \bibinfo {author} {\bibfnamefont
  {M.}~\bibnamefont {Bandres}}, \bibinfo {author} {\bibfnamefont
  {M.}~\bibnamefont {Emmerling}}, \bibinfo {author} {\bibfnamefont
  {L.}~\bibnamefont {Worschech}}, \bibinfo {author} {\bibfnamefont
  {T.}~\bibnamefont {Liew}}, \bibinfo {author} {\bibfnamefont {M.}~\bibnamefont
  {Segev}}, \bibinfo {author} {\bibfnamefont {C.}~\bibnamefont {Schneider}}, \
  and\ \bibinfo {author} {\bibfnamefont {S.}~\bibnamefont {Hoefling}},\ }\href
  {https://www.nature.com/articles/s41586-018-0601-5} {\bibfield  {journal}
  {\bibinfo  {journal} {Nature}\ }\textbf {\bibinfo {volume} {562}},\ \bibinfo
  {pages} {552} (\bibinfo {year} {2018})}\BibitemShut {NoStop}%
\bibitem [{\citenamefont {Ravets}\ \emph {et~al.}(2018)\citenamefont {Ravets},
  \citenamefont {Kn\"uppel}, \citenamefont {Faelt}, \citenamefont {Cotlet},
  \citenamefont {Kroner}, \citenamefont {Wegscheider},\ and\ \citenamefont
  {Imamoglu}}]{RS18}%
  \BibitemOpen
  \bibfield  {author} {\bibinfo {author} {\bibfnamefont {S.}~\bibnamefont
  {Ravets}}, \bibinfo {author} {\bibfnamefont {P.}~\bibnamefont {Kn\"uppel}},
  \bibinfo {author} {\bibfnamefont {S.}~\bibnamefont {Faelt}}, \bibinfo
  {author} {\bibfnamefont {O.}~\bibnamefont {Cotlet}}, \bibinfo {author}
  {\bibfnamefont {M.}~\bibnamefont {Kroner}}, \bibinfo {author} {\bibfnamefont
  {W.}~\bibnamefont {Wegscheider}}, \ and\ \bibinfo {author} {\bibfnamefont
  {A.}~\bibnamefont {Imamoglu}},\ }\href {\doibase
  10.1103/PhysRevLett.120.057401} {\bibfield  {journal} {\bibinfo  {journal}
  {Phys. Rev. Lett.}\ }\textbf {\bibinfo {volume} {120}},\ \bibinfo {pages}
  {057401} (\bibinfo {year} {2018})}\BibitemShut {NoStop}%
\bibitem [{\citenamefont {Giles}\ \emph {et~al.}(2018)\citenamefont {Giles},
  \citenamefont {Dai}, \citenamefont {Vurgaftman}, \citenamefont {Hoffman},
  \citenamefont {Liu}, \citenamefont {Lindsay}, \citenamefont {Ellis},
  \citenamefont {Assefa}, \citenamefont {Chatzakis}, \citenamefont {Reinecke},
  \citenamefont {Tischler}, \citenamefont {Fogler}, \citenamefont {Edgar},
  \citenamefont {Basov},\ and\ \citenamefont {Caldwell}}]{GA18}%
  \BibitemOpen
  \bibfield  {author} {\bibinfo {author} {\bibfnamefont {A.~J.}\ \bibnamefont
  {Giles}}, \bibinfo {author} {\bibfnamefont {S.}~\bibnamefont {Dai}}, \bibinfo
  {author} {\bibfnamefont {I.}~\bibnamefont {Vurgaftman}}, \bibinfo {author}
  {\bibfnamefont {T.}~\bibnamefont {Hoffman}}, \bibinfo {author} {\bibfnamefont
  {S.}~\bibnamefont {Liu}}, \bibinfo {author} {\bibfnamefont {L.}~\bibnamefont
  {Lindsay}}, \bibinfo {author} {\bibfnamefont {C.~T.}\ \bibnamefont {Ellis}},
  \bibinfo {author} {\bibfnamefont {N.}~\bibnamefont {Assefa}}, \bibinfo
  {author} {\bibfnamefont {I.}~\bibnamefont {Chatzakis}}, \bibinfo {author}
  {\bibfnamefont {T.~L.}\ \bibnamefont {Reinecke}}, \bibinfo {author}
  {\bibfnamefont {J.~G.}\ \bibnamefont {Tischler}}, \bibinfo {author}
  {\bibfnamefont {M.~M.}\ \bibnamefont {Fogler}}, \bibinfo {author}
  {\bibfnamefont {J.~H.}\ \bibnamefont {Edgar}}, \bibinfo {author}
  {\bibfnamefont {D.~N.}\ \bibnamefont {Basov}}, \ and\ \bibinfo {author}
  {\bibfnamefont {J.~D.}\ \bibnamefont {Caldwell}},\ }\href
  {https://www.nature.com/articles/nmat5047} {\bibfield  {journal} {\bibinfo
  {journal} {Nat. Mater.}\ }\textbf {\bibinfo {volume} {17}},\ \bibinfo {pages}
  {134} (\bibinfo {year} {2018})}\BibitemShut {NoStop}%
\bibitem [{\citenamefont {Keller}\ \emph {et~al.}(2020)\citenamefont {Keller},
  \citenamefont {Scalari}, \citenamefont {Appugliese}, \citenamefont
  {Rajabali}, \citenamefont {Beck}, \citenamefont {Haase}, \citenamefont
  {Lehner}, \citenamefont {Wegscheider}, \citenamefont {Failla}, \citenamefont
  {Myronov}, \citenamefont {Leadley}, \citenamefont {Lloyd-Hughes},
  \citenamefont {Nataf},\ and\ \citenamefont {Faist}}]{KJ20}%
  \BibitemOpen
  \bibfield  {author} {\bibinfo {author} {\bibfnamefont {J.}~\bibnamefont
  {Keller}}, \bibinfo {author} {\bibfnamefont {G.}~\bibnamefont {Scalari}},
  \bibinfo {author} {\bibfnamefont {F.}~\bibnamefont {Appugliese}}, \bibinfo
  {author} {\bibfnamefont {S.}~\bibnamefont {Rajabali}}, \bibinfo {author}
  {\bibfnamefont {M.}~\bibnamefont {Beck}}, \bibinfo {author} {\bibfnamefont
  {J.}~\bibnamefont {Haase}}, \bibinfo {author} {\bibfnamefont {C.~A.}\
  \bibnamefont {Lehner}}, \bibinfo {author} {\bibfnamefont {W.}~\bibnamefont
  {Wegscheider}}, \bibinfo {author} {\bibfnamefont {M.}~\bibnamefont {Failla}},
  \bibinfo {author} {\bibfnamefont {M.}~\bibnamefont {Myronov}}, \bibinfo
  {author} {\bibfnamefont {D.~R.}\ \bibnamefont {Leadley}}, \bibinfo {author}
  {\bibfnamefont {J.}~\bibnamefont {Lloyd-Hughes}}, \bibinfo {author}
  {\bibfnamefont {P.}~\bibnamefont {Nataf}}, \ and\ \bibinfo {author}
  {\bibfnamefont {J.}~\bibnamefont {Faist}},\ }\href {\doibase
  10.1103/PhysRevB.101.075301} {\bibfield  {journal} {\bibinfo  {journal}
  {Phys. Rev. B}\ }\textbf {\bibinfo {volume} {101}},\ \bibinfo {pages}
  {075301} (\bibinfo {year} {2020})}\BibitemShut {NoStop}%
\bibitem [{\citenamefont {Chervy}\ \emph {et~al.}(2020)\citenamefont {Chervy},
  \citenamefont {Kn\"uppel}, \citenamefont {Abbaspour}, \citenamefont
  {Lupatini}, \citenamefont {F\"alt}, \citenamefont {Wegscheider},
  \citenamefont {Kroner},\ and\ \citenamefont {Imamoglu}}]{CT20}%
  \BibitemOpen
  \bibfield  {author} {\bibinfo {author} {\bibfnamefont {T.}~\bibnamefont
  {Chervy}}, \bibinfo {author} {\bibfnamefont {P.}~\bibnamefont {Kn\"uppel}},
  \bibinfo {author} {\bibfnamefont {H.}~\bibnamefont {Abbaspour}}, \bibinfo
  {author} {\bibfnamefont {M.}~\bibnamefont {Lupatini}}, \bibinfo {author}
  {\bibfnamefont {S.}~\bibnamefont {F\"alt}}, \bibinfo {author} {\bibfnamefont
  {W.}~\bibnamefont {Wegscheider}}, \bibinfo {author} {\bibfnamefont
  {M.}~\bibnamefont {Kroner}}, \ and\ \bibinfo {author} {\bibfnamefont
  {A.}~\bibnamefont {Imamoglu}},\ }\href {\doibase 10.1103/PhysRevX.10.011040}
  {\bibfield  {journal} {\bibinfo  {journal} {Phys. Rev. X}\ }\textbf {\bibinfo
  {volume} {10}},\ \bibinfo {pages} {011040} (\bibinfo {year}
  {2020})}\BibitemShut {NoStop}%
\bibitem [{\citenamefont {Mueller}\ \emph {et~al.}(2020)\citenamefont
  {Mueller}, \citenamefont {Okamura}, \citenamefont {Vieira}, \citenamefont
  {Juergensen}, \citenamefont {Lange}, \citenamefont {Barros}, \citenamefont
  {Schulz},\ and\ \citenamefont {Reich}}]{MNS20}%
  \BibitemOpen
  \bibfield  {author} {\bibinfo {author} {\bibfnamefont {N.~S.}\ \bibnamefont
  {Mueller}}, \bibinfo {author} {\bibfnamefont {Y.}~\bibnamefont {Okamura}},
  \bibinfo {author} {\bibfnamefont {B.~G.}\ \bibnamefont {Vieira}}, \bibinfo
  {author} {\bibfnamefont {S.}~\bibnamefont {Juergensen}}, \bibinfo {author}
  {\bibfnamefont {H.}~\bibnamefont {Lange}}, \bibinfo {author} {\bibfnamefont
  {E.~B.}\ \bibnamefont {Barros}}, \bibinfo {author} {\bibfnamefont
  {F.}~\bibnamefont {Schulz}}, \ and\ \bibinfo {author} {\bibfnamefont
  {S.}~\bibnamefont {Reich}},\ }\href
  {https://www.nature.com/articles/s41586-020-2508-1} {\bibfield  {journal}
  {\bibinfo  {journal} {Nature}\ }\textbf {\bibinfo {volume} {583}},\ \bibinfo
  {pages} {780} (\bibinfo {year} {2020})}\BibitemShut {NoStop}%
\bibitem [{\citenamefont {Ruggenthaler}\ \emph {et~al.}(2014)\citenamefont
  {Ruggenthaler}, \citenamefont {Flick}, \citenamefont {Pellegrini},
  \citenamefont {Appel}, \citenamefont {Tokatly},\ and\ \citenamefont
  {Rubio}}]{RM14}%
  \BibitemOpen
  \bibfield  {author} {\bibinfo {author} {\bibfnamefont {M.}~\bibnamefont
  {Ruggenthaler}}, \bibinfo {author} {\bibfnamefont {J.}~\bibnamefont {Flick}},
  \bibinfo {author} {\bibfnamefont {C.}~\bibnamefont {Pellegrini}}, \bibinfo
  {author} {\bibfnamefont {H.}~\bibnamefont {Appel}}, \bibinfo {author}
  {\bibfnamefont {I.~V.}\ \bibnamefont {Tokatly}}, \ and\ \bibinfo {author}
  {\bibfnamefont {A.}~\bibnamefont {Rubio}},\ }\href {\doibase
  10.1103/PhysRevA.90.012508} {\bibfield  {journal} {\bibinfo  {journal} {Phys.
  Rev. A}\ }\textbf {\bibinfo {volume} {90}},\ \bibinfo {pages} {012508}
  (\bibinfo {year} {2014})}\BibitemShut {NoStop}%
\bibitem [{\citenamefont {Schachenmayer}\ \emph {et~al.}(2015)\citenamefont
  {Schachenmayer}, \citenamefont {Genes}, \citenamefont {Tignone},\ and\
  \citenamefont {Pupillo}}]{SJ15}%
  \BibitemOpen
  \bibfield  {author} {\bibinfo {author} {\bibfnamefont {J.}~\bibnamefont
  {Schachenmayer}}, \bibinfo {author} {\bibfnamefont {C.}~\bibnamefont
  {Genes}}, \bibinfo {author} {\bibfnamefont {E.}~\bibnamefont {Tignone}}, \
  and\ \bibinfo {author} {\bibfnamefont {G.}~\bibnamefont {Pupillo}},\ }\href
  {\doibase 10.1103/PhysRevLett.114.196403} {\bibfield  {journal} {\bibinfo
  {journal} {Phys. Rev. Lett.}\ }\textbf {\bibinfo {volume} {114}},\ \bibinfo
  {pages} {196403} (\bibinfo {year} {2015})}\BibitemShut {NoStop}%
\bibitem [{\citenamefont {Hagenm\"uller}\ \emph {et~al.}(2017)\citenamefont
  {Hagenm\"uller}, \citenamefont {Schachenmayer}, \citenamefont {Sch\"utz},
  \citenamefont {Genes},\ and\ \citenamefont {Pupillo}}]{HD17}%
  \BibitemOpen
  \bibfield  {author} {\bibinfo {author} {\bibfnamefont {D.}~\bibnamefont
  {Hagenm\"uller}}, \bibinfo {author} {\bibfnamefont {J.}~\bibnamefont
  {Schachenmayer}}, \bibinfo {author} {\bibfnamefont {S.}~\bibnamefont
  {Sch\"utz}}, \bibinfo {author} {\bibfnamefont {C.}~\bibnamefont {Genes}}, \
  and\ \bibinfo {author} {\bibfnamefont {G.}~\bibnamefont {Pupillo}},\ }\href
  {\doibase 10.1103/PhysRevLett.119.223601} {\bibfield  {journal} {\bibinfo
  {journal} {Phys. Rev. Lett.}\ }\textbf {\bibinfo {volume} {119}},\ \bibinfo
  {pages} {223601} (\bibinfo {year} {2017})}\BibitemShut {NoStop}%
\bibitem [{\citenamefont {Sentef}\ \emph {et~al.}(2018)\citenamefont {Sentef},
  \citenamefont {Ruggenthaler},\ and\ \citenamefont {Rubio}}]{SMA18}%
  \BibitemOpen
  \bibfield  {author} {\bibinfo {author} {\bibfnamefont {M.~A.}\ \bibnamefont
  {Sentef}}, \bibinfo {author} {\bibfnamefont {M.}~\bibnamefont
  {Ruggenthaler}}, \ and\ \bibinfo {author} {\bibfnamefont {A.}~\bibnamefont
  {Rubio}},\ }\href {https://advances.sciencemag.org/content/4/11/eaau6969}
  {\bibfield  {journal} {\bibinfo  {journal} {Sci. Adv.}\ }\textbf {\bibinfo
  {volume} {4}} (\bibinfo {year} {2018})}\BibitemShut {NoStop}%
\bibitem [{\citenamefont {Schlawin}\ \emph {et~al.}(2019)\citenamefont
  {Schlawin}, \citenamefont {Cavalleri},\ and\ \citenamefont {Jaksch}}]{SF19}%
  \BibitemOpen
  \bibfield  {author} {\bibinfo {author} {\bibfnamefont {F.}~\bibnamefont
  {Schlawin}}, \bibinfo {author} {\bibfnamefont {A.}~\bibnamefont {Cavalleri}},
  \ and\ \bibinfo {author} {\bibfnamefont {D.}~\bibnamefont {Jaksch}},\ }\href
  {\doibase 10.1103/PhysRevLett.122.133602} {\bibfield  {journal} {\bibinfo
  {journal} {Phys. Rev. Lett.}\ }\textbf {\bibinfo {volume} {122}},\ \bibinfo
  {pages} {133602} (\bibinfo {year} {2019})}\BibitemShut {NoStop}%
\bibitem [{\citenamefont {Curtis}\ \emph {et~al.}(2019)\citenamefont {Curtis},
  \citenamefont {Raines}, \citenamefont {Allocca}, \citenamefont {Hafezi},\
  and\ \citenamefont {Galitski}}]{CJB19}%
  \BibitemOpen
  \bibfield  {author} {\bibinfo {author} {\bibfnamefont {J.~B.}\ \bibnamefont
  {Curtis}}, \bibinfo {author} {\bibfnamefont {Z.~M.}\ \bibnamefont {Raines}},
  \bibinfo {author} {\bibfnamefont {A.~A.}\ \bibnamefont {Allocca}}, \bibinfo
  {author} {\bibfnamefont {M.}~\bibnamefont {Hafezi}}, \ and\ \bibinfo {author}
  {\bibfnamefont {V.~M.}\ \bibnamefont {Galitski}},\ }\href {\doibase
  10.1103/PhysRevLett.122.167002} {\bibfield  {journal} {\bibinfo  {journal}
  {Phys. Rev. Lett.}\ }\textbf {\bibinfo {volume} {122}},\ \bibinfo {pages}
  {167002} (\bibinfo {year} {2019})}\BibitemShut {NoStop}%
\bibitem [{\citenamefont {Rokaj}\ \emph {et~al.}(2019)\citenamefont {Rokaj},
  \citenamefont {Penz}, \citenamefont {Sentef}, \citenamefont {Ruggenthaler},\
  and\ \citenamefont {Rubio}}]{RV19}%
  \BibitemOpen
  \bibfield  {author} {\bibinfo {author} {\bibfnamefont {V.}~\bibnamefont
  {Rokaj}}, \bibinfo {author} {\bibfnamefont {M.}~\bibnamefont {Penz}},
  \bibinfo {author} {\bibfnamefont {M.~A.}\ \bibnamefont {Sentef}}, \bibinfo
  {author} {\bibfnamefont {M.}~\bibnamefont {Ruggenthaler}}, \ and\ \bibinfo
  {author} {\bibfnamefont {A.}~\bibnamefont {Rubio}},\ }\href {\doibase
  10.1103/PhysRevLett.123.047202} {\bibfield  {journal} {\bibinfo  {journal}
  {Phys. Rev. Lett.}\ }\textbf {\bibinfo {volume} {123}},\ \bibinfo {pages}
  {047202} (\bibinfo {year} {2019})}\BibitemShut {NoStop}%
\bibitem [{\citenamefont {Mazza}\ and\ \citenamefont {Georges}(2019)}]{MG19}%
  \BibitemOpen
  \bibfield  {author} {\bibinfo {author} {\bibfnamefont {G.}~\bibnamefont
  {Mazza}}\ and\ \bibinfo {author} {\bibfnamefont {A.}~\bibnamefont
  {Georges}},\ }\href {\doibase 10.1103/PhysRevLett.122.017401} {\bibfield
  {journal} {\bibinfo  {journal} {Phys. Rev. Lett.}\ }\textbf {\bibinfo
  {volume} {122}},\ \bibinfo {pages} {017401} (\bibinfo {year}
  {2019})}\BibitemShut {NoStop}%
\bibitem [{\citenamefont {Ashida}\ \emph
  {et~al.}(2020{\natexlab{a}})\citenamefont {Ashida}, \citenamefont {Imamoglu},
  \citenamefont {Faist}, \citenamefont {Jaksch}, \citenamefont {Cavalleri},\
  and\ \citenamefont {Demler}}]{YA20}%
  \BibitemOpen
  \bibfield  {author} {\bibinfo {author} {\bibfnamefont {Y.}~\bibnamefont
  {Ashida}}, \bibinfo {author} {\bibfnamefont {A.}~\bibnamefont {Imamoglu}},
  \bibinfo {author} {\bibfnamefont {J.}~\bibnamefont {Faist}}, \bibinfo
  {author} {\bibfnamefont {D.}~\bibnamefont {Jaksch}}, \bibinfo {author}
  {\bibfnamefont {A.}~\bibnamefont {Cavalleri}}, \ and\ \bibinfo {author}
  {\bibfnamefont {E.}~\bibnamefont {Demler}},\ }\href {\doibase
  10.1103/PhysRevX.10.041027} {\bibfield  {journal} {\bibinfo  {journal} {Phys.
  Rev. X}\ }\textbf {\bibinfo {volume} {10}},\ \bibinfo {pages} {041027}
  (\bibinfo {year} {2020}{\natexlab{a}})}\BibitemShut {NoStop}%
\bibitem [{\citenamefont {Pilar}\ \emph {et~al.}(2020)\citenamefont {Pilar},
  \citenamefont {De~Bernardis},\ and\ \citenamefont {Rabl}}]{PP20}%
  \BibitemOpen
  \bibfield  {author} {\bibinfo {author} {\bibfnamefont {P.}~\bibnamefont
  {Pilar}}, \bibinfo {author} {\bibfnamefont {D.}~\bibnamefont {De~Bernardis}},
  \ and\ \bibinfo {author} {\bibfnamefont {P.}~\bibnamefont {Rabl}},\ }\href
  {\doibase 10.22331/q-2020-09-28-335} {\bibfield  {journal} {\bibinfo
  {journal} {{Quantum}}\ }\textbf {\bibinfo {volume} {4}},\ \bibinfo {pages}
  {335} (\bibinfo {year} {2020})}\BibitemShut {NoStop}%
\bibitem [{\citenamefont {Latini}\ \emph {et~al.}(2021)\citenamefont {Latini},
  \citenamefont {Shin}, \citenamefont {Sato}, \citenamefont {Schäfer},
  \citenamefont {Giovannini}, \citenamefont {Hübener},\ and\ \citenamefont
  {Rubio}}]{SL21}%
  \BibitemOpen
  \bibfield  {author} {\bibinfo {author} {\bibfnamefont {S.}~\bibnamefont
  {Latini}}, \bibinfo {author} {\bibfnamefont {D.}~\bibnamefont {Shin}},
  \bibinfo {author} {\bibfnamefont {S.~A.}\ \bibnamefont {Sato}}, \bibinfo
  {author} {\bibfnamefont {C.}~\bibnamefont {Schäfer}}, \bibinfo {author}
  {\bibfnamefont {U.~D.}\ \bibnamefont {Giovannini}}, \bibinfo {author}
  {\bibfnamefont {H.}~\bibnamefont {Hübener}}, \ and\ \bibinfo {author}
  {\bibfnamefont {A.}~\bibnamefont {Rubio}},\ }\href {\doibase
  10.1073/pnas.2105618118} {\bibfield  {journal} {\bibinfo  {journal} {Proc.
  Natl. Acad. Sci. U.S.A.}\ }\textbf {\bibinfo {volume} {118}},\ \bibinfo
  {pages} {e2105618118} (\bibinfo {year} {2021})}\BibitemShut {NoStop}%
\bibitem [{\citenamefont {Dmytruk}\ and\ \citenamefont
  {Schir\'o}(2021)}]{DO21}%
  \BibitemOpen
  \bibfield  {author} {\bibinfo {author} {\bibfnamefont {O.}~\bibnamefont
  {Dmytruk}}\ and\ \bibinfo {author} {\bibfnamefont {M.}~\bibnamefont
  {Schir\'o}},\ }\href {\doibase 10.1103/PhysRevB.103.075131} {\bibfield
  {journal} {\bibinfo  {journal} {Phys. Rev. B}\ }\textbf {\bibinfo {volume}
  {103}},\ \bibinfo {pages} {075131} (\bibinfo {year} {2021})}\BibitemShut
  {NoStop}%
\bibitem [{\citenamefont {Ashida}\ \emph {et~al.}(2021)\citenamefont {Ashida},
  \citenamefont {Imamoglu},\ and\ \citenamefont {Demler}}]{YA21}%
  \BibitemOpen
  \bibfield  {author} {\bibinfo {author} {\bibfnamefont {Y.}~\bibnamefont
  {Ashida}}, \bibinfo {author} {\bibfnamefont {A.}~\bibnamefont {Imamoglu}}, \
  and\ \bibinfo {author} {\bibfnamefont {E.}~\bibnamefont {Demler}},\ }\href
  {\doibase 10.1103/PhysRevLett.126.153603} {\bibfield  {journal} {\bibinfo
  {journal} {Phys. Rev. Lett.}\ }\textbf {\bibinfo {volume} {126}},\ \bibinfo
  {pages} {153603} (\bibinfo {year} {2021})}\BibitemShut {NoStop}%
\bibitem [{\citenamefont {Chiocchetta}\ \emph {et~al.}(2021)\citenamefont
  {Chiocchetta}, \citenamefont {Kiese}, \citenamefont {Zelle}, \citenamefont
  {Piazza},\ and\ \citenamefont {Diehl}}]{CA21}%
  \BibitemOpen
  \bibfield  {author} {\bibinfo {author} {\bibfnamefont {A.}~\bibnamefont
  {Chiocchetta}}, \bibinfo {author} {\bibfnamefont {D.}~\bibnamefont {Kiese}},
  \bibinfo {author} {\bibfnamefont {C.~P.}\ \bibnamefont {Zelle}}, \bibinfo
  {author} {\bibfnamefont {F.}~\bibnamefont {Piazza}}, \ and\ \bibinfo {author}
  {\bibfnamefont {S.}~\bibnamefont {Diehl}},\ }\href
  {https://www.nature.com/articles/s41467-021-26076-3} {\bibfield  {journal}
  {\bibinfo  {journal} {Nat. Commun.}\ }\textbf {\bibinfo {volume} {12}},\
  \bibinfo {pages} {1} (\bibinfo {year} {2021})}\BibitemShut {NoStop}%
\bibitem [{\citenamefont {Basov}\ \emph {et~al.}(2016)\citenamefont {Basov},
  \citenamefont {Fogler},\ and\ \citenamefont {Garc{\'\i}a~de
  Abajo}}]{Basovaag1992}%
  \BibitemOpen
  \bibfield  {author} {\bibinfo {author} {\bibfnamefont {D.~N.}\ \bibnamefont
  {Basov}}, \bibinfo {author} {\bibfnamefont {M.~M.}\ \bibnamefont {Fogler}}, \
  and\ \bibinfo {author} {\bibfnamefont {F.~J.}\ \bibnamefont {Garc{\'\i}a~de
  Abajo}},\ }\href {https://science.sciencemag.org/content/354/6309/aag1992}
  {\bibfield  {journal} {\bibinfo  {journal} {Science}\ }\textbf {\bibinfo
  {volume} {354}} (\bibinfo {year} {2016})}\BibitemShut {NoStop}%
\bibitem [{\citenamefont {Baumberg}\ \emph {et~al.}(2019)\citenamefont
  {Baumberg}, \citenamefont {Aizpurua}, \citenamefont {Mikkelsen},\ and\
  \citenamefont {Smith}}]{BJ19}%
  \BibitemOpen
  \bibfield  {author} {\bibinfo {author} {\bibfnamefont {J.~J.}\ \bibnamefont
  {Baumberg}}, \bibinfo {author} {\bibfnamefont {J.}~\bibnamefont {Aizpurua}},
  \bibinfo {author} {\bibfnamefont {M.~H.}\ \bibnamefont {Mikkelsen}}, \ and\
  \bibinfo {author} {\bibfnamefont {D.~R.}\ \bibnamefont {Smith}},\ }\href
  {https://www.nature.com/articles/s41563-019-0290-y} {\bibfield  {journal}
  {\bibinfo  {journal} {Nat. Mater.}\ }\textbf {\bibinfo {volume} {18}},\
  \bibinfo {pages} {668} (\bibinfo {year} {2019})}\BibitemShut {NoStop}%
\bibitem [{\citenamefont {Garcia-Vidal}\ \emph {et~al.}(2021)\citenamefont
  {Garcia-Vidal}, \citenamefont {Ciuti},\ and\ \citenamefont
  {Ebbesen}}]{FJGV21}%
  \BibitemOpen
  \bibfield  {author} {\bibinfo {author} {\bibfnamefont {F.~J.}\ \bibnamefont
  {Garcia-Vidal}}, \bibinfo {author} {\bibfnamefont {C.}~\bibnamefont {Ciuti}},
  \ and\ \bibinfo {author} {\bibfnamefont {T.~W.}\ \bibnamefont {Ebbesen}},\
  }\href {\doibase 10.1126/science.abd0336} {\bibfield  {journal} {\bibinfo
  {journal} {Science}\ }\textbf {\bibinfo {volume} {373}},\ \bibinfo {pages}
  {eabd0336} (\bibinfo {year} {2021})}\BibitemShut {NoStop}%
\bibitem [{\citenamefont {Schlawin}\ \emph {et~al.}(2022)\citenamefont
  {Schlawin}, \citenamefont {Kennes},\ and\ \citenamefont {Sentef}}]{SF22}%
  \BibitemOpen
  \bibfield  {author} {\bibinfo {author} {\bibfnamefont {F.}~\bibnamefont
  {Schlawin}}, \bibinfo {author} {\bibfnamefont {D.~M.}\ \bibnamefont
  {Kennes}}, \ and\ \bibinfo {author} {\bibfnamefont {M.~A.}\ \bibnamefont
  {Sentef}},\ }\href {\doibase 10.1063/5.0083825} {\bibfield  {journal}
  {\bibinfo  {journal} {Appl. Phys. Rev.}\ }\textbf {\bibinfo {volume} {9}},\
  \bibinfo {pages} {011312} (\bibinfo {year} {2022})}\BibitemShut {NoStop}%
\bibitem [{\citenamefont {Bloch}\ \emph {et~al.}(2022)\citenamefont {Bloch},
  \citenamefont {Cavalleri}, \citenamefont {Galitski}, \citenamefont {Hafezi},\
  and\ \citenamefont {Rubio}}]{BJ22}%
  \BibitemOpen
  \bibfield  {author} {\bibinfo {author} {\bibfnamefont {J.}~\bibnamefont
  {Bloch}}, \bibinfo {author} {\bibfnamefont {A.}~\bibnamefont {Cavalleri}},
  \bibinfo {author} {\bibfnamefont {V.}~\bibnamefont {Galitski}}, \bibinfo
  {author} {\bibfnamefont {M.}~\bibnamefont {Hafezi}}, \ and\ \bibinfo {author}
  {\bibfnamefont {A.}~\bibnamefont {Rubio}},\ }\href
  {https://www.nature.com/articles/s41586-022-04726-w} {\bibfield  {journal}
  {\bibinfo  {journal} {Nature}\ }\textbf {\bibinfo {volume} {606}},\ \bibinfo
  {pages} {41} (\bibinfo {year} {2022})}\BibitemShut {NoStop}%
\bibitem [{\citenamefont {Hutchison}\ \emph {et~al.}(2012)\citenamefont
  {Hutchison}, \citenamefont {Schwartz}, \citenamefont {Genet}, \citenamefont
  {Devaux},\ and\ \citenamefont {Ebbesen}}]{HJA12}%
  \BibitemOpen
  \bibfield  {author} {\bibinfo {author} {\bibfnamefont {J.~A.}\ \bibnamefont
  {Hutchison}}, \bibinfo {author} {\bibfnamefont {T.}~\bibnamefont {Schwartz}},
  \bibinfo {author} {\bibfnamefont {C.}~\bibnamefont {Genet}}, \bibinfo
  {author} {\bibfnamefont {E.}~\bibnamefont {Devaux}}, \ and\ \bibinfo {author}
  {\bibfnamefont {T.~W.}\ \bibnamefont {Ebbesen}},\ }\href {\doibase
  10.1002/anie.201107033} {\bibfield  {journal} {\bibinfo  {journal} {Angew.
  Chem. Int. Ed.}\ }\textbf {\bibinfo {volume} {51}},\ \bibinfo {pages} {1592}
  (\bibinfo {year} {2012})}\BibitemShut {NoStop}%
\bibitem [{\citenamefont {Galego}\ \emph {et~al.}(2015)\citenamefont {Galego},
  \citenamefont {Garcia-Vidal},\ and\ \citenamefont {Feist}}]{GJ15}%
  \BibitemOpen
  \bibfield  {author} {\bibinfo {author} {\bibfnamefont {J.}~\bibnamefont
  {Galego}}, \bibinfo {author} {\bibfnamefont {F.~J.}\ \bibnamefont
  {Garcia-Vidal}}, \ and\ \bibinfo {author} {\bibfnamefont {J.}~\bibnamefont
  {Feist}},\ }\href {\doibase 10.1103/PhysRevX.5.041022} {\bibfield  {journal}
  {\bibinfo  {journal} {Phys. Rev. X}\ }\textbf {\bibinfo {volume} {5}},\
  \bibinfo {pages} {041022} (\bibinfo {year} {2015})}\BibitemShut {NoStop}%
\bibitem [{\citenamefont {Flick}\ \emph {et~al.}(2015)\citenamefont {Flick},
  \citenamefont {Ruggenthaler}, \citenamefont {Appel},\ and\ \citenamefont
  {Rubio}}]{FJ15}%
  \BibitemOpen
  \bibfield  {author} {\bibinfo {author} {\bibfnamefont {J.}~\bibnamefont
  {Flick}}, \bibinfo {author} {\bibfnamefont {M.}~\bibnamefont {Ruggenthaler}},
  \bibinfo {author} {\bibfnamefont {H.}~\bibnamefont {Appel}}, \ and\ \bibinfo
  {author} {\bibfnamefont {A.}~\bibnamefont {Rubio}},\ }\href {\doibase
  10.1073/pnas.1518224112} {\bibfield  {journal} {\bibinfo  {journal} {Proc.
  Natl. Acad. Sci. U.S.A.}\ }\textbf {\bibinfo {volume} {112}},\ \bibinfo
  {pages} {15285} (\bibinfo {year} {2015})}\BibitemShut {NoStop}%
\bibitem [{\citenamefont {Herrera}\ and\ \citenamefont {Spano}(2016)}]{HF16}%
  \BibitemOpen
  \bibfield  {author} {\bibinfo {author} {\bibfnamefont {F.}~\bibnamefont
  {Herrera}}\ and\ \bibinfo {author} {\bibfnamefont {F.~C.}\ \bibnamefont
  {Spano}},\ }\href {\doibase 10.1103/PhysRevLett.116.238301} {\bibfield
  {journal} {\bibinfo  {journal} {Phys. Rev. Lett.}\ }\textbf {\bibinfo
  {volume} {116}},\ \bibinfo {pages} {238301} (\bibinfo {year}
  {2016})}\BibitemShut {NoStop}%
\bibitem [{\citenamefont {Feist}\ \emph {et~al.}(2017)\citenamefont {Feist},
  \citenamefont {Galego},\ and\ \citenamefont {Garcia-Vidal}}]{FJ17}%
  \BibitemOpen
  \bibfield  {author} {\bibinfo {author} {\bibfnamefont {J.}~\bibnamefont
  {Feist}}, \bibinfo {author} {\bibfnamefont {J.}~\bibnamefont {Galego}}, \
  and\ \bibinfo {author} {\bibfnamefont {F.~J.}\ \bibnamefont {Garcia-Vidal}},\
  }\href {https://pubs.acs.org/doi/abs/10.1021/acsphotonics.7b00680} {\bibfield
   {journal} {\bibinfo  {journal} {ACS Photonics}\ }\textbf {\bibinfo {volume}
  {5}},\ \bibinfo {pages} {205} (\bibinfo {year} {2017})}\BibitemShut {NoStop}%
\bibitem [{\citenamefont {Thomas}\ \emph {et~al.}(2019)\citenamefont {Thomas},
  \citenamefont {Lethuillier-Karl}, \citenamefont {Nagarajan}, \citenamefont
  {Vergauwe}, \citenamefont {George}, \citenamefont {Chervy}, \citenamefont
  {Shalabney}, \citenamefont {Devaux}, \citenamefont {Genet}, \citenamefont
  {Moran},\ and\ \citenamefont {Ebbesen}}]{TA19}%
  \BibitemOpen
  \bibfield  {author} {\bibinfo {author} {\bibfnamefont {A.}~\bibnamefont
  {Thomas}}, \bibinfo {author} {\bibfnamefont {L.}~\bibnamefont
  {Lethuillier-Karl}}, \bibinfo {author} {\bibfnamefont {K.}~\bibnamefont
  {Nagarajan}}, \bibinfo {author} {\bibfnamefont {R.~M.~A.}\ \bibnamefont
  {Vergauwe}}, \bibinfo {author} {\bibfnamefont {J.}~\bibnamefont {George}},
  \bibinfo {author} {\bibfnamefont {T.}~\bibnamefont {Chervy}}, \bibinfo
  {author} {\bibfnamefont {A.}~\bibnamefont {Shalabney}}, \bibinfo {author}
  {\bibfnamefont {E.}~\bibnamefont {Devaux}}, \bibinfo {author} {\bibfnamefont
  {C.}~\bibnamefont {Genet}}, \bibinfo {author} {\bibfnamefont
  {J.}~\bibnamefont {Moran}}, \ and\ \bibinfo {author} {\bibfnamefont {T.~W.}\
  \bibnamefont {Ebbesen}},\ }\href {\doibase 10.1126/science.aau7742}
  {\bibfield  {journal} {\bibinfo  {journal} {Science}\ }\textbf {\bibinfo
  {volume} {363}},\ \bibinfo {pages} {615} (\bibinfo {year}
  {2019})}\BibitemShut {NoStop}%
\bibitem [{\citenamefont {Du}\ and\ \citenamefont {Yuen-Zhou}(2022)}]{DM22}%
  \BibitemOpen
  \bibfield  {author} {\bibinfo {author} {\bibfnamefont {M.}~\bibnamefont
  {Du}}\ and\ \bibinfo {author} {\bibfnamefont {J.}~\bibnamefont {Yuen-Zhou}},\
  }\href {\doibase 10.1103/PhysRevLett.128.096001} {\bibfield  {journal}
  {\bibinfo  {journal} {Phys. Rev. Lett.}\ }\textbf {\bibinfo {volume} {128}},\
  \bibinfo {pages} {096001} (\bibinfo {year} {2022})}\BibitemShut {NoStop}%
\bibitem [{\citenamefont {Blais}\ \emph {et~al.}(2021)\citenamefont {Blais},
  \citenamefont {Grimsmo}, \citenamefont {Girvin},\ and\ \citenamefont
  {Wallraff}}]{BA21}%
  \BibitemOpen
  \bibfield  {author} {\bibinfo {author} {\bibfnamefont {A.}~\bibnamefont
  {Blais}}, \bibinfo {author} {\bibfnamefont {A.~L.}\ \bibnamefont {Grimsmo}},
  \bibinfo {author} {\bibfnamefont {S.~M.}\ \bibnamefont {Girvin}}, \ and\
  \bibinfo {author} {\bibfnamefont {A.}~\bibnamefont {Wallraff}},\ }\href
  {\doibase 10.1103/RevModPhys.93.025005} {\bibfield  {journal} {\bibinfo
  {journal} {Rev. Mod. Phys.}\ }\textbf {\bibinfo {volume} {93}},\ \bibinfo
  {pages} {025005} (\bibinfo {year} {2021})}\BibitemShut {NoStop}%
\bibitem [{\citenamefont {Clerk}\ \emph {et~al.}(2020)\citenamefont {Clerk},
  \citenamefont {Lehnert}, \citenamefont {Bertet}, \citenamefont {Petta},\ and\
  \citenamefont {Nakamura}}]{CAA20}%
  \BibitemOpen
  \bibfield  {author} {\bibinfo {author} {\bibfnamefont {A.}~\bibnamefont
  {Clerk}}, \bibinfo {author} {\bibfnamefont {K.}~\bibnamefont {Lehnert}},
  \bibinfo {author} {\bibfnamefont {P.}~\bibnamefont {Bertet}}, \bibinfo
  {author} {\bibfnamefont {J.}~\bibnamefont {Petta}}, \ and\ \bibinfo {author}
  {\bibfnamefont {Y.}~\bibnamefont {Nakamura}},\ }\href
  {https://www.nature.com/articles/s41567-020-0797-9} {\bibfield  {journal}
  {\bibinfo  {journal} {Nat. Phys.}\ }\textbf {\bibinfo {volume} {16}},\
  \bibinfo {pages} {257} (\bibinfo {year} {2020})}\BibitemShut {NoStop}%
\bibitem [{\citenamefont {Forn-D{\'\i}az}\ \emph {et~al.}(2017)\citenamefont
  {Forn-D{\'\i}az}, \citenamefont {Garc{\'\i}a-Ripoll}, \citenamefont
  {Peropadre}, \citenamefont {Orgiazzi}, \citenamefont {Yurtalan},
  \citenamefont {Belyansky}, \citenamefont {Wilson},\ and\ \citenamefont
  {Lupascu}}]{FDP17}%
  \BibitemOpen
  \bibfield  {author} {\bibinfo {author} {\bibfnamefont {P.}~\bibnamefont
  {Forn-D{\'\i}az}}, \bibinfo {author} {\bibfnamefont {J.~J.}\ \bibnamefont
  {Garc{\'\i}a-Ripoll}}, \bibinfo {author} {\bibfnamefont {B.}~\bibnamefont
  {Peropadre}}, \bibinfo {author} {\bibfnamefont {J.-L.}\ \bibnamefont
  {Orgiazzi}}, \bibinfo {author} {\bibfnamefont {M.}~\bibnamefont {Yurtalan}},
  \bibinfo {author} {\bibfnamefont {R.}~\bibnamefont {Belyansky}}, \bibinfo
  {author} {\bibfnamefont {C.~M.}\ \bibnamefont {Wilson}}, \ and\ \bibinfo
  {author} {\bibfnamefont {A.}~\bibnamefont {Lupascu}},\ }\href
  {https://www.nature.com/articles/nphys3905} {\bibfield  {journal} {\bibinfo
  {journal} {Nat. Phys.}\ }\textbf {\bibinfo {volume} {13}},\ \bibinfo {pages}
  {39} (\bibinfo {year} {2017})}\BibitemShut {NoStop}%
\bibitem [{\citenamefont {Mart{\'\i}nez}\ \emph {et~al.}(2019)\citenamefont
  {Mart{\'\i}nez}, \citenamefont {L{\'e}ger}, \citenamefont {Gheeraert},
  \citenamefont {Dassonneville}, \citenamefont {Planat}, \citenamefont
  {Foroughi}, \citenamefont {Krupko}, \citenamefont {Buisson}, \citenamefont
  {Naud}, \citenamefont {Hasch-Guichard}, \citenamefont {Florens},
  \citenamefont {Snyman},\ and\ \citenamefont {Roch}}]{MJP19}%
  \BibitemOpen
  \bibfield  {author} {\bibinfo {author} {\bibfnamefont {J.~P.}\ \bibnamefont
  {Mart{\'\i}nez}}, \bibinfo {author} {\bibfnamefont {S.}~\bibnamefont
  {L{\'e}ger}}, \bibinfo {author} {\bibfnamefont {N.}~\bibnamefont
  {Gheeraert}}, \bibinfo {author} {\bibfnamefont {R.}~\bibnamefont
  {Dassonneville}}, \bibinfo {author} {\bibfnamefont {L.}~\bibnamefont
  {Planat}}, \bibinfo {author} {\bibfnamefont {F.}~\bibnamefont {Foroughi}},
  \bibinfo {author} {\bibfnamefont {Y.}~\bibnamefont {Krupko}}, \bibinfo
  {author} {\bibfnamefont {O.}~\bibnamefont {Buisson}}, \bibinfo {author}
  {\bibfnamefont {C.}~\bibnamefont {Naud}}, \bibinfo {author} {\bibfnamefont
  {W.}~\bibnamefont {Hasch-Guichard}}, \bibinfo {author} {\bibfnamefont
  {S.}~\bibnamefont {Florens}}, \bibinfo {author} {\bibfnamefont
  {I.}~\bibnamefont {Snyman}}, \ and\ \bibinfo {author} {\bibfnamefont
  {N.}~\bibnamefont {Roch}},\ }\href
  {https://www.nature.com/articles/s41534-018-0104-0} {\bibfield  {journal}
  {\bibinfo  {journal} {npj Quantum Inf.}\ }\textbf {\bibinfo {volume} {5}},\
  \bibinfo {pages} {1} (\bibinfo {year} {2019})}\BibitemShut {NoStop}%
\bibitem [{\citenamefont {L{\'e}ger}\ \emph {et~al.}(2019)\citenamefont
  {L{\'e}ger}, \citenamefont {Puertas-Mart{\'\i}nez}, \citenamefont
  {Bharadwaj}, \citenamefont {Dassonneville}, \citenamefont {Delaforce},
  \citenamefont {Foroughi}, \citenamefont {Milchakov}, \citenamefont {Planat},
  \citenamefont {Buisson}, \citenamefont {Naud}, \citenamefont
  {Hasch-Guichard}, \citenamefont {Florens}, \citenamefont {Snyman},\ and\
  \citenamefont {Roch}}]{LS19}%
  \BibitemOpen
  \bibfield  {author} {\bibinfo {author} {\bibfnamefont {S.}~\bibnamefont
  {L{\'e}ger}}, \bibinfo {author} {\bibfnamefont {J.}~\bibnamefont
  {Puertas-Mart{\'\i}nez}}, \bibinfo {author} {\bibfnamefont {K.}~\bibnamefont
  {Bharadwaj}}, \bibinfo {author} {\bibfnamefont {R.}~\bibnamefont
  {Dassonneville}}, \bibinfo {author} {\bibfnamefont {J.}~\bibnamefont
  {Delaforce}}, \bibinfo {author} {\bibfnamefont {F.}~\bibnamefont {Foroughi}},
  \bibinfo {author} {\bibfnamefont {V.}~\bibnamefont {Milchakov}}, \bibinfo
  {author} {\bibfnamefont {L.}~\bibnamefont {Planat}}, \bibinfo {author}
  {\bibfnamefont {O.}~\bibnamefont {Buisson}}, \bibinfo {author} {\bibfnamefont
  {C.}~\bibnamefont {Naud}}, \bibinfo {author} {\bibfnamefont {W.}~\bibnamefont
  {Hasch-Guichard}}, \bibinfo {author} {\bibfnamefont {S.}~\bibnamefont
  {Florens}}, \bibinfo {author} {\bibfnamefont {I.}~\bibnamefont {Snyman}}, \
  and\ \bibinfo {author} {\bibfnamefont {N.}~\bibnamefont {Roch}},\ }\href
  {https://www.nature.com/articles/s41467-019-13199-x} {\bibfield  {journal}
  {\bibinfo  {journal} {Nat. Commun.}\ }\textbf {\bibinfo {volume} {10}},\
  \bibinfo {pages} {1} (\bibinfo {year} {2019})}\BibitemShut {NoStop}%
\bibitem [{\citenamefont {Kuzmin}\ \emph {et~al.}(2019)\citenamefont {Kuzmin},
  \citenamefont {Mehta}, \citenamefont {Grabon}, \citenamefont {Mencia},\ and\
  \citenamefont {Manucharyan}}]{KR19}%
  \BibitemOpen
  \bibfield  {author} {\bibinfo {author} {\bibfnamefont {R.}~\bibnamefont
  {Kuzmin}}, \bibinfo {author} {\bibfnamefont {N.}~\bibnamefont {Mehta}},
  \bibinfo {author} {\bibfnamefont {N.}~\bibnamefont {Grabon}}, \bibinfo
  {author} {\bibfnamefont {R.}~\bibnamefont {Mencia}}, \ and\ \bibinfo {author}
  {\bibfnamefont {V.~E.}\ \bibnamefont {Manucharyan}},\ }\href
  {https://www.nature.com/articles/s41534-019-0134-2} {\bibfield  {journal}
  {\bibinfo  {journal} {npj Quantum Inf.}\ }\textbf {\bibinfo {volume} {5}},\
  \bibinfo {pages} {1} (\bibinfo {year} {2019})}\BibitemShut {NoStop}%
\bibitem [{\citenamefont {Peruzzo}\ \emph {et~al.}(2020)\citenamefont
  {Peruzzo}, \citenamefont {Trioni}, \citenamefont {Hassani}, \citenamefont
  {Zemlicka},\ and\ \citenamefont {Fink}}]{PM20}%
  \BibitemOpen
  \bibfield  {author} {\bibinfo {author} {\bibfnamefont {M.}~\bibnamefont
  {Peruzzo}}, \bibinfo {author} {\bibfnamefont {A.}~\bibnamefont {Trioni}},
  \bibinfo {author} {\bibfnamefont {F.}~\bibnamefont {Hassani}}, \bibinfo
  {author} {\bibfnamefont {M.}~\bibnamefont {Zemlicka}}, \ and\ \bibinfo
  {author} {\bibfnamefont {J.~M.}\ \bibnamefont {Fink}},\ }\href {\doibase
  10.1103/PhysRevApplied.14.044055} {\bibfield  {journal} {\bibinfo  {journal}
  {Phys. Rev. Applied}\ }\textbf {\bibinfo {volume} {14}},\ \bibinfo {pages}
  {044055} (\bibinfo {year} {2020})}\BibitemShut {NoStop}%
\bibitem [{\citenamefont {Kuzmin}\ \emph {et~al.}(2021)\citenamefont {Kuzmin},
  \citenamefont {Grabon}, \citenamefont {Mehta}, \citenamefont {Burshtein},
  \citenamefont {Goldstein}, \citenamefont {Houzet}, \citenamefont {Glazman},\
  and\ \citenamefont {Manucharyan}}]{KR21}%
  \BibitemOpen
  \bibfield  {author} {\bibinfo {author} {\bibfnamefont {R.}~\bibnamefont
  {Kuzmin}}, \bibinfo {author} {\bibfnamefont {N.}~\bibnamefont {Grabon}},
  \bibinfo {author} {\bibfnamefont {N.}~\bibnamefont {Mehta}}, \bibinfo
  {author} {\bibfnamefont {A.}~\bibnamefont {Burshtein}}, \bibinfo {author}
  {\bibfnamefont {M.}~\bibnamefont {Goldstein}}, \bibinfo {author}
  {\bibfnamefont {M.}~\bibnamefont {Houzet}}, \bibinfo {author} {\bibfnamefont
  {L.~I.}\ \bibnamefont {Glazman}}, \ and\ \bibinfo {author} {\bibfnamefont
  {V.~E.}\ \bibnamefont {Manucharyan}},\ }\href {\doibase
  10.1103/PhysRevLett.126.197701} {\bibfield  {journal} {\bibinfo  {journal}
  {Phys. Rev. Lett.}\ }\textbf {\bibinfo {volume} {126}},\ \bibinfo {pages}
  {197701} (\bibinfo {year} {2021})}\BibitemShut {NoStop}%
\bibitem [{\citenamefont {L{\'e}ger}\ \emph {et~al.}(2022)\citenamefont
  {L{\'e}ger}, \citenamefont {S{\'e}pulcre}, \citenamefont {Fraudet},
  \citenamefont {Buisson}, \citenamefont {Naud}, \citenamefont
  {Hasch-Guichard}, \citenamefont {Florens}, \citenamefont {Snyman},
  \citenamefont {Basko},\ and\ \citenamefont {Roch}}]{LS22}%
  \BibitemOpen
  \bibfield  {author} {\bibinfo {author} {\bibfnamefont {S.}~\bibnamefont
  {L{\'e}ger}}, \bibinfo {author} {\bibfnamefont {T.}~\bibnamefont
  {S{\'e}pulcre}}, \bibinfo {author} {\bibfnamefont {D.}~\bibnamefont
  {Fraudet}}, \bibinfo {author} {\bibfnamefont {O.}~\bibnamefont {Buisson}},
  \bibinfo {author} {\bibfnamefont {C.}~\bibnamefont {Naud}}, \bibinfo {author}
  {\bibfnamefont {W.}~\bibnamefont {Hasch-Guichard}}, \bibinfo {author}
  {\bibfnamefont {S.}~\bibnamefont {Florens}}, \bibinfo {author} {\bibfnamefont
  {I.}~\bibnamefont {Snyman}}, \bibinfo {author} {\bibfnamefont {D.~M.}\
  \bibnamefont {Basko}}, \ and\ \bibinfo {author} {\bibfnamefont
  {N.}~\bibnamefont {Roch}},\ }\href {https://arxiv.org/abs/2208.03053}
  {\bibfield  {journal} {\bibinfo  {journal} {arXiv:2208.03053}\ } (\bibinfo
  {year} {2022})}\BibitemShut {NoStop}%
\bibitem [{\citenamefont {Goldstein}\ \emph {et~al.}(2013)\citenamefont
  {Goldstein}, \citenamefont {Devoret}, \citenamefont {Houzet},\ and\
  \citenamefont {Glazman}}]{Goldstein13}%
  \BibitemOpen
  \bibfield  {author} {\bibinfo {author} {\bibfnamefont {M.}~\bibnamefont
  {Goldstein}}, \bibinfo {author} {\bibfnamefont {M.~H.}\ \bibnamefont
  {Devoret}}, \bibinfo {author} {\bibfnamefont {M.}~\bibnamefont {Houzet}}, \
  and\ \bibinfo {author} {\bibfnamefont {L.~I.}\ \bibnamefont {Glazman}},\
  }\href {\doibase 10.1103/PhysRevLett.110.017002} {\bibfield  {journal}
  {\bibinfo  {journal} {Phys. Rev. Lett.}\ }\textbf {\bibinfo {volume} {110}},\
  \bibinfo {pages} {017002} (\bibinfo {year} {2013})}\BibitemShut {NoStop}%
\bibitem [{\citenamefont {Sanchez-Burillo}\ \emph {et~al.}(2014)\citenamefont
  {Sanchez-Burillo}, \citenamefont {Zueco}, \citenamefont {Garcia-Ripoll},\
  and\ \citenamefont {Martin-Moreno}}]{SBE14}%
  \BibitemOpen
  \bibfield  {author} {\bibinfo {author} {\bibfnamefont {E.}~\bibnamefont
  {Sanchez-Burillo}}, \bibinfo {author} {\bibfnamefont {D.}~\bibnamefont
  {Zueco}}, \bibinfo {author} {\bibfnamefont {J.~J.}\ \bibnamefont
  {Garcia-Ripoll}}, \ and\ \bibinfo {author} {\bibfnamefont {L.}~\bibnamefont
  {Martin-Moreno}},\ }\href {\doibase 10.1103/PhysRevLett.113.263604}
  {\bibfield  {journal} {\bibinfo  {journal} {Phys. Rev. Lett.}\ }\textbf
  {\bibinfo {volume} {113}},\ \bibinfo {pages} {263604} (\bibinfo {year}
  {2014})}\BibitemShut {NoStop}%
\bibitem [{\citenamefont {Snyman}\ and\ \citenamefont {Florens}(2015)}]{SI15}%
  \BibitemOpen
  \bibfield  {author} {\bibinfo {author} {\bibfnamefont {I.}~\bibnamefont
  {Snyman}}\ and\ \bibinfo {author} {\bibfnamefont {S.}~\bibnamefont
  {Florens}},\ }\href {\doibase 10.1103/PhysRevB.92.085131} {\bibfield
  {journal} {\bibinfo  {journal} {Phys. Rev. B}\ }\textbf {\bibinfo {volume}
  {92}},\ \bibinfo {pages} {085131} (\bibinfo {year} {2015})}\BibitemShut
  {NoStop}%
\bibitem [{\citenamefont {Lepp\"akangas}\ \emph {et~al.}(2018)\citenamefont
  {Lepp\"akangas}, \citenamefont {Braum\"uller}, \citenamefont {Hauck},
  \citenamefont {Reiner}, \citenamefont {Schwenk}, \citenamefont {Zanker},
  \citenamefont {Fritz}, \citenamefont {Ustinov}, \citenamefont {Weides},\ and\
  \citenamefont {Marthaler}}]{LJ18}%
  \BibitemOpen
  \bibfield  {author} {\bibinfo {author} {\bibfnamefont {J.}~\bibnamefont
  {Lepp\"akangas}}, \bibinfo {author} {\bibfnamefont {J.}~\bibnamefont
  {Braum\"uller}}, \bibinfo {author} {\bibfnamefont {M.}~\bibnamefont {Hauck}},
  \bibinfo {author} {\bibfnamefont {J.-M.}\ \bibnamefont {Reiner}}, \bibinfo
  {author} {\bibfnamefont {I.}~\bibnamefont {Schwenk}}, \bibinfo {author}
  {\bibfnamefont {S.}~\bibnamefont {Zanker}}, \bibinfo {author} {\bibfnamefont
  {L.}~\bibnamefont {Fritz}}, \bibinfo {author} {\bibfnamefont {A.~V.}\
  \bibnamefont {Ustinov}}, \bibinfo {author} {\bibfnamefont {M.}~\bibnamefont
  {Weides}}, \ and\ \bibinfo {author} {\bibfnamefont {M.}~\bibnamefont
  {Marthaler}},\ }\href {\doibase 10.1103/PhysRevA.97.052321} {\bibfield
  {journal} {\bibinfo  {journal} {Phys. Rev. A}\ }\textbf {\bibinfo {volume}
  {97}},\ \bibinfo {pages} {052321} (\bibinfo {year} {2018})}\BibitemShut
  {NoStop}%
\bibitem [{\citenamefont {Magazz{\`u}}\ \emph {et~al.}(2018)\citenamefont
  {Magazz{\`u}}, \citenamefont {Forn-D{\'\i}az}, \citenamefont {Belyansky},
  \citenamefont {Orgiazzi}, \citenamefont {Yurtalan}, \citenamefont {Otto},
  \citenamefont {Lupascu}, \citenamefont {Wilson},\ and\ \citenamefont
  {Grifoni}}]{ML18}%
  \BibitemOpen
  \bibfield  {author} {\bibinfo {author} {\bibfnamefont {L.}~\bibnamefont
  {Magazz{\`u}}}, \bibinfo {author} {\bibfnamefont {P.}~\bibnamefont
  {Forn-D{\'\i}az}}, \bibinfo {author} {\bibfnamefont {R.}~\bibnamefont
  {Belyansky}}, \bibinfo {author} {\bibfnamefont {J.-L.}\ \bibnamefont
  {Orgiazzi}}, \bibinfo {author} {\bibfnamefont {M.}~\bibnamefont {Yurtalan}},
  \bibinfo {author} {\bibfnamefont {M.~R.}\ \bibnamefont {Otto}}, \bibinfo
  {author} {\bibfnamefont {A.}~\bibnamefont {Lupascu}}, \bibinfo {author}
  {\bibfnamefont {C.}~\bibnamefont {Wilson}}, \ and\ \bibinfo {author}
  {\bibfnamefont {M.}~\bibnamefont {Grifoni}},\ }\href
  {https://www.nature.com/articles/s41467-018-03626-w} {\bibfield  {journal}
  {\bibinfo  {journal} {Nat. Commun.}\ }\textbf {\bibinfo {volume} {9}},\
  \bibinfo {pages} {1} (\bibinfo {year} {2018})}\BibitemShut {NoStop}%
\bibitem [{\citenamefont {{Le Hur}}\ \emph {et~al.}(2018)\citenamefont {{Le
  Hur}}, \citenamefont {Henriet}, \citenamefont {Herviou}, \citenamefont
  {Plekhanov}, \citenamefont {Petrescu}, \citenamefont {Goren}, \citenamefont
  {Schiro}, \citenamefont {Mora},\ and\ \citenamefont {Orth}}]{KLK18}%
  \BibitemOpen
  \bibfield  {author} {\bibinfo {author} {\bibfnamefont {K.}~\bibnamefont {{Le
  Hur}}}, \bibinfo {author} {\bibfnamefont {L.}~\bibnamefont {Henriet}},
  \bibinfo {author} {\bibfnamefont {L.}~\bibnamefont {Herviou}}, \bibinfo
  {author} {\bibfnamefont {K.}~\bibnamefont {Plekhanov}}, \bibinfo {author}
  {\bibfnamefont {A.}~\bibnamefont {Petrescu}}, \bibinfo {author}
  {\bibfnamefont {T.}~\bibnamefont {Goren}}, \bibinfo {author} {\bibfnamefont
  {M.}~\bibnamefont {Schiro}}, \bibinfo {author} {\bibfnamefont
  {C.}~\bibnamefont {Mora}}, \ and\ \bibinfo {author} {\bibfnamefont {P.~P.}\
  \bibnamefont {Orth}},\ }\href {\doibase
  https://doi.org/10.1016/j.crhy.2018.04.003} {\bibfield  {journal} {\bibinfo
  {journal} {C. R. Phys.}\ }\textbf {\bibinfo {volume} {19}},\ \bibinfo {pages}
  {451} (\bibinfo {year} {2018})}\BibitemShut {NoStop}%
\bibitem [{\citenamefont {Gheeraert}\ \emph {et~al.}(2018)\citenamefont
  {Gheeraert}, \citenamefont {Zhang}, \citenamefont {S\'epulcre}, \citenamefont
  {Bera}, \citenamefont {Roch}, \citenamefont {Baranger},\ and\ \citenamefont
  {Florens}}]{GN18}%
  \BibitemOpen
  \bibfield  {author} {\bibinfo {author} {\bibfnamefont {N.}~\bibnamefont
  {Gheeraert}}, \bibinfo {author} {\bibfnamefont {X.~H.~H.}\ \bibnamefont
  {Zhang}}, \bibinfo {author} {\bibfnamefont {T.}~\bibnamefont {S\'epulcre}},
  \bibinfo {author} {\bibfnamefont {S.}~\bibnamefont {Bera}}, \bibinfo {author}
  {\bibfnamefont {N.}~\bibnamefont {Roch}}, \bibinfo {author} {\bibfnamefont
  {H.~U.}\ \bibnamefont {Baranger}}, \ and\ \bibinfo {author} {\bibfnamefont
  {S.}~\bibnamefont {Florens}},\ }\href {\doibase 10.1103/PhysRevA.98.043816}
  {\bibfield  {journal} {\bibinfo  {journal} {Phys. Rev. A}\ }\textbf {\bibinfo
  {volume} {98}},\ \bibinfo {pages} {043816} (\bibinfo {year}
  {2018})}\BibitemShut {NoStop}%
\bibitem [{\citenamefont {Houzet}\ and\ \citenamefont {Glazman}(2020)}]{HM20}%
  \BibitemOpen
  \bibfield  {author} {\bibinfo {author} {\bibfnamefont {M.}~\bibnamefont
  {Houzet}}\ and\ \bibinfo {author} {\bibfnamefont {L.~I.}\ \bibnamefont
  {Glazman}},\ }\href {\doibase 10.1103/PhysRevLett.125.267701} {\bibfield
  {journal} {\bibinfo  {journal} {Phys. Rev. Lett.}\ }\textbf {\bibinfo
  {volume} {125}},\ \bibinfo {pages} {267701} (\bibinfo {year}
  {2020})}\BibitemShut {NoStop}%
\bibitem [{\citenamefont {Yamamoto}\ \emph {et~al.}(2021)\citenamefont
  {Yamamoto}, \citenamefont {Glazman},\ and\ \citenamefont {Houzet}}]{YT21}%
  \BibitemOpen
  \bibfield  {author} {\bibinfo {author} {\bibfnamefont {T.}~\bibnamefont
  {Yamamoto}}, \bibinfo {author} {\bibfnamefont {L.~I.}\ \bibnamefont
  {Glazman}}, \ and\ \bibinfo {author} {\bibfnamefont {M.}~\bibnamefont
  {Houzet}},\ }\href {\doibase 10.1103/PhysRevB.103.224211} {\bibfield
  {journal} {\bibinfo  {journal} {Phys. Rev. B}\ }\textbf {\bibinfo {volume}
  {103}},\ \bibinfo {pages} {224211} (\bibinfo {year} {2021})}\BibitemShut
  {NoStop}%
\bibitem [{\citenamefont {Burshtein}\ \emph {et~al.}(2021)\citenamefont
  {Burshtein}, \citenamefont {Kuzmin}, \citenamefont {Manucharyan},\ and\
  \citenamefont {Goldstein}}]{BAm21}%
  \BibitemOpen
  \bibfield  {author} {\bibinfo {author} {\bibfnamefont {A.}~\bibnamefont
  {Burshtein}}, \bibinfo {author} {\bibfnamefont {R.}~\bibnamefont {Kuzmin}},
  \bibinfo {author} {\bibfnamefont {V.~E.}\ \bibnamefont {Manucharyan}}, \ and\
  \bibinfo {author} {\bibfnamefont {M.}~\bibnamefont {Goldstein}},\ }\href
  {\doibase 10.1103/PhysRevLett.126.137701} {\bibfield  {journal} {\bibinfo
  {journal} {Phys. Rev. Lett.}\ }\textbf {\bibinfo {volume} {126}},\ \bibinfo
  {pages} {137701} (\bibinfo {year} {2021})}\BibitemShut {NoStop}%
\bibitem [{\citenamefont {Kaur}\ \emph {et~al.}(2021)\citenamefont {Kaur},
  \citenamefont {S\'epulcre}, \citenamefont {Roch}, \citenamefont {Snyman},
  \citenamefont {Florens},\ and\ \citenamefont {Bera}}]{KK21}%
  \BibitemOpen
  \bibfield  {author} {\bibinfo {author} {\bibfnamefont {K.}~\bibnamefont
  {Kaur}}, \bibinfo {author} {\bibfnamefont {T.}~\bibnamefont {S\'epulcre}},
  \bibinfo {author} {\bibfnamefont {N.}~\bibnamefont {Roch}}, \bibinfo {author}
  {\bibfnamefont {I.}~\bibnamefont {Snyman}}, \bibinfo {author} {\bibfnamefont
  {S.}~\bibnamefont {Florens}}, \ and\ \bibinfo {author} {\bibfnamefont
  {S.}~\bibnamefont {Bera}},\ }\href {\doibase 10.1103/PhysRevLett.127.237702}
  {\bibfield  {journal} {\bibinfo  {journal} {Phys. Rev. Lett.}\ }\textbf
  {\bibinfo {volume} {127}},\ \bibinfo {pages} {237702} (\bibinfo {year}
  {2021})}\BibitemShut {NoStop}%
\bibitem [{\citenamefont {Gonz\'alez-Guti\'errez}\ \emph
  {et~al.}(2021)\citenamefont {Gonz\'alez-Guti\'errez}, \citenamefont
  {Rom\'an-Roche},\ and\ \citenamefont {Zueco}}]{GGC21}%
  \BibitemOpen
  \bibfield  {author} {\bibinfo {author} {\bibfnamefont {C.~A.}\ \bibnamefont
  {Gonz\'alez-Guti\'errez}}, \bibinfo {author} {\bibfnamefont {J.}~\bibnamefont
  {Rom\'an-Roche}}, \ and\ \bibinfo {author} {\bibfnamefont {D.}~\bibnamefont
  {Zueco}},\ }\href {\doibase 10.1103/PhysRevA.104.053701} {\bibfield
  {journal} {\bibinfo  {journal} {Phys. Rev. A}\ }\textbf {\bibinfo {volume}
  {104}},\ \bibinfo {pages} {053701} (\bibinfo {year} {2021})}\BibitemShut
  {NoStop}%
\bibitem [{\citenamefont {Ashida}\ \emph {et~al.}(2022)\citenamefont {Ashida},
  \citenamefont {Yokota}, \citenamefont {Imamoglu},\ and\ \citenamefont
  {Demler}}]{YA22}%
  \BibitemOpen
  \bibfield  {author} {\bibinfo {author} {\bibfnamefont {Y.}~\bibnamefont
  {Ashida}}, \bibinfo {author} {\bibfnamefont {T.}~\bibnamefont {Yokota}},
  \bibinfo {author} {\bibfnamefont {A.}~\bibnamefont {Imamoglu}}, \ and\
  \bibinfo {author} {\bibfnamefont {E.}~\bibnamefont {Demler}},\ }\href
  {\doibase 10.1103/PhysRevResearch.4.023194} {\bibfield  {journal} {\bibinfo
  {journal} {Phys. Rev. Research}\ }\textbf {\bibinfo {volume} {4}},\ \bibinfo
  {pages} {023194} (\bibinfo {year} {2022})}\BibitemShut {NoStop}%
\bibitem [{\citenamefont {Masuki}\ \emph {et~al.}(2022)\citenamefont {Masuki},
  \citenamefont {Sudo}, \citenamefont {Oshikawa},\ and\ \citenamefont
  {Ashida}}]{KM22}%
  \BibitemOpen
  \bibfield  {author} {\bibinfo {author} {\bibfnamefont {K.}~\bibnamefont
  {Masuki}}, \bibinfo {author} {\bibfnamefont {H.}~\bibnamefont {Sudo}},
  \bibinfo {author} {\bibfnamefont {M.}~\bibnamefont {Oshikawa}}, \ and\
  \bibinfo {author} {\bibfnamefont {Y.}~\bibnamefont {Ashida}},\ }\href
  {\doibase 10.1103/PhysRevLett.129.087001} {\bibfield  {journal} {\bibinfo
  {journal} {Phys. Rev. Lett.}\ }\textbf {\bibinfo {volume} {129}},\ \bibinfo
  {pages} {087001} (\bibinfo {year} {2022})}\BibitemShut {NoStop}%
\bibitem [{\citenamefont {Murani}\ \emph {et~al.}(2020)\citenamefont {Murani},
  \citenamefont {Bourlet}, \citenamefont {le~Sueur}, \citenamefont {Portier},
  \citenamefont {Altimiras}, \citenamefont {Esteve}, \citenamefont {Grabert},
  \citenamefont {Stockburger}, \citenamefont {Ankerhold},\ and\ \citenamefont
  {Joyez}}]{Murami20}%
  \BibitemOpen
  \bibfield  {author} {\bibinfo {author} {\bibfnamefont {A.}~\bibnamefont
  {Murani}}, \bibinfo {author} {\bibfnamefont {N.}~\bibnamefont {Bourlet}},
  \bibinfo {author} {\bibfnamefont {H.}~\bibnamefont {le~Sueur}}, \bibinfo
  {author} {\bibfnamefont {F.}~\bibnamefont {Portier}}, \bibinfo {author}
  {\bibfnamefont {C.}~\bibnamefont {Altimiras}}, \bibinfo {author}
  {\bibfnamefont {D.}~\bibnamefont {Esteve}}, \bibinfo {author} {\bibfnamefont
  {H.}~\bibnamefont {Grabert}}, \bibinfo {author} {\bibfnamefont
  {J.}~\bibnamefont {Stockburger}}, \bibinfo {author} {\bibfnamefont
  {J.}~\bibnamefont {Ankerhold}}, \ and\ \bibinfo {author} {\bibfnamefont
  {P.}~\bibnamefont {Joyez}},\ }\href {\doibase 10.1103/PhysRevX.10.021003}
  {\bibfield  {journal} {\bibinfo  {journal} {Phys. Rev. X}\ }\textbf {\bibinfo
  {volume} {10}},\ \bibinfo {pages} {021003} (\bibinfo {year}
  {2020})}\BibitemShut {NoStop}%
\bibitem [{\citenamefont {Hakonen}\ and\ \citenamefont
  {Sonin}(2021)}]{Hakonen21}%
  \BibitemOpen
  \bibfield  {author} {\bibinfo {author} {\bibfnamefont {P.~J.}\ \bibnamefont
  {Hakonen}}\ and\ \bibinfo {author} {\bibfnamefont {E.~B.}\ \bibnamefont
  {Sonin}},\ }\href {\doibase 10.1103/PhysRevX.11.018001} {\bibfield  {journal}
  {\bibinfo  {journal} {Phys. Rev. X}\ }\textbf {\bibinfo {volume} {11}},\
  \bibinfo {pages} {018001} (\bibinfo {year} {2021})}\BibitemShut {NoStop}%
\bibitem [{\citenamefont {Murani}\ \emph {et~al.}(2021)\citenamefont {Murani},
  \citenamefont {Bourlet}, \citenamefont {le~Sueur}, \citenamefont {Portier},
  \citenamefont {Altimiras}, \citenamefont {Esteve}, \citenamefont {Grabert},
  \citenamefont {Stockburger}, \citenamefont {Ankerhold},\ and\ \citenamefont
  {Joyez}}]{Murani21}%
  \BibitemOpen
  \bibfield  {author} {\bibinfo {author} {\bibfnamefont {A.}~\bibnamefont
  {Murani}}, \bibinfo {author} {\bibfnamefont {N.}~\bibnamefont {Bourlet}},
  \bibinfo {author} {\bibfnamefont {H.}~\bibnamefont {le~Sueur}}, \bibinfo
  {author} {\bibfnamefont {F.}~\bibnamefont {Portier}}, \bibinfo {author}
  {\bibfnamefont {C.}~\bibnamefont {Altimiras}}, \bibinfo {author}
  {\bibfnamefont {D.}~\bibnamefont {Esteve}}, \bibinfo {author} {\bibfnamefont
  {H.}~\bibnamefont {Grabert}}, \bibinfo {author} {\bibfnamefont
  {J.}~\bibnamefont {Stockburger}}, \bibinfo {author} {\bibfnamefont
  {J.}~\bibnamefont {Ankerhold}}, \ and\ \bibinfo {author} {\bibfnamefont
  {P.}~\bibnamefont {Joyez}},\ }\href {\doibase 10.1103/PhysRevX.11.018002}
  {\bibfield  {journal} {\bibinfo  {journal} {Phys. Rev. X}\ }\textbf {\bibinfo
  {volume} {11}},\ \bibinfo {pages} {018002} (\bibinfo {year}
  {2021})}\BibitemShut {NoStop}%
\bibitem [{\citenamefont {Caldeira}\ and\ \citenamefont
  {Leggett}(1981)}]{Caldeira81}%
  \BibitemOpen
  \bibfield  {author} {\bibinfo {author} {\bibfnamefont {A.~O.}\ \bibnamefont
  {Caldeira}}\ and\ \bibinfo {author} {\bibfnamefont {A.~J.}\ \bibnamefont
  {Leggett}},\ }\href {\doibase 10.1103/PhysRevLett.46.211} {\bibfield
  {journal} {\bibinfo  {journal} {Phys. Rev. Lett.}\ }\textbf {\bibinfo
  {volume} {46}},\ \bibinfo {pages} {211} (\bibinfo {year} {1981})}\BibitemShut
  {NoStop}%
\bibitem [{\citenamefont {Caldeira}\ and\ \citenamefont
  {Leggett}(1983{\natexlab{a}})}]{Caldeira83a}%
  \BibitemOpen
  \bibfield  {author} {\bibinfo {author} {\bibfnamefont {A.~O.}\ \bibnamefont
  {Caldeira}}\ and\ \bibinfo {author} {\bibfnamefont {A.~J.}\ \bibnamefont
  {Leggett}},\ }\href {\doibase https://doi.org/10.1016/0003-4916(83)90202-6}
  {\bibfield  {journal} {\bibinfo  {journal} {Ann. Phys.}\ }\textbf {\bibinfo
  {volume} {149}},\ \bibinfo {pages} {374} (\bibinfo {year}
  {1983}{\natexlab{a}})}\BibitemShut {NoStop}%
\bibitem [{\citenamefont {Caldeira}\ and\ \citenamefont
  {Leggett}(1983{\natexlab{b}})}]{Caldeira83b}%
  \BibitemOpen
  \bibfield  {author} {\bibinfo {author} {\bibfnamefont {A.~O.}\ \bibnamefont
  {Caldeira}}\ and\ \bibinfo {author} {\bibfnamefont {A.~J.}\ \bibnamefont
  {Leggett}},\ }\href {\doibase https://doi.org/10.1016/0378-4371(83)90013-4}
  {\bibfield  {journal} {\bibinfo  {journal} {Physica A}\ }\textbf {\bibinfo
  {volume} {121}},\ \bibinfo {pages} {587} (\bibinfo {year}
  {1983}{\natexlab{b}})}\BibitemShut {NoStop}%
\bibitem [{\citenamefont {Leggett}\ \emph {et~al.}(1987)\citenamefont
  {Leggett}, \citenamefont {Chakravarty}, \citenamefont {Dorsey}, \citenamefont
  {Fisher}, \citenamefont {Garg},\ and\ \citenamefont {Zwerger}}]{Leggett87}%
  \BibitemOpen
  \bibfield  {author} {\bibinfo {author} {\bibfnamefont {A.~J.}\ \bibnamefont
  {Leggett}}, \bibinfo {author} {\bibfnamefont {S.}~\bibnamefont
  {Chakravarty}}, \bibinfo {author} {\bibfnamefont {A.~T.}\ \bibnamefont
  {Dorsey}}, \bibinfo {author} {\bibfnamefont {M.~P.~A.}\ \bibnamefont
  {Fisher}}, \bibinfo {author} {\bibfnamefont {A.}~\bibnamefont {Garg}}, \ and\
  \bibinfo {author} {\bibfnamefont {W.}~\bibnamefont {Zwerger}},\ }\href
  {\doibase 10.1103/RevModPhys.59.1} {\bibfield  {journal} {\bibinfo  {journal}
  {Rev. Mod. Phys.}\ }\textbf {\bibinfo {volume} {59}},\ \bibinfo {pages} {1}
  (\bibinfo {year} {1987})}\BibitemShut {NoStop}%
\bibitem [{\citenamefont {Kane}\ and\ \citenamefont {Fisher}(1992)}]{Kane92}%
  \BibitemOpen
  \bibfield  {author} {\bibinfo {author} {\bibfnamefont {C.~L.}\ \bibnamefont
  {Kane}}\ and\ \bibinfo {author} {\bibfnamefont {M.~P.~A.}\ \bibnamefont
  {Fisher}},\ }\href {\doibase 10.1103/PhysRevB.46.15233} {\bibfield  {journal}
  {\bibinfo  {journal} {Phys. Rev. B}\ }\textbf {\bibinfo {volume} {46}},\
  \bibinfo {pages} {15233} (\bibinfo {year} {1992})}\BibitemShut {NoStop}%
\bibitem [{\citenamefont {Furusaki}\ and\ \citenamefont
  {Nagaosa}(1993)}]{Furusaki93}%
  \BibitemOpen
  \bibfield  {author} {\bibinfo {author} {\bibfnamefont {A.}~\bibnamefont
  {Furusaki}}\ and\ \bibinfo {author} {\bibfnamefont {N.}~\bibnamefont
  {Nagaosa}},\ }\href {\doibase 10.1103/PhysRevB.47.4631} {\bibfield  {journal}
  {\bibinfo  {journal} {Phys. Rev. B}\ }\textbf {\bibinfo {volume} {47}},\
  \bibinfo {pages} {4631} (\bibinfo {year} {1993})}\BibitemShut {NoStop}%
\bibitem [{\citenamefont {Affleck}\ \emph {et~al.}(2001)\citenamefont
  {Affleck}, \citenamefont {Oshikawa},\ and\ \citenamefont
  {Saleur}}]{Affleck01}%
  \BibitemOpen
  \bibfield  {author} {\bibinfo {author} {\bibfnamefont {I.}~\bibnamefont
  {Affleck}}, \bibinfo {author} {\bibfnamefont {M.}~\bibnamefont {Oshikawa}}, \
  and\ \bibinfo {author} {\bibfnamefont {H.}~\bibnamefont {Saleur}},\ }\href
  {\doibase https://doi.org/10.1016/S0550-3213(00)00499-5} {\bibfield
  {journal} {\bibinfo  {journal} {Nucl. Phys. B}\ }\textbf {\bibinfo {volume}
  {594}},\ \bibinfo {pages} {535} (\bibinfo {year} {2001})}\BibitemShut
  {NoStop}%
\bibitem [{\citenamefont {Giamarchi}(2003)}]{GT03}%
  \BibitemOpen
  \bibfield  {author} {\bibinfo {author} {\bibfnamefont {T.}~\bibnamefont
  {Giamarchi}},\ }\href@noop {} {\emph {\bibinfo {title} {Quantum physics in
  one dimension}}},\ Vol.\ \bibinfo {volume} {121}\ (\bibinfo  {publisher}
  {Clarendon press},\ \bibinfo {year} {2003})\BibitemShut {NoStop}%
\bibitem [{\citenamefont {Werner}\ and\ \citenamefont
  {Troyer}(2005)}]{Werner05a}%
  \BibitemOpen
  \bibfield  {author} {\bibinfo {author} {\bibfnamefont {P.}~\bibnamefont
  {Werner}}\ and\ \bibinfo {author} {\bibfnamefont {M.}~\bibnamefont
  {Troyer}},\ }\href {\doibase 10.1103/PhysRevLett.95.060201} {\bibfield
  {journal} {\bibinfo  {journal} {Phys. Rev. Lett.}\ }\textbf {\bibinfo
  {volume} {95}},\ \bibinfo {pages} {060201} (\bibinfo {year}
  {2005})}\BibitemShut {NoStop}%
\bibitem [{\citenamefont {Oshikawa}\ \emph {et~al.}(2006)\citenamefont
  {Oshikawa}, \citenamefont {Chamon},\ and\ \citenamefont
  {Affleck}}]{Oshikawa06}%
  \BibitemOpen
  \bibfield  {author} {\bibinfo {author} {\bibfnamefont {M.}~\bibnamefont
  {Oshikawa}}, \bibinfo {author} {\bibfnamefont {C.}~\bibnamefont {Chamon}}, \
  and\ \bibinfo {author} {\bibfnamefont {I.}~\bibnamefont {Affleck}},\ }\href
  {\doibase 10.1088/1742-5468/2006/02/p02008} {\bibfield  {journal} {\bibinfo
  {journal} {J. Stat. Mech.}\ }\textbf {\bibinfo {volume} {2006}},\ \bibinfo
  {pages} {P02008} (\bibinfo {year} {2006})}\BibitemShut {NoStop}%
\bibitem [{\citenamefont {Refael}\ \emph {et~al.}(2007)\citenamefont {Refael},
  \citenamefont {Demler}, \citenamefont {Oreg},\ and\ \citenamefont
  {Fisher}}]{RG07}%
  \BibitemOpen
  \bibfield  {author} {\bibinfo {author} {\bibfnamefont {G.}~\bibnamefont
  {Refael}}, \bibinfo {author} {\bibfnamefont {E.}~\bibnamefont {Demler}},
  \bibinfo {author} {\bibfnamefont {Y.}~\bibnamefont {Oreg}}, \ and\ \bibinfo
  {author} {\bibfnamefont {D.~S.}\ \bibnamefont {Fisher}},\ }\href {\doibase
  10.1103/PhysRevB.75.014522} {\bibfield  {journal} {\bibinfo  {journal} {Phys.
  Rev. B}\ }\textbf {\bibinfo {volume} {75}},\ \bibinfo {pages} {014522}
  (\bibinfo {year} {2007})}\BibitemShut {NoStop}%
\bibitem [{\citenamefont {Hur}(2008)}]{HUR20082208}%
  \BibitemOpen
  \bibfield  {author} {\bibinfo {author} {\bibfnamefont {K.~L.}\ \bibnamefont
  {Hur}},\ }\href {\doibase https://doi.org/10.1016/j.aop.2007.12.003}
  {\bibfield  {journal} {\bibinfo  {journal} {Ann. Phys.}\ }\textbf {\bibinfo
  {volume} {323}},\ \bibinfo {pages} {2208} (\bibinfo {year}
  {2008})}\BibitemShut {NoStop}%
\bibitem [{\citenamefont {Halperin}\ \emph {et~al.}(2011)\citenamefont
  {Halperin}, \citenamefont {Refael},\ and\ \citenamefont
  {Demler}}]{Halperin11_2}%
  \BibitemOpen
  \bibfield  {author} {\bibinfo {author} {\bibfnamefont {B.~I.}\ \bibnamefont
  {Halperin}}, \bibinfo {author} {\bibfnamefont {G.}~\bibnamefont {Refael}}, \
  and\ \bibinfo {author} {\bibfnamefont {E.}~\bibnamefont {Demler}},\
  }\href@noop {} {\emph {\bibinfo {title} {Resistance in superconductors}}}\
  (\bibinfo  {publisher} {World Scientific},\ \bibinfo {year} {2011})\ pp.\
  \bibinfo {pages} {185--226}\BibitemShut {NoStop}%
\bibitem [{\citenamefont {Weiss}(2008)}]{Weiss12}%
  \BibitemOpen
  \bibfield  {author} {\bibinfo {author} {\bibfnamefont {U.}~\bibnamefont
  {Weiss}},\ }\href {\doibase 10.1142/6738} {\emph {\bibinfo {title} {Quantum
  Dissipative Systems}}},\ \bibinfo {edition} {3rd}\ ed.\ (\bibinfo
  {publisher} {WORLD SCIENTIFIC},\ \bibinfo {address} {Singapore},\ \bibinfo
  {year} {2008})\BibitemShut {NoStop}%
\bibitem [{\citenamefont {Kapitulnik}\ \emph {et~al.}(2019)\citenamefont
  {Kapitulnik}, \citenamefont {Kivelson},\ and\ \citenamefont {Spivak}}]{KA19}%
  \BibitemOpen
  \bibfield  {author} {\bibinfo {author} {\bibfnamefont {A.}~\bibnamefont
  {Kapitulnik}}, \bibinfo {author} {\bibfnamefont {S.~A.}\ \bibnamefont
  {Kivelson}}, \ and\ \bibinfo {author} {\bibfnamefont {B.}~\bibnamefont
  {Spivak}},\ }\href {\doibase 10.1103/RevModPhys.91.011002} {\bibfield
  {journal} {\bibinfo  {journal} {Rev. Mod. Phys.}\ }\textbf {\bibinfo {volume}
  {91}},\ \bibinfo {pages} {011002} (\bibinfo {year} {2019})}\BibitemShut
  {NoStop}%
\bibitem [{\citenamefont {Anderson}\ \emph {et~al.}(1970)\citenamefont
  {Anderson}, \citenamefont {Yuval},\ and\ \citenamefont
  {Hamann}}]{anderson_exact_1970}%
  \BibitemOpen
  \bibfield  {author} {\bibinfo {author} {\bibfnamefont {P.~W.}\ \bibnamefont
  {Anderson}}, \bibinfo {author} {\bibfnamefont {G.}~\bibnamefont {Yuval}}, \
  and\ \bibinfo {author} {\bibfnamefont {D.~R.}\ \bibnamefont {Hamann}},\
  }\href {\doibase 10.1103/PhysRevB.1.4464} {\bibfield  {journal} {\bibinfo
  {journal} {Phys. Rev. B}\ }\textbf {\bibinfo {volume} {1}},\ \bibinfo {pages}
  {4464} (\bibinfo {year} {1970})}\BibitemShut {NoStop}%
\bibitem [{\citenamefont {Guinea}\ \emph
  {et~al.}(1985{\natexlab{a}})\citenamefont {Guinea}, \citenamefont {Hakim},\
  and\ \citenamefont {Muramatsu}}]{GuF85}%
  \BibitemOpen
  \bibfield  {author} {\bibinfo {author} {\bibfnamefont {F.}~\bibnamefont
  {Guinea}}, \bibinfo {author} {\bibfnamefont {V.}~\bibnamefont {Hakim}}, \
  and\ \bibinfo {author} {\bibfnamefont {A.}~\bibnamefont {Muramatsu}},\ }\href
  {\doibase 10.1103/PhysRevB.32.4410} {\bibfield  {journal} {\bibinfo
  {journal} {Phys. Rev. B}\ }\textbf {\bibinfo {volume} {32}},\ \bibinfo
  {pages} {4410} (\bibinfo {year} {1985}{\natexlab{a}})}\BibitemShut {NoStop}%
\bibitem [{\citenamefont {Schmid}(1983)}]{Schmid83}%
  \BibitemOpen
  \bibfield  {author} {\bibinfo {author} {\bibfnamefont {A.}~\bibnamefont
  {Schmid}},\ }\href {\doibase 10.1103/PhysRevLett.51.1506} {\bibfield
  {journal} {\bibinfo  {journal} {Phys. Rev. Lett.}\ }\textbf {\bibinfo
  {volume} {51}},\ \bibinfo {pages} {1506} (\bibinfo {year}
  {1983})}\BibitemShut {NoStop}%
\bibitem [{\citenamefont {Bulgadaev}(1984)}]{Bulgadaev84}%
  \BibitemOpen
  \bibfield  {author} {\bibinfo {author} {\bibfnamefont {S.}~\bibnamefont
  {Bulgadaev}},\ }\href {http://jetpletters.ru/ps/1289/article_19477.shtml}
  {\bibfield  {journal} {\bibinfo  {journal} {JETP Lett.}\ }\textbf {\bibinfo
  {volume} {39}},\ \bibinfo {pages} {264} (\bibinfo {year} {1984})}\BibitemShut
  {NoStop}%
\bibitem [{\citenamefont {Guinea}\ \emph
  {et~al.}(1985{\natexlab{b}})\citenamefont {Guinea}, \citenamefont {Hakim},\
  and\ \citenamefont {Muramatsu}}]{Guinea85}%
  \BibitemOpen
  \bibfield  {author} {\bibinfo {author} {\bibfnamefont {F.}~\bibnamefont
  {Guinea}}, \bibinfo {author} {\bibfnamefont {V.}~\bibnamefont {Hakim}}, \
  and\ \bibinfo {author} {\bibfnamefont {A.}~\bibnamefont {Muramatsu}},\ }\href
  {\doibase 10.1103/PhysRevLett.54.263} {\bibfield  {journal} {\bibinfo
  {journal} {Phys. Rev. Lett.}\ }\textbf {\bibinfo {volume} {54}},\ \bibinfo
  {pages} {263} (\bibinfo {year} {1985}{\natexlab{b}})}\BibitemShut {NoStop}%
\bibitem [{\citenamefont {Fisher}\ and\ \citenamefont
  {Zwerger}(1985)}]{Fisher85}%
  \BibitemOpen
  \bibfield  {author} {\bibinfo {author} {\bibfnamefont {M.~P.~A.}\
  \bibnamefont {Fisher}}\ and\ \bibinfo {author} {\bibfnamefont
  {W.}~\bibnamefont {Zwerger}},\ }\href {\doibase 10.1103/PhysRevB.32.6190}
  {\bibfield  {journal} {\bibinfo  {journal} {Phys. Rev. B}\ }\textbf {\bibinfo
  {volume} {32}},\ \bibinfo {pages} {6190} (\bibinfo {year}
  {1985})}\BibitemShut {NoStop}%
\bibitem [{\citenamefont {Sch\"{o}n}\ and\ \citenamefont
  {Zaikin}(1990)}]{Schon90}%
  \BibitemOpen
  \bibfield  {author} {\bibinfo {author} {\bibfnamefont {G.}~\bibnamefont
  {Sch\"{o}n}}\ and\ \bibinfo {author} {\bibfnamefont {A.}~\bibnamefont
  {Zaikin}},\ }\href {\doibase https://doi.org/10.1016/0370-1573(90)90156-V}
  {\bibfield  {journal} {\bibinfo  {journal} {Phys. Rep.}\ }\textbf {\bibinfo
  {volume} {198}},\ \bibinfo {pages} {237} (\bibinfo {year}
  {1990})}\BibitemShut {NoStop}%
\bibitem [{\citenamefont {Feynman}\ \emph {et~al.}(2010)\citenamefont
  {Feynman}, \citenamefont {Hibbs},\ and\ \citenamefont
  {Styer}}]{feynman_quantum_2010}%
  \BibitemOpen
  \bibfield  {author} {\bibinfo {author} {\bibfnamefont {R.~P.}\ \bibnamefont
  {Feynman}}, \bibinfo {author} {\bibfnamefont {A.~R.}\ \bibnamefont {Hibbs}},
  \ and\ \bibinfo {author} {\bibfnamefont {D.~F.}\ \bibnamefont {Styer}},\
  }\href@noop {} {\emph {\bibinfo {title} {Quantum mechanics and path
  integrals}}},\ \bibinfo {edition} {emended ed}\ ed.,\ Dover books on physics\
  (\bibinfo  {publisher} {Dover publ},\ \bibinfo {address} {Mineola},\ \bibinfo
  {year} {2010})\BibitemShut {NoStop}%
\bibitem [{\citenamefont {Altland}\ and\ \citenamefont
  {Simons}(2010)}]{altland_condensed_2010}%
  \BibitemOpen
  \bibfield  {author} {\bibinfo {author} {\bibfnamefont {A.}~\bibnamefont
  {Altland}}\ and\ \bibinfo {author} {\bibfnamefont {B.~D.}\ \bibnamefont
  {Simons}},\ }\href {http://site.ebrary.com/id/10406686} {\emph {\bibinfo
  {title} {Condensed matter field theory}}}\ (\bibinfo {year} {2010})\ \bibinfo
  {note} {oCLC: 651601524}\BibitemShut {NoStop}%
\bibitem [{\citenamefont {Peskin}\ and\ \citenamefont
  {Schroeder}(2019)}]{peskin_introduction_2019}%
  \BibitemOpen
  \bibfield  {author} {\bibinfo {author} {\bibfnamefont {M.~E.}\ \bibnamefont
  {Peskin}}\ and\ \bibinfo {author} {\bibfnamefont {D.~V.}\ \bibnamefont
  {Schroeder}},\ }\href@noop {} {\emph {\bibinfo {title} {An introduction to
  quantum field theory}}},\ The advanced book program\ (\bibinfo  {publisher}
  {CRC Press, Taylor \& Francis Group},\ \bibinfo {address} {Boca Raton London
  New York},\ \bibinfo {year} {2019})\BibitemShut {NoStop}%
\bibitem [{\citenamefont {Wetterich}(1993)}]{wetterich_exact_1993}%
  \BibitemOpen
  \bibfield  {author} {\bibinfo {author} {\bibfnamefont {C.}~\bibnamefont
  {Wetterich}},\ }\href {\doibase 10.1016/0370-2693(93)90726-X} {\bibfield
  {journal} {\bibinfo  {journal} {Phys. Lett. B}\ }\textbf {\bibinfo {volume}
  {301}},\ \bibinfo {pages} {90} (\bibinfo {year} {1993})}\BibitemShut
  {NoStop}%
\bibitem [{\citenamefont {Aoki}\ and\ \citenamefont
  {Horikoshi}(2002)}]{aoki_nonperturbative_2002}%
  \BibitemOpen
  \bibfield  {author} {\bibinfo {author} {\bibfnamefont {K.-I.}\ \bibnamefont
  {Aoki}}\ and\ \bibinfo {author} {\bibfnamefont {A.}~\bibnamefont
  {Horikoshi}},\ }\href {\doibase 10.1103/PhysRevA.66.042105} {\bibfield
  {journal} {\bibinfo  {journal} {Phys. Rev. A}\ }\textbf {\bibinfo {volume}
  {66}},\ \bibinfo {pages} {042105} (\bibinfo {year} {2002})}\BibitemShut
  {NoStop}%
\bibitem [{\citenamefont {Kovacs}\ \emph {et~al.}(2017)\citenamefont {Kovacs},
  \citenamefont {Fazekas}, \citenamefont {Nagy},\ and\ \citenamefont
  {Sailer}}]{kovacs_quantum-classical_2017}%
  \BibitemOpen
  \bibfield  {author} {\bibinfo {author} {\bibfnamefont {J.}~\bibnamefont
  {Kovacs}}, \bibinfo {author} {\bibfnamefont {B.}~\bibnamefont {Fazekas}},
  \bibinfo {author} {\bibfnamefont {S.}~\bibnamefont {Nagy}}, \ and\ \bibinfo
  {author} {\bibfnamefont {K.}~\bibnamefont {Sailer}},\ }\href {\doibase
  10.1016/j.aop.2016.12.010} {\bibfield  {journal} {\bibinfo  {journal} {Ann.
  Phys.}\ }\textbf {\bibinfo {volume} {376}},\ \bibinfo {pages} {372} (\bibinfo
  {year} {2017})}\BibitemShut {NoStop}%
\bibitem [{\citenamefont {Aslangul}\ \emph {et~al.}(1985)\citenamefont
  {Aslangul}, \citenamefont {Pottier},\ and\ \citenamefont
  {Saint-James}}]{aslangul_quantum_1985}%
  \BibitemOpen
  \bibfield  {author} {\bibinfo {author} {\bibfnamefont {C.}~\bibnamefont
  {Aslangul}}, \bibinfo {author} {\bibfnamefont {N.}~\bibnamefont {Pottier}}, \
  and\ \bibinfo {author} {\bibfnamefont {D.}~\bibnamefont {Saint-James}},\
  }\href {\doibase 10.1016/0375-9601(85)90570-5} {\bibfield  {journal}
  {\bibinfo  {journal} {Phys. Lett. A}\ }\textbf {\bibinfo {volume} {111}},\
  \bibinfo {pages} {175} (\bibinfo {year} {1985})}\BibitemShut {NoStop}%
\bibitem [{\citenamefont {Zaikin}\ and\ \citenamefont
  {Panyukov}(1987)}]{zaikin_dynamics_1987}%
  \BibitemOpen
  \bibfield  {author} {\bibinfo {author} {\bibfnamefont {A.}~\bibnamefont
  {Zaikin}}\ and\ \bibinfo {author} {\bibfnamefont {S.}~\bibnamefont
  {Panyukov}},\ }\href {\doibase 10.1016/0375-9601(87)90677-3} {\bibfield
  {journal} {\bibinfo  {journal} {Phys. Lett. A}\ }\textbf {\bibinfo {volume}
  {120}},\ \bibinfo {pages} {306} (\bibinfo {year} {1987})}\BibitemShut
  {NoStop}%
\bibitem [{\citenamefont {Nakamura}\ \emph {et~al.}(1999)\citenamefont
  {Nakamura}, \citenamefont {Pashkin},\ and\ \citenamefont {Tsai}}]{YN99}%
  \BibitemOpen
  \bibfield  {author} {\bibinfo {author} {\bibfnamefont {Y.}~\bibnamefont
  {Nakamura}}, \bibinfo {author} {\bibfnamefont {Y.~A.}\ \bibnamefont
  {Pashkin}}, \ and\ \bibinfo {author} {\bibfnamefont {J.}~\bibnamefont
  {Tsai}},\ }\href {https://www.nature.com/articles/19718} {\bibfield
  {journal} {\bibinfo  {journal} {Nature}\ }\textbf {\bibinfo {volume} {398}},\
  \bibinfo {pages} {786} (\bibinfo {year} {1999})}\BibitemShut {NoStop}%
\bibitem [{\citenamefont {Makhlin}\ \emph {et~al.}(2001)\citenamefont
  {Makhlin}, \citenamefont {Sch\"on},\ and\ \citenamefont {Shnirman}}]{MY01}%
  \BibitemOpen
  \bibfield  {author} {\bibinfo {author} {\bibfnamefont {Y.}~\bibnamefont
  {Makhlin}}, \bibinfo {author} {\bibfnamefont {G.}~\bibnamefont {Sch\"on}}, \
  and\ \bibinfo {author} {\bibfnamefont {A.}~\bibnamefont {Shnirman}},\ }\href
  {\doibase 10.1103/RevModPhys.73.357} {\bibfield  {journal} {\bibinfo
  {journal} {Rev. Mod. Phys.}\ }\textbf {\bibinfo {volume} {73}},\ \bibinfo
  {pages} {357} (\bibinfo {year} {2001})}\BibitemShut {NoStop}%
\bibitem [{\citenamefont {Ashida}\ \emph
  {et~al.}(2020{\natexlab{b}})\citenamefont {Ashida}, \citenamefont {Gong},\
  and\ \citenamefont {Ueda}}]{YAreview}%
  \BibitemOpen
  \bibfield  {author} {\bibinfo {author} {\bibfnamefont {Y.}~\bibnamefont
  {Ashida}}, \bibinfo {author} {\bibfnamefont {Z.}~\bibnamefont {Gong}}, \ and\
  \bibinfo {author} {\bibfnamefont {M.}~\bibnamefont {Ueda}},\ }\href
  {https://doi.org/10.1080/00018732.2021.1876991} {\bibfield  {journal}
  {\bibinfo  {journal} {Adv. Phys.}\ }\textbf {\bibinfo {volume} {69}},\
  \bibinfo {pages} {249} (\bibinfo {year} {2020}{\natexlab{b}})}\BibitemShut
  {NoStop}%
\bibitem [{\citenamefont {Anders}\ \emph {et~al.}(2007)\citenamefont {Anders},
  \citenamefont {Bulla},\ and\ \citenamefont {Vojta}}]{AFB07}%
  \BibitemOpen
  \bibfield  {author} {\bibinfo {author} {\bibfnamefont {F.~B.}\ \bibnamefont
  {Anders}}, \bibinfo {author} {\bibfnamefont {R.}~\bibnamefont {Bulla}}, \
  and\ \bibinfo {author} {\bibfnamefont {M.}~\bibnamefont {Vojta}},\ }\href
  {\doibase 10.1103/PhysRevLett.98.210402} {\bibfield  {journal} {\bibinfo
  {journal} {Phys. Rev. Lett.}\ }\textbf {\bibinfo {volume} {98}},\ \bibinfo
  {pages} {210402} (\bibinfo {year} {2007})}\BibitemShut {NoStop}%
\bibitem [{\citenamefont {Bulla}\ \emph {et~al.}(2008)\citenamefont {Bulla},
  \citenamefont {Costi},\ and\ \citenamefont {Pruschke}}]{RB08}%
  \BibitemOpen
  \bibfield  {author} {\bibinfo {author} {\bibfnamefont {R.}~\bibnamefont
  {Bulla}}, \bibinfo {author} {\bibfnamefont {T.~A.}\ \bibnamefont {Costi}}, \
  and\ \bibinfo {author} {\bibfnamefont {T.}~\bibnamefont {Pruschke}},\ }\href
  {\doibase 10.1103/RevModPhys.80.395} {\bibfield  {journal} {\bibinfo
  {journal} {Rev. Mod. Phys.}\ }\textbf {\bibinfo {volume} {80}},\ \bibinfo
  {pages} {395} (\bibinfo {year} {2008})}\BibitemShut {NoStop}%
\end{thebibliography}%

\end{document}